\documentclass{aa}  

\usepackage{graphicx}
\usepackage{lscape}
\usepackage{rotating}
\usepackage[labelfont=bf]{caption}
\usepackage{subfig}
\usepackage{natbib}
\usepackage{hyperref}  
\usepackage{definitions}

\graphicspath{{figures/}}
\pdfoutput=1
\usepackage{txfonts}
\usepackage{hyperref}
\hypersetup{colorlinks=true,citecolor=blue,linkcolor=black}
\begin{document} 

   \title{Classifying the embedded young stellar population in Perseus and Taurus \& the LOMASS database}
   
   %\subtitle{Separating the embedded from the non-embedded using molecular line and dust continuum maps}

   \author{M.T. Carney \inst{1}, U.A. Y{\i}ld{\i}z \inst{1,2}, J.C. Mottram \inst{1}, E.F. van Dishoeck \inst{1,3}, J. Ramchandani \inst{1},  J.K. J{{\o}}rgensen \inst{4,5}
          }
   
   \institute{Leiden Observatory, Leiden University, PO Box 9513, 2300 RA, The Netherlands. \\
              \email{masoncarney@strw.leidenuniv.nl}
         \and
            Jet Propulsion Laboratory, California Institute of Technology, 4800 Oak Grove Drive, Pasadena CA, 91109, USA
         \and
	     Max-Planck Institut fur Extraterrestriche Physik, Giessenbachstrasse 1, 85748 Garching, Germany
	 \and
	     Niels Bohr Institute, University of Copenhagen, Juliane Maries Vej 30, DK-2100 Copenhagen {{\O}}., Denmark
	 \and
	     Centre for Star and Planet Formation, Natural History Museum of Denmark, University of Copenhagen, {{\O}}ster Voldgade 5-7, DK-1350 Copenhagen K., Denmark
             }

   \date{Received 14 April 2015 / Accepted 11 September 2015 \\
	 }

% \abstract{}{}{}{}{} 
% 5 {} token are mandatory
 
  \abstract
   {The classification of young stellar objects (YSOs) is typically done using the 
   infrared spectral slope or bolometric temperature, 
   but either can result in contamination of samples. More accurate methods to 
   determine the evolutionary stage of YSOs will improve the reliability of statistics for 
   the embedded YSO population and provide more robust stage lifetimes.}
   {We aim to separate the truly embedded YSOs from more 
   evolved sources.}
   {Maps of HCO$^{+}$ $J$=4--3 and 
   C$^{18}$O $J$=3--2 were observed with HARP on 
   the James Clerk Maxwell Telescope (JCMT) for a sample of 56 candidate YSOs in Perseus and Taurus
   in order to characterize the presence and morphology of emission from 
   high density ($n_{\mathrm{crit}}$ >10$^{6}$ cm$^{-3}$) and high column density gas, respectively.
   These are supplemented with archival dust continuum maps observed with SCUBA on the JCMT
   and {\it Herschel} PACS to compare the morphology of the gas and dust in the 
   protostellar envelopes. The spatial concentration of 
   HCO$^{+}$ $J$=4--3 and 850 $\mu$m dust emission are used to classify 
   the embedded nature of YSOs.}
   {Approximately 30\% of Class 0+I sources in Perseus and Taurus are not Stage I, 
   but are likely to be more evolved Stage II pre-main sequence (PMS) stars with 
   disks. An additional 16\% are confused sources with an uncertain evolutionary 
   stage. Outflows are found to make a negligible contribution
   to the integrated HCO$^+$ intensity for the majority of sources in this study.
   }
   {Separating classifications by cloud reveals that a high percentage of 
   the Class 0+I sources in the Perseus star forming region are truly embedded Stage I sources 
   (71\%), while the Taurus cloud hosts a majority of evolved PMS stars 
   with disks (68\%). The concentration factor method is useful to correct misidentified 
   embedded YSOs, yielding higher accuracy for YSO population statistics and Stage timescales. 
   Current estimates (0.54 Myr) may overpredict the Stage I lifetime
   on the order of 30\%, resulting in timescales of 0.38 Myr for the embedded phase.}

  %\keywords{}
   \authorrunning{Carney, M.~T. et al.}
   \titlerunning{Embedded protostars in Perseus and Taurus}
   \maketitle
%
%________________________________________________________________

\section{Introduction}
\label{sec:intro}

Low-mass young stellar objects (YSOs) have traditionally been
classified based on their observed infrared slope 
($\alpha_{\mathrm{IR}}$) in the wavelength range from 2 to 20 
$\mu$m \citep{Lada1984} or their bolometric temperature 
\citep[$T_\mathrm{bol}$,][]{Myers1993}. In the earliest phases of 
star formation, most of the emission appears at far-IR and
submillimeter wavelengths. Such observationally based evolutionary 
schemes start with Class 0 \citep{Andre2000}, the earliest phase of low-mass star formation, where the protostar
is still deeply embedded. This phase only lasts a short time, $\sim$0.15~Myr according to
\citet{Evans2009}. Class 0 is followed by Class I, with a combined Class 0+I halflife of $\sim$0.5 Myr 
representing the embedded phase of the protostar, then Class II, and Class III 
as the spectral energy distribution (SED) of the source shifts towards
optical wavelengths and gradually loses intensity at far-IR/sub-mm wavelengths.
The term ``Class'' refers only to an observational phase defined by 
$\alpha_{\mathrm{IR}}$ and/or $T_\mathrm{bol}$, whereas
``Stage'' refers to a phase of low-mass star formation based on physical parameters.

Physical definitions of YSOs use
parameters such as star, disk, and envelope mass ratios to determine
the evolutionary stage \citep{Adams1987,Whitney2003a, Whitney2003b,
  Robitaille2006}. In Stage I, the protostar is surrounded by a
collapsing envelope and a circumstellar disk through which material is
accreted onto the growing star ($M_{\mathrm{env}} > M_{\mathrm{star}}$ at very early times). 
This stage is critical for the subsequent evolution
since the mass of the star and the physical and chemical structure of
the circumstellar disk are determined here. These protostars also
power bipolar outflows with a very high degree of collimation, and
there is evidence of shock processing of molecular gas even in cases of very
low stellar luminosity \citep{Tafalla2000,Dunham2008}. As the YSO evolves
the envelope is dispersed by the outflow, the disk grows, the object becomes brighter at
IR wavelengths, and the outflows diminish in force.  Stage I
embedded sources ($M_{\mathrm{disk}}/M_{\mathrm{envelope}} < 2$ and
$[M_{\mathrm{disk}} + M_{\mathrm{envelope}}] < M_{\mathrm{star}}$)
differ from Stage II gas-rich classical T Tauri stars
($M_{\mathrm{envelope}} \lesssim 0.2 M_{\odot} \ \mathrm{and} \
M_{\mathrm{disk}}/M_{\mathrm{star}} \ll 1$), which are more evolved
pre-main sequence (PMS) objects and often have an associated circumstellar disk.
Stage I sources may also be distinguished from Stage II sources through
enhanced continuum veiling of the young central star, as seen in their near-infrared spectra
\citep[e.g.,][]{Casali1992, Greene1996, White2004}.

Lifetimes of astronomical objects are generally inferred by counting
the number of objects in each phase and taking the age of one of these
phases to be known \citep[e.g.,][]{Evans2009,Mottram2011}. In star 
formation, the reference for low-mass protostars is usually the lifetime 
of the Class II T Tauri phase, 
generally corresponding to Stage II physical parameters, which is $\sim$2--3
Myr based on current estimates \citep{Haisch2001,Spezzi2008,Muench2007,Evans2009}. However, this method
assumes that the classification methods for all phases are
correct and complete. Previous studies have shown that the above-mentioned 
criteria based on $\alpha_{\mathrm{IR}}$ or $T_\mathrm{bol}$ lead
to samples with significant contamination: some Class 0+I sources turn
out to be edge-on Stage II PMS stars with disks, or sources obscured by cloud material due to
projection effects \citep{Brandner2000, Lahuis2006}.  Models indicate
that such confusion may be quite common and can include up to 30\% of
the total embedded sources \citep{Whitney2003a, Robitaille2006,
  Crapsi2008}. Comparison between single-dish and interferometric
(sub-)mm observations is able to correctly identify such sources
\citep{Lommen2008,Crapsi2008}. However, the capabilities of current
interferometers make this method quite time consuming, and it is
limited to only a small number of the brightest sub-mm sources.

Using the 345 GHz Heterodyne Array Receiver Program
\citep[HARP,][]{Buckle2009} and Submillimeter Common-User Bolometer Array
\citep[SCUBA,][]{Holland1999} instruments on the James Clerk Maxwell Telescope
\footnote{The James Clerk Maxwell Telescope has historically
  been operated by the Joint Astronomy Centre on behalf of the Science
  and Technology Facilities Council of the United Kingdom, the
  National Research Council of Canada and the Netherlands Organisation
  for Scientific Research.} (JCMT), \citet{vanKempen2009} identified
truly embedded sources without the need for interferometry, by instead
comparing maps of HCO$^{+}$ $J$=4--3, C$^{18}$O $J$=3--2, and 850~$\mu$m dust
emission over an approximately 2$\arcmin$ region around each
source at a spatial resolution of $\sim$15$\arcsec$. In 
these maps, HCO$^{+}$ 4--3 ($n_{\mathrm{crit}}$ $>$ 10$^{6}$ cm$^{-3}$ ) emission traces the presence
of dense gas in the envelope , and C$^{18}$O
3--2 ($n_{\mathrm{crit}}$ >10$^{4}$ cm$^{-3}$) traces the column density. HCO$^{+}$ 4--3 can also be an 
effective tracer of outflows in early phase YSOs and over large spatial regions
\citep{Walker-Smith2014}. Both the absolute line strengths and
their variation over the region (concentration) play a crucial role in
the identification of embedded sources. \citet{vanKempen2009} applied this method to all
sources in Ophiuchus that were thought to be Class I based on the
traditional criteria, and they found that all tracers are centrally concentrated 
for truly embedded sources, with the possible exception of
C$^{18}$O. Some sources appear unresolved, but are still bright in
both HCO$^{+}$ 4--3 and C$^{18}$O 3--2 ($>$1 K km s$^{-1}$).  In
contrast, edge-on disks or obscured sources have weak
point-like and/or flat sub-mm dust emission, respectively, 
when observed with SCUBA. In molecular emission they have a
flat C$^{18}$O 3--2 emission distribution (if any) and/or a
HCO$^{+}$ 4--3 intensity of at most 0.3 K km s$^{-1}$. By using this
method, 16 out of 38 Class I and flat SED (intermediate between Class I and II) sources were
found to be truly embedded Stage I YSOs in Ophiuchus, whereas the
remainder were characterized as Stage II evolved PMS stars with disks 
and/or obscured sources \citep{vanKempen2009}. An
earlier single pointing HCO$^+$ 3--2 study by \citet{Hogerheijde1997}
of a number of Taurus Class I sources also found a significant
fraction of them to be Stage II objects.

These new statistics clearly affect the lifetime of the embedded
phase, implying that the Stage I lifetime has been overestimated, at
least for Ophiuchus. This paper aims to apply these same criteria to
sources in the Perseus and Taurus star-forming regions. In addition,
the robustness of using just a single HCO$^+$ spectrum at the source
position versus complete maps to classify sources is investigated:
only central source spectra are available for the bulk of the many
hundreds of Class 0+I and flat-spectrum  sources found in nearby
star-forming regions \citep{Heiderman2015}.  

The structure of
the paper is as follows: Section~\ref{sec:obs} outlines the telescopes
and instruments used during observations and the data collected from
previous surveys. Section~\ref{sec:res} provides details on the
properties of central spectra extracted from the observations and
categorizes the morphology of the spatial data at each wavelength. 
Section~\ref{sec:class} describes the method to classify the objects
in the sample as truly embedded YSOs versus more evolved sources.
Section~\ref{sec:lomass} describes the LOMASS database.
Section~\ref{sec:concl} provides the conclusions.

%__________________________________________________________________

\section{Observations}
\label{sec:obs}

\begin{table*}[!htbp]
\caption{Source coordinates, infrared slope values, and central spectra properties.}
\label{tab:data_props}
\centering 
\resizebox{!}{9cm}{
\begin{tabular}{l l c c c c c c c c c c c c c}     % 7 columns 
\hline\hline

 &  &  &  &  &  &  &  & \multicolumn{3}{c}{HCO$^{+}$ 4--3} & \multicolumn{3}{c}{C$^{18}$O 3--2} \\ 
Object Name & Alternate Name(s)\tablefootmark{a} & Cloud & RA [J2000] & DEC [J2000] & Class\tablefootmark{b} & $\alpha_{\mathrm{IR}}$\tablefootmark{c} & $V_{\mathrm{lsr}}$\tablefootmark{d} & $\int T_{\mathrm{{mb}}} \mathrm{d}V$ & $T_{\mathrm{mb}}$ & ${\rm FWHM}$ & $\int T_{\mathrm{{mb}}} \mathrm{d}V$ & $T_{\mathrm{mb}}$ & ${\rm FWHM}$ \\
 &  &  & [hh:mm:ss] & [dd:mm:ss] &  &  & [km s$^{-1}$] & [K km s$^{-1}$] & [K] & [km s$^{-1}$] & [K km s$^{-1}$] & [K] & [km s$^{-1}$] \\
 \hline
 
J032522.3+304513& L1448-IRS 2, Pers01 & Perseus & 03:25:22.4& 30:45:13.6 & I (I) & 2.3 (2.2) & 4.1 & 2.9$\pm$0.09 & 1.9 & 1.5 & 2.4$\pm$0.06 & 2.8 & 0.7 \\  
J032536.4+304523& L1448-N, Pers02 & Perseus & 03:25:36.5 & 30:45:23.2 & I (I) & 2.6 (2.6) & 4.6 & 14.8$\pm$0.1 & 6.2 & 2.2 & 6.7$\pm$0.2 & 4.0 & 1.4 \\  
J032637.4+301528& Pers-4, Pers04 & Perseus & 03:26:37.5 & 30:15:28.2 & I (I) & 1.1 (1.0) & 5.2 & 1.1$\pm$0.04 & 1.1 & 0.5 & 0.8$\pm$0.08 & 1.1 & 0.8 \\  
J032738.2+301358& L1455-FIR 2 & Perseus & 03:27:38.3 & 30:13:58.5 & F (II) & -0.2 (-0.4) & 4.4 & 1.7$\pm$0.06 & 1.0 & 1.7 & 0.8$\pm$0.09 & 0.7 & 1.0 \\  
J032743.2+301228& L1455-IRS 4 & Perseus & 03:27:43.3 & 30:12:28.9 & I (I) & 2.4 (2.2) & 5.0 & 2.2$\pm$0.06 & 2.0 & 0.9 & 1.2$\pm$0.1 & 1.3 & 1.2 \\  
J032832.5+311104& Pers-9 & Perseus & 03:28:32.6 & 31:11:04.8 & I (I) & 0.8 (0.5) & 7.5 & 2.5$\pm$0.1 & 1.4 & 1.9 & E/A & E/A & E/A \\  
J032834.5+310705& Pers-10 & Perseus & 03:28:34.5 & 31:07:05.5 & I (I) & 0.5 (0.3)  & $-$ & E/A & E/A & E/A & A & A & A \\  
J032837.1+311328& IRAS03255+3103, Pers05 & Perseus & 03:28:37.1 & 31:13:28.3 & $-$ & $-$ & 7.3 & 4.0$\pm$0.08 & 4.0 & 1.2 & 2.1$\pm$0.08 & 3.0 & 0.7 \\  
J032839.1+310601& Pers-12 & Perseus & 03:28:39.1 & 31:06:01.6 & I (I) & 1.7 (1.6) & 7.2 & 0.6$\pm$0.05 & 1.2 & 0.5 & 0.5$\pm$0.07 & 0.8 & 0.4 \\  
J032840.6+311756& Pers-13 & Perseus & 03:28:40.6 & 31:17:56.5 & I (I) & 1.0 (1.0) & 8.1 & 1.3$\pm$0.1 & 1.4 & 1.0 & 1.0$\pm$0.06 & 1.6 & 0.8 \\  
J032845.3+310541& IRAS03256+3055 & Perseus & 03:28:45.3 & 31:05:41.9 & I (I) & 1.1 (1.1) & 7.4 & 1.0$\pm$0.04 & 1.4 & 0.7 & E/A & E/A & E/A \\  
J032859.5+312146& Pers-17 & Perseus & 03:28:59.3 & 31:21:46.7 & II (II) & -0.8 (-1.1) & 7.7 & 6.4$\pm$0.06 & 4.0 & 1.5 & 7.2$\pm$0.1 & 4.5 & 1.6 \\  
J032900.6+311200& Pers-18, Pers07 & Perseus & 03:29:00.6 & 31:12:00.4 & I (I) & 2.2 (2.0) & 7.4 & 2.6$\pm$0.06 & 2.4 & 1.0 & 1.5$\pm$0.1 & 1.1 & 1.3 \\  
J032901.6+312028& Pers-19, Pers08 & Perseus & 03:29:01.7 & 31:20:28.5 & $-$ & $-$ & 7.9 & 12.1$\pm$0.06 & 8.0 & 1.5 & 9.2$\pm$0.1 & 6.0 & 1.3 \\  
J032903.3+311555& SVS13 & Perseus & 03:29:03.3 & 31:15:55.5 & $-$ & $-$ & 8.2 & 12.4$\pm$0.08 & 7.0 & 1.6 & 8.3$\pm$0.1 & 4.7 & 1.5 \\  
J032904.0+311446& HH 7-11 MMS6 & Perseus & 03:29:04.1 & 31:14:46.6 & I (I) & 1.4 (1.3) & 9.0 & 8.3$\pm$0.08 & 4.4 & 1.9 & 4.7$\pm$0.1 & 2.5 & 1.9 \\  
J032910.7+311820& Pers-23, Pers10 & Perseus & 03:29:10.7 & 31:18:20.5 & I (I) & 2.0 (1.9) & 8.6 & 5.7$\pm$0.1 & 3.7 & 1.4 & 4.8$\pm$0.09 & 3.2 & 1.3 \\  
J032917.2+312746& Pers-27 & Perseus & 03:29:17.2 & 31:27:46.2 & I (I) & 1.8 (1.7) & 7.6 & 2.6$\pm$0.06 & 2.7 & 0.9 & 2.7$\pm$0.07 & 3.3 & 0.7 \\  
J032918.2+312319& Pers-28 & Perseus & 03:29:18.3 & 31:23:19.9 & I (I) & 1.3 (1.1) & 7.7 & 1.2$\pm$0.07 & 2.1 & 0.6 & 3.6$\pm$0.08 & 4.2 & 0.9 \\  
J032923.5+313329& IRAS03262+3123 & Perseus & 03:29:23.5 & 31:33:29.4 & I (I) & 1.5 (1.5) & 7.5 & 1.0$\pm$0.05 & 1.7 & 0.6 & 0.9$\pm$0.08 & 1.9 & 0.4 \\  
J032951.8+313905& IRAS03267+3128, Pers13 & Perseus & 03:29:51.9 & 31:39:05.6 & I (I) & 3.4 (3.4) & 8.0 & 3.5$\pm$0.05 & 3.5 & 1.0 & 1.8$\pm$0.1 & 2.0 & 0.7 \\  
J033121.0+304530& IRAS03282+3035, Pers15 & Perseus & 03:31:21.0 & 30:45:30.0 & I (I) & 1.0 (1.5) & 7.0 & 3.3$\pm$0.07 & 2.8 & 1.3 & 1.0$\pm$0.2 & 1.0 & 1.4 \\  
J033218.0+304946& IRAS03292+3039, Pers16 & Perseus & 03:32:18.0 & 30:49:46.9 & I (I) & 1.1 (0.9) & 7.0 & 3.9$\pm$0.07 & 3.8 & 1.2 & 2.8$\pm$0.2 & 2.0 & 1.0 \\  
J033313.8+312005& Pers-34 & Perseus & 03:33:13.8 & 31:20:05.2 & I (I) & 1.4 (1.3) & 7.0 & 0.5$\pm$0.05 & 0.7 & 0.7 & E/A & E/A & E/A \\  
J033314.4+310710& B1-SMM3, Pers17 & Perseus & 03:33:14.4 & 31:07:10.8 & I (I) & 2.2 (2.7) & 6.6 & 1.9$\pm$0.06 & 1.9 & 0.8 & 2.9$\pm$0.1 & 2.1 & 1.2 \\  
J033320.3+310721& B1-b & Perseus & 03:33:20.3 & 31:07:21.4 & I (I) & 0.9 (0.6) & 6.6 & 2.7$\pm$0.1 & 1.9 & 1.3 & 2.9$\pm$0.1 & 2.4 & 1.2 \\  
J033327.3+310710& Pers-40, Pers19 & Perseus & 03:33:27.3 & 31:07:10.2 & I (I)  & 1.9 (1.8) & 6.9 & 1.5$\pm$0.06 & 1.5 & 0.9 & 2.3$\pm$0.2 & 2.8 & 0.8 \\  
J034350.9+320324& Pers-41 & Perseus & 03:43:50.9 & 32:03:24.7 & I (I) & 1.5 (1.4) & 8.6 & 2.9$\pm$0.1 & 2.8 & 1.0 & 3.1$\pm$0.1 & 2.7 & 0.9 \\  
J034356.9+320304& IC 348-MMS, Pers21 & Perseus & 03:43:56.9 & 32:03:04.2 & I (I) & 1.4 (1.8) & 8.8 & 4.7$\pm$0.07 & 3.8 & 1.1 & 4.3$\pm$0.1 & 4.8 & 0.8 \\  
J034357.3+320047& HH 211-FIR & Perseus & 03:43:57.3 & 32:00:47.6 & F (F)  & 0.02 (-0.07) & 9.0 & 7.0$\pm$0.1 & 3.2 & 2.1 & 3.7$\pm$0.1 & 4.5 & 0.8 \\  
J034359.4+320035& Pers-46 & Perseus & 03:43:59.4 & 32:00:35.5 & F (F) & 0.1 (-0.2) & 8.6 & 1.5$\pm$0.09 & 0.5 & 1.7 & 1.9$\pm$0.2 & 1.8 & 1.1 \\  
J034402.4+320204& Pers-47 & Perseus & 03:44:02.4 & 32:02:04.7 & I (I) & 1.5 (1.5) & 8.9 & 4.2$\pm$0.06 & 4.0 & 1.0 & 5.3$\pm$0.2 & 4.3 & 1.1 \\  
J034443.3+320131& IRAS03415+3152, Pers22 & Perseus & 03:44:43.3 & 32:01:31.6 & I (I) & 0.5 (0.6) & 9.8 & 5.2$\pm$0.1 & 2.5 & 2.1 & 4.9$\pm$0.2 & 3.7 & 1.2 \\  
J034741.6+325143& B5-IRS 1 & Perseus & 03:47:41.6 & 32:51:43.9 & I (I) & 0.8 (1.3) & 10.1 & 1.7$\pm$0.07 & 1.8 & 0.6 & 1.7$\pm$0.1 & 2.0 & 0.8 \\  
J041354.7+281132& L1495-IRS & Taurus & 04:13:54.7 & 28:11:32.9 & I\tablefootmark{K} & $-$ & 6.8 & $<$0.38 & $-$ & $-$ & 0.9$\pm$0.2 & 0.9 & 1.0 \\  
J041412.3+280837& IRAS04111+2800 & Taurus & 04:14:12.3 & 28:08:37.3 & $-$ & $-$ & 7.1 & $<$0.35 & $-$ & $-$ & 0.9$\pm$0.2 & 0.9 & 0.9 \\  
J041534.5+291347& IRAS04123+2906 & Taurus & 04:15:34.5 & 29:13:47.0 & $-$ & $-$ & $-$ & $<$0.40 & $-$ & $-$ & $<$0.43 & $-$ & $-$ \\  
J041851.4+282026& CoKu Tau 1 & Taurus & 04:18:51.5 & 28:20:26.5 & II\tablefootmark{K} & $-$ & 7.3 & $<$0.34 & $-$ & $-$ & 0.8$\pm$0.1 & 1.4 & 0.6 \\  
J041942.5+271336& IRAS04166+2706 & Taurus & 04:19:42.5 & 27:13:36.4 & I\tablefootmark{K}& $-$ & 6.7 & 3.3$\pm$0.2$^\dagger$ & 2.5 & 1.4 & 0.8$\pm$0.2 & 1.3 & 0.5 \\  
J041959.2+270958& IRAS04169+2702, Tau01 & Taurus & 04:19:59.2 & 27:09:58.6 & I\tablefootmark{K}  & $-$ & 6.8 & 1.8$\pm$0.1 & 2.0 & 0.9 & 2.1$\pm$0.1 & 2.2 & 0.8 \\  
J042110.0+270142& IRAS04181+2655 & Taurus & 04:21:10.0 & 27:01:42.0 & I\tablefootmark{K} & $-$ & 6.9 & 0.5$\pm$0.1 & 1.3 & 0.5 & 1.2$\pm$0.09 & 1.3 & 0.6 \\  
J042111.4+270108& IRAS04181+2654AB, Tau02 & Taurus & 04:21:11.4 & 27:01:08.9 & I\tablefootmark{K} & $-$ & 6.6 & 1.1$\pm$0.2 & 1.4 & 1.0 & 1.5$\pm$0.1 & 2.1 & 0.5 \\  
J042656.3+244335& IRAS04239+2436 & Taurus & 04:26:56.4 & 24:43:35.9 & I\tablefootmark{K} & $-$ & 6.7 & 0.6$\pm$0.2 & 0.5 & 2.4 & 1.7$\pm$0.1 & 2.3 & 0.8 \\  
J042757.3+261918& IRAS04248+2612, Tau06 & Taurus & 04:27:57.3 & 26:19:18.3 & I\tablefootmark{K} & $-$ & 7.3 & 1.9$\pm$0.1 & 2.1 & 0.8 & 1.0$\pm$0.1 & 1.1 & 1.1 \\  
J042838.9+265135& L1521F-IRS & Taurus & 04:28:38.9 & 26:51:35.2 & I & 1.9 & 6.5 & 3.2$\pm$0.2$^\dagger$ & 2.7 & 1.6 & 0.4$\pm$0.1 & 1.4 & 0.4 \\  
J042905.0+264904& IRAS04260+2642 & Taurus & 04:29:05.0 & 26:49:04.4 & I\tablefootmark{K} & $-$ & $-$ & $<$0.45 & $-$ & $-$ & $<$0.65 & $-$ & $-$ \\  
J042907.6+244350& IRAS04264+2433, Tau07 & Taurus & 04:29:07.7 & 24:43:50.1 & I & 0.4 & $-$ & $<$0.45 & $-$ & $-$ & $<$0.41 & $-$ & $-$ \\  
J042923.6+243302& GV Tau & Taurus & 04:29:23.7 & 24:33:02.0 & I\tablefootmark{K} & $-$ & 6.4 & 3.6$\pm$0.1 & 2.2 & 1.1 & 3.2$\pm$0.1 & 3.0 & 0.8 \\  
J043150.6+242418& HK Tau & Taurus & 04:31:50.6 & 24:24:18.2 & II & -0.45 & $-$ & $<$0.49 & $-$ & $-$ & $<$0.52 & $-$ & $-$ \\ 
J043215.4+242903& Haro 6-13 & Taurus & 04:32:15.4 & 24:29:03.4 & II & -0.74 & $-$ & $<$0.44 & $-$ & $-$ & $<$0.45 & $-$ & $-$ \\  
J043232.0+225726& L1536-IRS, Tau08 & Taurus & 04:32:32.0 & 22:57:26.7 & I\tablefootmark{K} & $-$ & $-$ & $<$0.68 & $-$ & $-$ & $<$0.55 & $-$ & $-$ \\  
J043316.5+225320& IRAS04302+2247 & Taurus & 04:33:16.5 & 22:53:20.4 & I\tablefootmark{K} & $-$ & 5.5 & 1.6$\pm$0.2 & 0.4 & 4.6 & $<$0.50 & $-$ & $-$ \\  
J043535.0+240822& IRAS04325+2402, Tau09 & Taurus & 04:35:35.0 & 24:08:22.0 & I\tablefootmark{K} & $-$ & 5.6 & 1.0$\pm$0.1 & 0.9 & 1.2 & 1.0$\pm$0.2 & 1.7 & 0.8 \\  
J043556.7+225436& Haro 6-28 & Taurus & 04:35:56.8 & 22:54:36.5 & I\tablefootmark{K} & $-$ & $-$ & $<$0.34 & $-$ & $-$ & $<$0.44 & $-$ & $-$ \\  
J043953.9+260309& L1527 & Taurus & 04:39:53.9 & 26:03:09.7 & $-$ & $-$ & 5.8 & 8.2$\pm$0.2$^\dagger$ & 8.1 & 1.5 & 3.5$\pm$0.08 & 3.4 & 0.9 \\  
J044138.8+255626& XEST07-041 & Taurus & 04:41:38.8 & 25:56:26.8 & F & 0.1 & $-$ & $<$0.47 & $-$ & $-$ & $<$0.43 & $-$ & $-$ \\  
J182844.0+005303& Ser-2 & Serpens & 18:28:44.0 & 00:53:03.0 & I (I) & 0.5 (0.5) & 8.0 & $-$ & $-$ & $-$ & 0.6$\pm$0.05 & 1.1 & 0.6 \\
J182845.8+005132& Ser-3 & Serpens & 18:28:45.8 & 00:51:32.0 & I (I) & 1.3 (1.1) & 7.9 & $-$ & $-$ & $-$ & 0.5$\pm$0.05 & 0.8 & 0.6 \\
J182853.0+001904& Ser-7 & Serpens & 18:28:53.0 & 00:19:04.0 & I (I) & 0.5 (0.3) & 7.9 & $-$ & $-$ & $-$ & 0.4$\pm$0.06 & 0.6 & 0.4 \\
J182855.2+002928& Ser-8 & Serpens & 18:28:55.2 & 00:29:28.0 & I (I) & 1.9 (1.7) & 8.0 & $-$ & $-$ & $-$ & 3.0$\pm$0.08 & 2.8 & 1.1 \\
J182855.9+004830& Ser-9 & Serpens & 18:28:55.9 & 00:48:30.0 & $-$ &  & 8.2 & $-$ & $-$ & $-$ & 0.5$\pm$0.06 & 0.9 & 0.4 \\
J182900.2+003020& Ser-13 & Serpens & 18:29:00.2 & 00:30:20.0 & I (I) & 1.8 (1.6) & 8.0 & $-$ & $-$ & $-$ & 1.5$\pm$0.08 & 1.9 & 0.8 \\
J182907.0+003042& Ser-14 & Serpens & 18:29:07.0 & 00:30:42.0 & I (I) & 1.6 (1.4) & 7.8 & $-$ & $-$ & $-$ & 4.7$\pm$0.1 & 3.1 & 1.4 \\
J182909.6+003137& Ser-15 & Serpens & 18:29:09.6 & 00:31:37.0 & I (I) & 2.3 (2.1) & 7.8 & $-$ & $-$ & $-$ & 1.8$\pm$0.1 & 1.2 & 1.6 \\
J182916.4+001815& Ser-17 & Serpens & 18:29:16.4 & 00:18:15.0 & F (F) & -0.07 (0.06) & 9.1 & $-$ & $-$ & $-$ & 0.7$\pm$0.1 & 0.9 & 1.2 \\
\hline
\end{tabular}
}
\tablefoot{\scriptsize{Upper limits are calculated at the 3$\sigma_{I}$ level (see Section~\ref{sec:res_spec}). E/A indicates an emission-absorption 
profile and A indicates a pure absorption profile which are omitted from further analysis due to contamination, likely by emission at the reference position.\\ 
\tablefoottext{a}{First alternate names for Perseus sources are from Table 3 in \citet{Jorgensen2007}. Designators were taken 
from the Other Identifiers column or as Pers-\# where the number is taken from the Number column. 
First alternate names for Taurus are from \citet{Hartmann2002}, SIMBAD commonly used identifiers, or the IRAS 
Point Source Catalogue \citep{Beichman1988}. Second alternate names for Perseus and Taurus are from the ``William \textit{Herschel} Line Legacy'' (WILL) survey 
(Mottram et al., in prep.) overview paper. Serpens alternate names are from Table 1 of \citet{Enoch2007} as Ser-\# where the number is 
taken from the Bolocam ID column.} \\
\tablefoottext{b}{Class is assigned according to $\alpha_{\mathrm{IR}}$ classification \citep[e.g.,][]{Evans2009}. Extinction-corrected Class assignment appears parenthetically.
\tablefoottext{K}Class taken from SED designation in \citet{Kenyon1995}.} \\
\tablefoottext{c}{$\alpha_{\mathrm{IR}}$ taken from the \textit{Spitzer} c2d catalogue \citep{Evans2003}. Extinction-corrected values from Dunham et al. (subm.)
appear parenthetically.} \\
\tablefoottext{d}{$V_{\mathrm{lsr}}$ values are based on Gaussian fits to the C$^{18}$O 3--2 line.} \\
$\dagger$ Line profile of central spectrum shows asymmetry indicative of infall.}}
\end{table*}

An overview of the source list is given in
Table~\ref{tab:data_props}. Sources and their coordinates are taken from the following:
Perseus Class 0+I sources from \citet{Jorgensen2007}, Taurus targets
from the Class 0+I sample in \citet{Hartmann2002} with additional
sources derived from a comparison between SCUBA and \textit{Spitzer} data in
the region (J{{\o}}rgensen, priv. comm.), and Serpens sources
selected from \citet{Enoch2007}. The sources were chosen to include
known Class 0+I ($\alpha_{\mathrm{IR}}$ $\geq 0.3$) objects in these clouds at the time of
selection (2008), with some borderline flat spectrum sources ($0.3 > \alpha_{\mathrm{IR}} 
\geq -0.3$) included. Table~\ref{tab:data_props} includes
alternate, more common names for many of the objects, which are
sometimes included parenthetically in the text.

Infrared slope values ($\alpha_{\mathrm{IR}}$) obtained from the 
c2d catalogue are presented in Column 7 of Table~\ref{tab:data_props}
for all overlapping sources observed in the c2d survey. These values may differ
from $\alpha_{\mathrm{IR}}$ at the time of selection. Extinction-corrected slopes
obtained from Dunham et al. (subm.) are also shown. There is only
one source for which there is a change in Class assignment: J032738.2+301358
(L1455-FIR 2) moves from a Flat spectrum to a Class II source.
All other sources are unaffected.

Spatially resolved maps are obtained for 56 sources in the Perseus 
($d=250\mathrm{pc}$) and Taurus ($d=140\mathrm{pc}$) star forming
molecular clouds in both HCO$^{+}$ 4--3 at
356.734 GHz and C$^{18}$O 3--2 at 329.330 GHz with the JCMT HARP instrument \citep{Buckle2009} using the ACSIS backend
\citep{Dent2000} in jiggle position-switch mode. An additional nine
sources in the Serpens ($d=260\mathrm{pc}$) cloud were mapped in C$^{18}$O 3--2
only. HARP is a heterodyne array with 16 SIS
detectors in a 4 x 4 configuration with pixel separations of
30$\arcsec$.  Array pixels have typical single side-band system
temperatures of 300$-$350 K.  The total foot print of the receiver is
2$\arcmin \times$2$\arcmin$ with a diffraction limited beam of
$\sim$15$\arcsec$. The offset switch used a slew of 1 degree in RA for
sky background measurements.

The HARP observations were carried out between 2008 and 2011 in good atmospheric conditions (optical depth $\tau_{225\mathrm{GHz}} \leq 0.1$). Typical rms noise levels
are 0.15 K after resampling the spectra into 0.2 km s$^{-1}$ bins. To achieve these noise levels, typical integration times were $\sim$30 minutes on- and off-source to
scan the whole 2$\arcmin \times$2$\arcmin$ region for both lines. All HARP data were converted to main beam brightness temperature
$T_{\mathrm{mb}} = T^{*}_{\mathrm{A}} / \eta_{\mathrm{mb}}$, where the beam efficiency of HARP is set to $\eta_{\mathrm{mb}} = 0.63$. 
Calibration errors are estimated to be $\sim$20\% \citep{Buckle2009}. All HARP data were reduced using the \textsc{gildas-class}\footnote{http://www.iram.fr/IRAMFR/GILDAS} software package.

In order to compare spectral observations to the dust continuum, data were taken from archival surveys of the JCMT SCUBA and 
{\it Herschel}\footnote{{\it Herschel} is an ESA space observatory with science instruments provided by European-led Principal Investigator consortia
and with important participation from NASA.} PACS instruments.
The SCUBA Legacy Catalogue survey \citep{diFrancesco2008} provides data for dust emission at 450 $\mu$m and 850 $\mu$m with beam sizes of 9$\arcsec$ and 15$\arcsec$, respectively. 
Uncertainties in flux values are about 20\% at 850 $\mu$m and up to 50\% at 450 $\mu$m.
{\it Herschel} PACS dust continuum maps at 70 $\mu$m and 160 $\mu$m were obtained from Herschel science archive observations
taken as part of {\it Herschel} Gould Belt Survey \citep{Andre2010}. 
Beam sizes for {\it Herschel} PACS are 
estimated to be about 12$\arcsec$ at 160 $\mu$m and 6$\arcsec$ at 70 $\mu$m in PACS/SPIRE parallel mode 
with respective uncertainties of order 20\% and 10\% \citep{Poglitsch2010}. 
Footprints of 2$\arcmin \times$2$\arcmin$ were extracted from the large scale SCUBA and PACS maps to match the spatial coverage of the HARP maps. 
All dust maps were converted to units of Jansky beam$^{-1}$, where required.

Combined 2$\arcmin \times$2$\arcmin$ field of view maps were assembled for JCMT HARP lines HCO$^{+}$ 4--3 and C$^{18}$O 3--2, {\it Herschel}
PACS 70 $\mu$m and 160 $\mu$m observations, and JCMT SCUBA 450 $\mu$m and 850 $\mu$m observations. 
These maps allow
characterization of the high density and high column density molecular line tracers and the warm and cool components of the dust continuum by
observing the morphology of the emission at each wavelength. Intensity scales for
all spatial maps are set to a percentage of the maximum value to create the best contrast between emission features and the background,
and for easy visual comparison from source to source.

All maps can be found in Appendix~\ref{app:C}. They have been
regridded to 100$\times$100 pixel images using a linear interpolation between data points.
HCO$^{+}$ 4--3 maps are available for 56 out of the 65 sources in our
sample - these maps exist for all sources in Perseus and Taurus but those in Serpens were not observed in HCO$^{+}$ 4--3. 
All sources were observed in C$^{18}$O 3--2.  We have a complete set
of PACS maps for all sources while 21 out of 65 sources ($\sim$30\%)
in the Perseus, Taurus, and Serpens sample do not have SCUBA observations. 
For HCO$^{+}$ 4--3 and C$^{18}$O 3--2 there are two maps in each
line. The first pair are spatial maps of integrated intensity in units of K
km s$^{-1}$, created by integrating the spectral emission with limits
of integration $V_{\mathrm{lsr}} \ \pm \ 6.0 \times \mathrm{FWHM}$ to
ensure that integration is done over all emission down to the rms
noise level. The second pair of maps in HCO$^{+}$ 4--3 and C$^{18}$O
3--2 display the main beam peak temperature in units of K. Peak
temperature maps are expected to have roughly the same morphology as the
integrated intensity maps, but the latter can potentially be contaminated by
outflow wings, particularly in HCO$^{+}$ 4--3.

Absorption is seen in the C$^{18}$O 3--2 spectra of J032832.5+311104, J032834.5+310705, J032845.3+310541, and J033313.8+312005. 
In J032834.5+310705, there is also strong absorption in HCO$^{+}$ 4--3. These features drop below the continuum level,
indicating that it is not self-absorption, but rather that the reference position contained emission in C$^{18}$O 3--2.
In the case of J032834.5+310705 the switch region was also emitting in
HCO$^{+}$ 4--3. The absorption features contaminate the data for these sources, and they are omitted from further analysis.

%__________________________________________________________________

\section{Results}
\label{sec:res}

\subsection{Central spectra}
\label{sec:res_spec}

Spectra are extracted from the central spatial pixel (spaxel) of the
jiggle mode maps for sources observed in HCO$^{+}$ 4--3 and
C$^{18}$O 3--2 (shown in Appendix \ref{app:B}). Integrated intensity, main beam peak
temperature, and local standard of rest velocity were extracted from
the central spectra for sources with detected emission. The majority
of sources can be fitted with a single Gaussian and do not exhibit a
broad component of emission in either HCO$^{+}$ 4--3 or C$^{18}$O
3--2, suggesting that the bulk of the on-source integrated intensity is
dominated by emission from the envelope with negligible contribution
from high velocity outflows. The properties of the central spectra for 
all sources are presented in Table~\ref{tab:data_props}. Column 6 shows the
local standard of rest velocity for each source based on the peak of the Gaussian fit 
to the C$^{18}$O 3--2 line. Subsequent columns
give the integrated intensity and peak main beam
temperature of HCO$^{+}$ 4--3 and C$^{18}$O 3--2.

Out of the 56 sources observed in both molecular lines, there are 45 detections in HCO$^{+}$ 4--3 and 43
detections in C$^{18}$O 3--2. Integrated intensities of HCO$^{+}$ 4--3 
range from 0.5 K km s$^{-1}$ to 14.8 K km s$^{-1}$ while 
peak temperatures range from 0.4 K to 8.1 K. For C$^{18}$O 3--2
the integrated intensities and peak temperatures range from 0.5 K km s$^{-1}$ to 9.2 K km s$^{-1}$ and 
0.6 K to 6.0 K, respectively. Additionally, C$^{18}$O 3--2 on 
source emission is detected for all nine sources in Serpens. On average
the C$^{18}$O 3--2 emission in the Serpens sources is weaker than those in Perseus and Taurus. 
The average full width half maximum (FWHM) of the lines are 0.9 km s$^{-1}$
and 1.3 km s$^{-1}$ for C$^{18}$O 3--2 and HCO$^{+}$ 4--3,
respectively. 

Among sources that cannot be
fitted easily with single Gaussians are J033314.4+310710 (B1-SMM3) and
J033320.3+310721 (B1-b), which show broadening in the red component, and 
J032901.6+312028, which shows a broadening in the blue component 
(see Figures in Appendix~\ref{app:B}). The resulting integrated intensity maps of these
sources may have some contribution from outflows that may cause
their morphology to appear less concentrated than if only emission
from the envelope were observed. 

There is evident self-absorption in
the HCO$^{+}$ 4--3 central spectra of J043953.9+260309 (L1527) in Taurus and its
asymmetric, blue-dominated profile indicates infall of high density envelope
material, in agreement with the inverse P Cygni line profile seen in \textit{Herschel} water spectra by \citet{Mottram2013}. 
Self-absorption and a blue-dominated, asymmetric infall signature in HCO$^{+}$ 4--3 can be seen to a lesser extent
in two other sources in Taurus: J041942.5+271336 and J042838.9+265135 (L1521F-IRS). These features
already provide strong evidence that these three objects in Taurus are likely Stage I YSOs.
Self-absorption is detected in J032832.5+311104 and J034357.3+320047 (HH 211-FIR), but it is not
accompanied by asymmetry. (See Section~\ref{sec:singlept} for further discussion.)

Ten out of the 22 sources in Taurus lack detections in HCO$^{+}$ 4--3
above 3$\sigma_{I}$. For the integrated intensity we define $\sigma_{I} = \sqrt{\delta v
  \Delta_{0} v} T_{\mathrm{rms}}$ where $\delta v$ is the velocity
resolution of the spectrum (0.2 km s$^{-1}$), $\Delta_{0} v$ is the
full line width down to zero intensity, and $T_{\mathrm{rms}}$ is the
rms noise level at the given resolution. Here $\Delta_{0} v$ is taken
to be three times the sample-averaged FWHM for each of the two spectral 
lines observed to ensure that zero intensity limits are reached in all cases.

In the Perseus cloud, only J032834.5+310705 lacks an obvious detection
due to the absorption features in its spectrum. There is an emission-absorption profile
in HCO$^{+}$ 4--3 and pure emission in
C$^{18}$O 3--2. These molecular line data are omitted
from further analysis as the absorption prevents proper stage classification in this
study. Uncontaminated data are needed to correctly characterize the source. 

Of the ten HCO$^{+}$ 4--3 non-detections in Taurus, seven also lack a
detection in C$^{18}$O 3--2.  The absence of both of these high (column)
density molecular tracers indicates that these sources are not likely
to be deeply embedded YSOs.  Two sources, J032845.3+310541 in Perseus
and J043316.5+225320 in Taurus, are detected in HCO$^{+}$ 4--3 but
have no detection in C$^{18}$O 3--2. This could be due to the
absence of any high column density envelope or cloud material along
the line of sight, strong freeze out of CO onto dust grains,
or a relatively more isolated environment for these sources.

\begin{figure*}[!htbp]
 \centering
 \resizebox{0.9\textwidth}{!}{
 \includegraphics[width=0.5\textwidth]{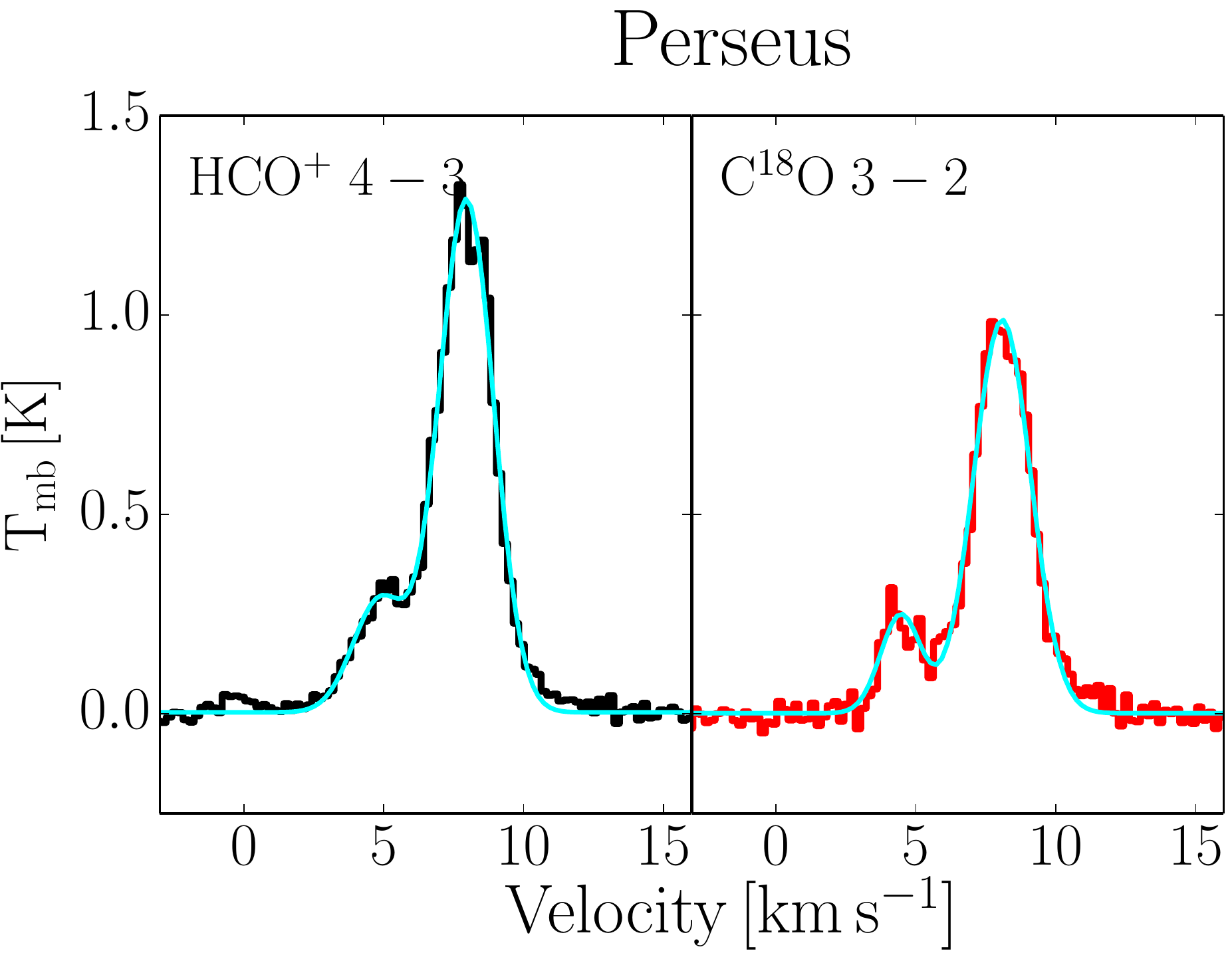}
 \includegraphics[width=0.5\textwidth]{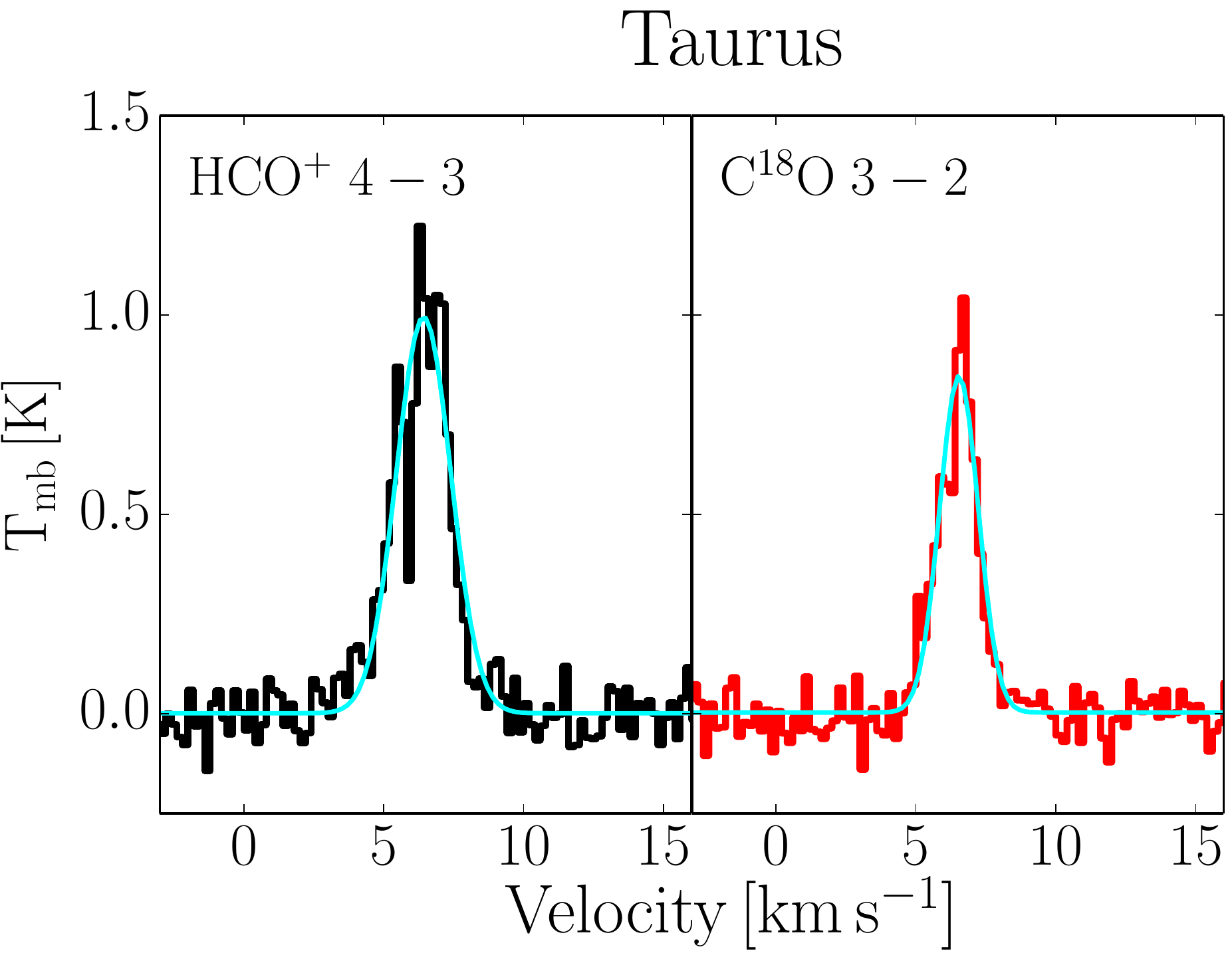}
 }
 \caption{Cloud-averaged spectra of all sources with detections in HCO$^{+}$ 4--3 and C$^{18}$O 3--2. Non-detections are omitted from
 averaging. The HCO$^{+}$ 4--3 data are shown in black, the C$^{18}$O 3--2 in red, and Gaussian fits in cyan. The sum of two
 1D Gaussian fits is shown for the Perseus cloud. Perseus shows a clear secondary feature while Taurus is well fit by single Gaussians in both lines. 
 Note that the y-axis scale is $0.25\times$ the figures presented in Appendix~\ref{app:B}.}
 \label{fig:avg_spectra}
\end{figure*}

\begin{table*}[!htbp]
 \caption{Cloud-average central spectra properties}
 \centering
 \label{tab:spec_fits}
 \begin{tabular}{lccccccc}
 \hline \hline
  &  & \multicolumn{3}{c}{HCO$^{+}$ 4--3} & \multicolumn{3}{c}{C$^{18}$O 3--2} \\
 \hline
 Cloud & $V_{\mathrm{lsr}}$ & $\int T_{\mathrm{{mb}}} \mathrm{d}V$ & $T_{\mathrm{mb}}$ & $\Delta v$ & $\int T_{\mathrm{{mb}}} \mathrm{d}V$ & $T_{\mathrm{mb}}$ & $\Delta v$ \\
 \hline
 Perseus\tablefootmark{a} & 4.7 & 0.7 & 0.3 & 2.3 & 0.4 & 0.3 & 1.7 \\
 Perseus\tablefootmark{b} & 8.0 & 3.2 & 1.3 & 2.3 & 2.5 & 1.0 & 2.4 \\
 Taurus & 6.4 & 2.4 & 1.2 & 2.3 & 1.5 & 1.0 & 1.6 \\
 \hline
 \end{tabular}
 \tablefoot{Central spectra are within a 15$\arcsec$ beam. Local standard of rest velocity, integrated intensity, and 
 line width are derived from the Gaussian fit. $\Delta v$ is the FWHM of the Gaussian fit. Peak main beam temperature is
 taken as the maximum value from the data. 
 \tablefoottext{a}{Fit to the weak blue component of Perseus spectra.} 
 \tablefoottext{b}{Fit to the strong red component of Perseus spectra.}}
\end{table*}

The central spectra for each source with detections in HCO$^{+}$ 4--3
and C$^{18}$O 3--2 were averaged by cloud, as shown in
Figure~\ref{fig:avg_spectra}.  Unlike the individual source spectra,
the cloud-averaged Perseus spectra show a secondary feature at 4.4 km
s$^{-1}$ on the blue side of the cloud-average $V_{\mathrm{lsr}}$ of 8.0 km s$^{-1}$ in both lines.
Four sources in the L1448
and L1455 regions taken together have an average $V_{\mathrm{lsr}}$ of
4.6 km s$^{-1}$, consistent with the weaker component of the Perseus
spectra. One source, J032536.4+304523 (L1448-N), has the strongest HCO$^{+}$
4--3 and fourth strongest C$^{18}$O 3--2 integrated intensity of
all sources in Perseus. The remaining 29 sources in Perseus have an
average $V_{\mathrm{lsr}}$ of 7.8 km s$^{-1}$ and make up the bulk of
emission for the cloud-averaged spectra, 
consistent with the stronger component at higher
velocity. Thus sources in the L1448 and L1455 regions are shifted in velocity with respect to the rest of the Perseus molecular cloud.

The Taurus cloud-averaged spectra are well-fit by a single
Gaussian in both lines, with no indication of any blending of different
velocity components of sources within the cloud. Given that the broad
nature and double peak of the line in Perseus can be explained by
blending, and no broadening is seen in Taurus, there is no evidence
for significant outflow contribution to the HCO$^{+}$ 4--3 and
C$^{18}$O 3--2 molecular emission in the averaged central spectra of either cloud.
Details of the fit properties for the cloud-averaged spectra are found in
Table~\ref{tab:spec_fits}. Although the outflow component does not 
contribute much (typically $\lesssim$5\%) to the on-source integrated HCO$^{+}$ 4--3 
intensity, it is possible to map the outflows in the HCO$^{+}$ line wings, 
as shown by \citet{Walker-Smith2014}. Their study maps regions of the 
Perseus cloud, therefore capturing outflow contribution on large scales.
Here the mapping of individual sources restricts the identification of outflows
to the immediate environment surrounding the YSOs. The possibility remains that off-source 
regions of the field of view exhibit broadening of the molecular line 
due to outflows. This is discussed further in Section~\ref{sec:res_morph}.

\subsection{Categorizing emission morphology}
\label{sec:res_morph}

\begin{figure*}[!htbp]
 \begin{minipage}{\textwidth}
 \subfloat[\large{J032837.1+311328 $\ $  HCO$^{+}$ 4--3}]{\includegraphics[width=0.5\textwidth,height=0.3\textheight]{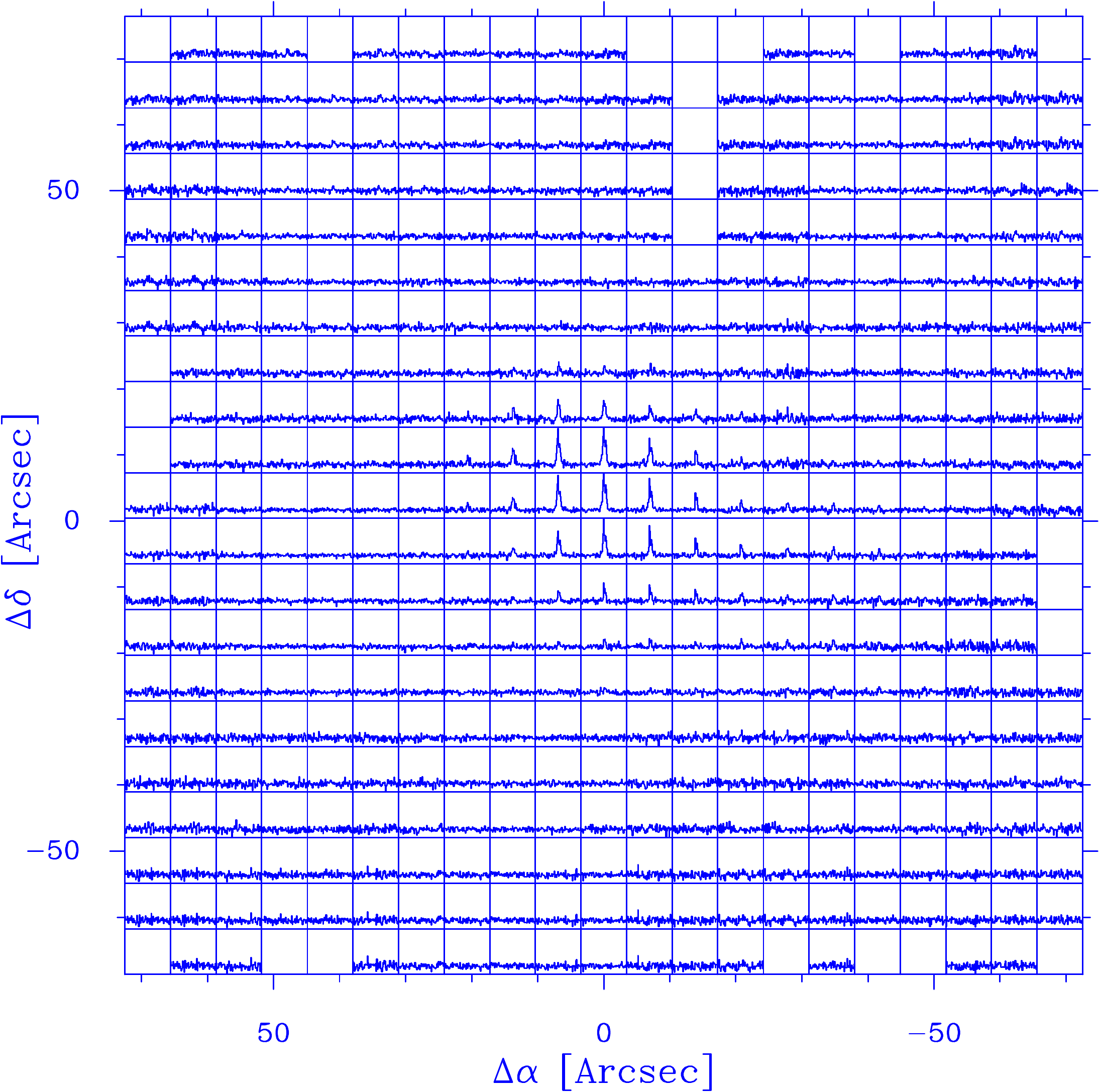}}
 \subfloat[\large{J032837.1+311328 $\ $  C$^{18}$O 3--2}]{\includegraphics[width=0.5\textwidth,height=0.3\textheight]{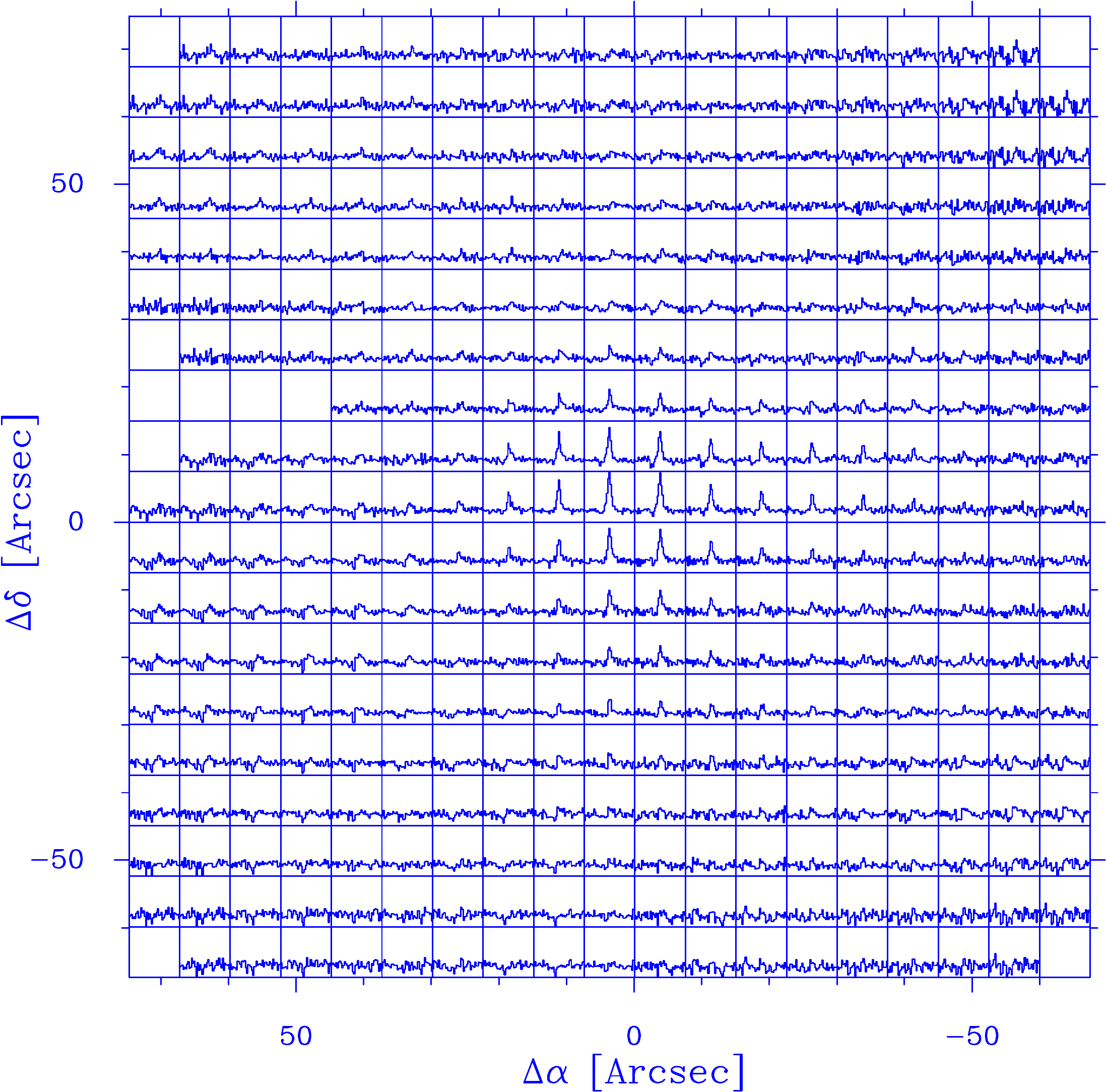}}\\
 \subfloat[\large{J032840.6+311756 $\ $  HCO$^{+}$ 4--3}]{\includegraphics[width=0.5\textwidth,height=0.3\textheight]{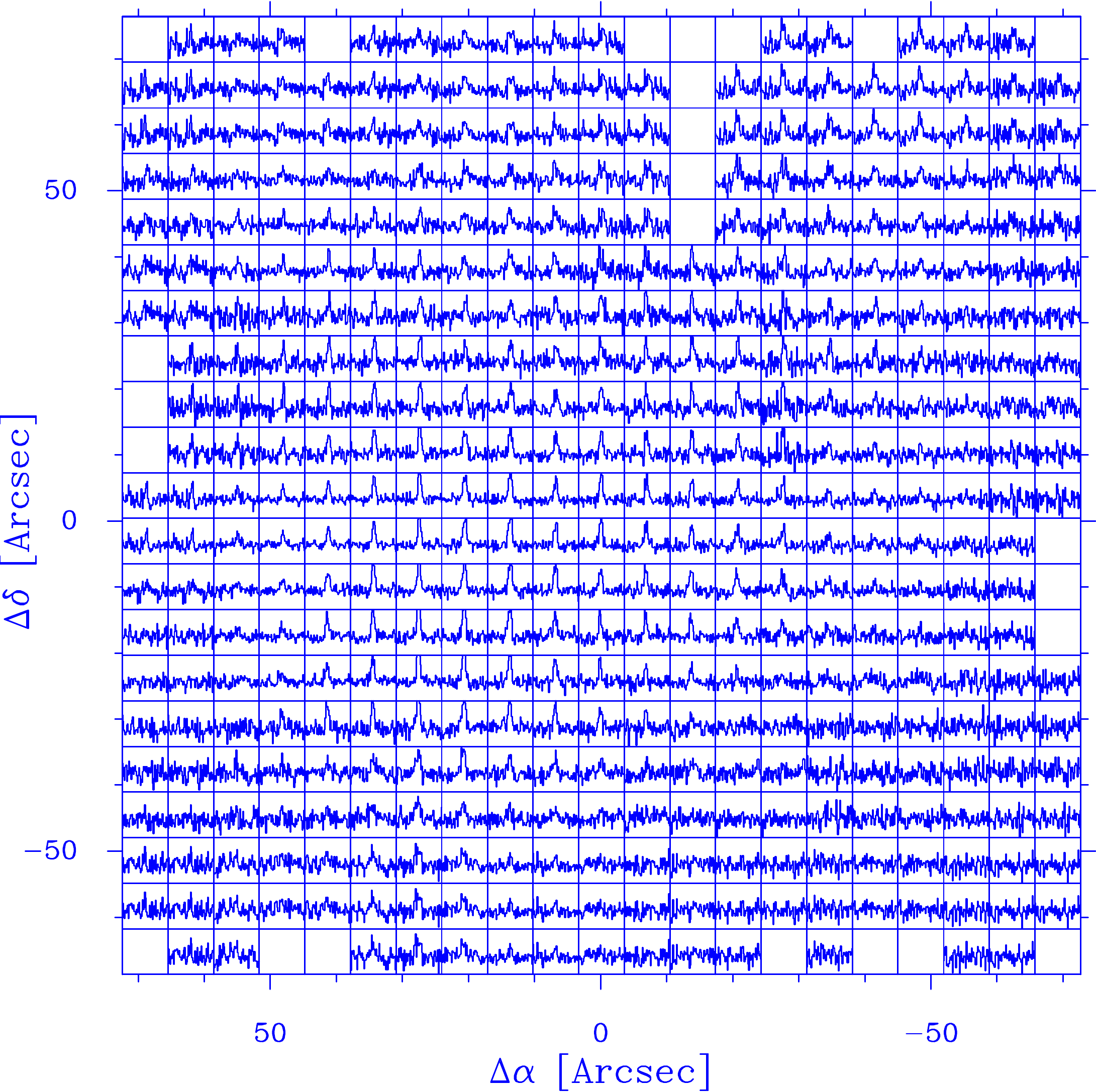}}
 \subfloat[\large{J032840.6+311756 $\ $  C$^{18}$O 3--2}]{\includegraphics[width=0.5\textwidth,height=0.3\textheight]{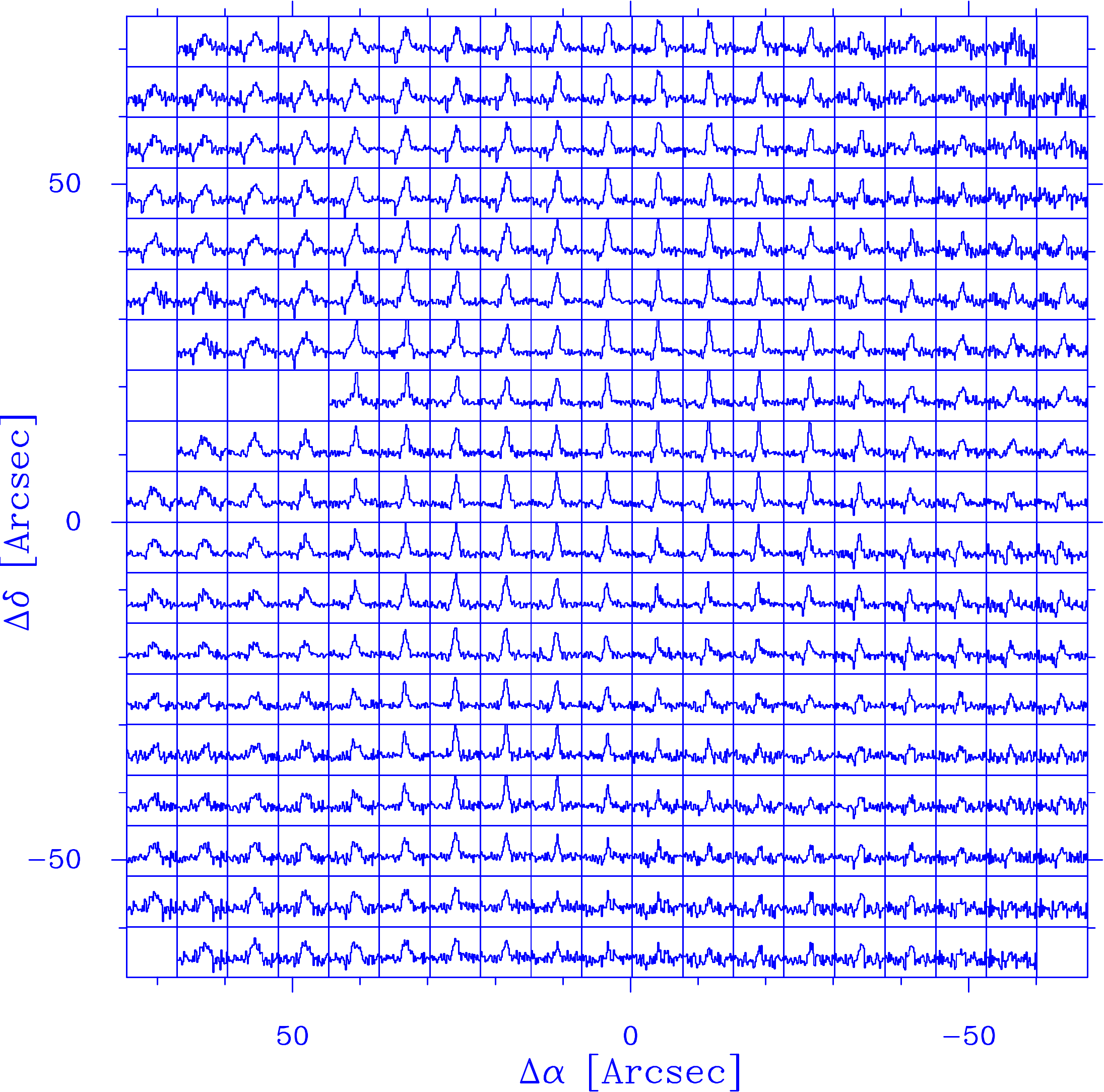}}
  \caption{Example of spectral maps obtained by the HARP instrument, visualized in Appendix~\ref{app:C}. The
  $T_{\mathrm{mb}}$ maximum value of the central spectrum dictates the scale of the y-axis in each map. 
  {\it (a)} and {\it (b)} show the distribution of spectra for J032837.1+311328, a bonafide Stage I source (see Section~\ref{sec:evstage}) with central concentration of emission. 
  {\it (c)} and {\it (d)} show the spectral map for a misfit source, J032840.6+311756, with widespread emission in both lines that result in a Stage II classification.} 
 \label{fig:spec_maps}
 \end{minipage}
\end{figure*}

Examples of the reduced spectral maps are shown in Figure~\ref{fig:spec_maps}. All spectra have been 
resampled to 0.2 km s$^{-1}$ velocity channels. Maps in 
HCO$^{+}$ 4--3 and C$^{18}$O 3--2 are presented for a bonafide
Stage I embedded source (J032837.1+311328, top) and a more evolved Stage II PMS with a disk (J032840.6+311756, bottom).
The Stage I source exhibits centrally concentrated 
emission with no contamination from nearby sources or clouds in 
the field. In contrast, the Stage II source
has widespread emission throughout the field.

The emission morphology can be categorized for all six wavelengths
observed. C$^{18}$O 3--2 serves as a tracer of the high column
density environment within the cloud. HCO$^{+}$ 4--3 has a higher
critical density ($n_{\mathrm{crit}}$ >10$^{6}$ cm$^{-3}$) and effectively probes the
inner regions of the protostellar envelope. The morphology of
integrated intensity maps for each of these molecular tracers provides
a qualitative indication of their embedded nature. Sources
with localized emission in HCO$^{+}$ 4--3 are good candidates for
truly embedded protostars while a lack of emission in HCO$^{+}$ 4--3
may indicate a non-embedded source such as an edge-on
disk. Accompanying peak temperature maps are included to check for
consistency with the morphology of the integrated intensity maps. For example, 
outflow emission would be revealed by more extended morphology 
in the integrated intensity maps compared to the peak temperature maps. The
dust continuum morphology at 70 $\mu$m and 160 $\mu$m from PACS reveal
warm dust originating from the region close to the protostar.
450 $\mu$m and 850 $\mu$m emission originates in the cooler parts of
the outer envelopes and partially from surrounding cloud material.

\begin{figure*}[!htbp]
  \centering
  \includegraphics[width=0.9\textwidth]{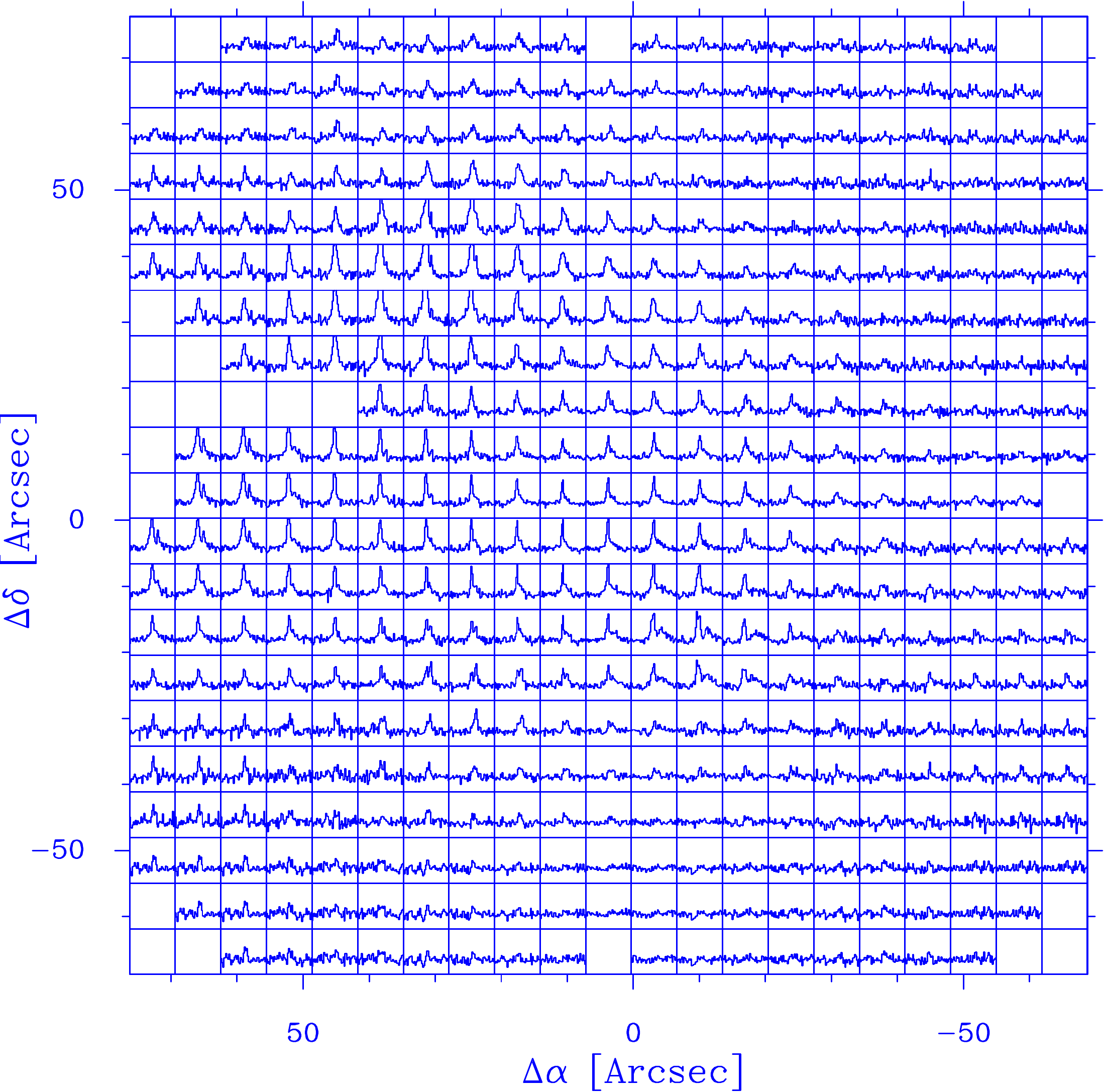}
  \caption{HCO$^{+}$ 4--3 spectral map for J033314.4+310710 (B1-SMM3) showing evidence of outflow wings. 
  Broadening of the line can be seen at low $T_\mathrm{mb}$ in the spectra off-source, e.g., 
  the red wing visible at offsets of about (-20$\arcsec$, +10$\arcsec$).
  The central spectrum can be found in Appendix~\ref{app:B}.} 
 \label{fig:outflow_map}
\end{figure*}

Based on the morphology of the molecular line and dust continuum maps
in Appendix \ref{app:C}, which are cross-checked with the spectral maps such as
shown in Figure~\ref{fig:spec_maps}, we define four distinct classes of emission.
A first group has concentrated emission that peaks at the source
position. Any emission offset of less than 5$\arcsec$ can be
attributed to pointing inaccuracies. J032837.1+311328 in Figure~\ref{fig:spec_maps} and Appendix~\ref{app:C}
is a good example of a source with emission peaking on-source in both
HCO$^{+}$ 4--3 and C$^{18}$O 3--2 as well as each wavelength of
dust continuum observations.

A second class has concentrated emission that is peaking off-source by
more than 5--10$\arcsec$. Such an offset from the mid-IR position in
HCO$^{+}$ 4--3 could be due to high density
material in outflows or binarity, which would have a significant effect on the spectrally
integrated intensity maps. J034350.9+320324 is an example of
off-source emission in HCO$^{+}$ 4--3 due to binarity. Other examples
include J032910.7+311820, J042110.0+270142, and J042111.4+270108. 
Molecular line emission in these sources is offset from the source position 
due to a companion within the same field of view. 
J042111.4+270108 has extended HCO$^{+}$ 4--3 throughout the field
due to the presence of a third source in the mapped area,
but with an off-source peak at the location of the binary companion.
Identification of these binaries is described in further detail at the 
end of this section. The J034359.4+320035
HCO$^{+}$ 4--3 map has emission peaking significantly off-source as
a result of another object seen in both molecular line and dust continuum
emission (see Appendix~\ref{app:A} for more detail).

The third class has extended emission, where a peak may be observed
but the emission is widespread within the field of view on scales of
20$\arcsec$ or more. Examples are J032839.1+310601 and 
J032840.6+311756, which show extended emission in
both molecular lines and in SCUBA dust continuum maps. Several 
sources with extended HCO$^{+}$ 4--3 emission exhibit off-source 
line broadening in their spectral maps of order a few km s$^{-1}$, suggesting outflow activity.
An example of a source with evidence for outflows, J033314.4+310710 (B1-SMM3), is shown in Figure~\ref{fig:outflow_map}.
Though present, the weak outflow wings do not constitute a significant 
contribution to the integrated intensity. Therefore, a likely cause of the extended
emission is contamination from another object in the field.
J032859.5+312146, J032900.6+311200, J032901.6+312028, 
J032903.3+311555, J032951.8+313905, and J033320.3+310721 (B1-b) also have spectral maps 
that indicate a slight broadening of the line off-source. In 
J033314.4+310710 (B1-SMM3) and J033320.3+310721 (B1-b) the weak outflow wing
can be seen in the central spectrum. In the case of J032859.5+312146, 
J032900.6+311200 and J032951.8+313905 there appear to be no other 
nearby YSOs or dense cores that might contaminate the field. 
The additional contribution from outflow wings alone is not enough 
to explain the widespread emission in these sources; there
may be distributed high density cloud material present in the field. 
J032901.6+312028, J032903.3+311555, J033314.4+310710 (B1-SMM3), and J033320.3+310721 (B1-b) 
all contain previously catalogued objects within their mapped regions
that are likely causes of the extended emission (see Appendix~\ref{app:A}).

Finally, the fourth class consists of all non-detections down to rms 
noise levels. 

These four groups are used to diagnose the morphology of 
emission for each source at each observed wavelength. The classification 
of emission for each source can be found in Table~\ref{tab:classification}. 
Concentrated emission that is peaking on-source
is indicated with a P, concentrated emission that is peaking
off-source is indicated with an O, extended emission is indicated with
an E, and a non-detection is indicated with an N. Subscripts for each
of these classifications correspond to the emission bands: H is
HCO$^{+}$ 4--3, C is C$^{18}$O 3--2, 7 is 70~$\mathrm{\mu}$m, 1 is
160~$\mathrm{\mu}$m, 4 is 450$\mathrm{\mu}$m and 8 is
850~$\mathrm{\mu}$m.

Four binary sources are identified in the sample: 
J032910.7+311820, J034350.9+320324, J042110.0+270142,
and J042111.4+270108. These are flagged as binaries based on their 
70~$\mathrm{\mu}$m morphology. The data at this wavelength has the highest spatial resolution and
the dust it traces is warm enough to reveal the presence of two objects while
emission at other wavelengths is too widespread to resolve two
objects and appears as emission on a larger scale from a single
object. The sources listed above show a secondary object within 20$\arcsec$ of
the primary target.

\section{Source Classification}
\label{sec:class}

\begin{table*}[!htbp]
\caption{Characterization of source emission, concentration factor parameters, and stage classifications based on JCMT HARP, JCMT SCUBA, and {\it Herschel} PACS spatial maps.}             % title of Table
\label{tab:classification}      % is used to refer this table in the text
\centering % used for centering table
\resizebox{18cm}{!}{
\begin{tabular}{l c c c c c c c c c c}        % centered columns (4 columns)
\hline\hline  % inserts double horizontal lines
 & Code\tablefootmark{a} & $S_{850}$\tablefootmark{b} & $F_{0}^{\mathrm{850}}$ & $C_{850}$ & $R_{\mathrm{obs}}^{850}$\tablefootmark{c} &  $S_{\mathrm{HCO^{+}}}$\tablefootmark{d} & $S_{\mathrm{HCO^{+}}}^{\mathrm{peak}}$ & $C_{\mathrm{HCO^{+}}}$ & $R_{\mathrm{obs}}^{\mathrm{HCO^{+}}}$\tablefootmark{c} & Stage \\    % table heading 
  &  & [Jy] & [Jy beam$^{-1}$] &  & [$\arcsec$] & [K km s$^{-1}$] & [K km s$^{-1}$ beam$^{-1}$] &  & [$\arcsec$] & \\
\hline                        % inserts single horizontal line
J032522.3+304513 & P$_{\mathrm{HC7148}}$ & 2.9 & 1.4 & 0.67 & 35 & 14.1 & 4.7 & 0.65 & 25 & I \\ 
J032536.4+304523 & P$_{\mathrm{HC7148}}$ & 11.0 & 5.4 & 0.61 & 32 & 160 & 17.6 & 0.63 & 45 & I \\ 
J032637.4+301528 & P$_{\mathrm{HC7148}}$ & 0.82 & 0.37 & 0.60 & 32 & 8.2 & 1.4 & 0.69 & 40 & I \\ 
J032738.2+301358 & P$_{\mathrm{H148}}$ E$_{\mathrm{C}}$ N$_{\mathrm{71}}$ & 0.50 & 0.29 & 0.48 & 25 & 10.4 & 2.8 & 0.66 & 30 & I \\ 
J032743.2+301228 & P$_{\mathrm{H7148}}$ E$_{\mathrm{C}}$ & 1.2 & 0.53 & 0.30 & 25 & 5.5 & 3.6 & 0.71 & 21 & I \\ 
J032832.5+311104 & P$_{\mathrm{H145}}$ N$_{\mathrm{C7}}$ & 1.3 & 0.30 & 0.47 & 40 & 7.3 & 3.6 & 0.61 & 21 & I \\ 
J032834.5+310705 & P$_{\mathrm{148}}$ N$_{\mathrm{HC7}}$ & 0.36 & 0.31 & 0.72 & 28 & $-$ & $-$ & $-$ & $-$ & C \\ 
J032837.1+311328 & P$_{\mathrm{HC7148}}$ & 1.7 & 0.87 & 0.59 & 30 & 12.5 & 5.2 & 0.61 & 22 & I \\ 
J032839.1+310601 & E$_{\mathrm{HC148}}$ N$_{\mathrm{7}}$ & 1.3 & 0.35 & 0.22 & 30 & 10.8 & 1.0 & 0.45 & 40 & II \\ 
J032840.6+311756 & E$_{\mathrm{HC148}}$ N$_{\mathrm{7}}$ & 2.0 & 0.66 & 0.41 & 40 & 24.0 & 2.4 & 0.49 & 40 & II \\ 
J032845.3+310541 & P$_{\mathrm{HC71}}$ E$_{\mathrm{48}}$ & 0.75 & 0.26 & 0.35 & 30 & 4.5 & 1.6 & 0.62 & 25 & I \\ 
J032859.5+312146 & O$_{\mathrm{48}}$ E$_{\mathrm{HC}}$ N$_{\mathrm{71}}$ & 3.6 & 1.3 & 0.42 & 30 & 92.0 & 11.9 & 0.61 & 40 & I \\ 
J032900.6+311200 & P$_{\mathrm{7148}}$ E$_{\mathrm{HC}}$ & 0.33 & 0.34 & 0.71 & 25 & 21.0 & 3.7 & 0.62 & 35 & I \\ 
J032901.6+312028 & P$_{\mathrm{71}}$ E$_{\mathrm{HC48}}$ & 7.3 & 2.1 & 0.45 & 35 & 179 & 25.2 & 0.70 & 44 & C \\ 
J032903.3+311555 & P$_{\mathrm{4871}}$ E$_{\mathrm{HC}}$ & 10.3 & 4.8 & 0.56 & 30 & 74.0 & 19.4 & 0.62 & 29 & I \\ 
J032904.0+311446 & E$_{\mathrm{HC48}}$ N$_{\mathrm{71}}$ & 3.6 & 1.1 & 0.31 & 30 & 40.2 & 10.4 & 0.50 & 25 & C \\ 
J032910.7+311820$^{\mathrm{B}}$ & O$_{\mathrm{HC7148}}$ & 6.9 & 1.8 & 0.64 & 45 & 69.7 & 9.2 & 0.63 & 32 & I \\ 
J032917.2+312746 & P$_{\mathrm{HC7148}}$ & 1.1 & 0.48 & 0.63 & 36 & 15.3 & 4.4 & 0.73 & 32 & I \\ 
J032918.2+312319 & O$_{\mathrm{7148}}$ E$_{\mathrm{HC}}$ & 1.3 & 0.60 & 0.53 & 30 & 19.9 & 2.6 & 0.64 & 35 & I \\ 
J032923.5+313329 & P$_{\mathrm{H71}}$ O$_{\mathrm{C}}$ & $-$ & $-$ & $-$ & $-$ & 9.7 & 1.6 & 0.66 & 37 & I \\ 
J032951.8+313905 & P$_{\mathrm{718}}$ O$_{\mathrm{4}}$ E$_{\mathrm{HC}}$ & 1.1 & 0.49 & 0.63 & 35 & 20.9 & 4.5 & 0.65 & 33 & I \\ 
J033121.0+304530 & P$_{\mathrm{HC7148}}$ & 1.9 & 1.3 & 0.62 & 27 & 11.8 & 4.1 & 0.68 & 27 & I \\ 
J033218.0+304946 & P$_{\mathrm{C7148}}$ O$_{\mathrm{H}}$ & 3.4 & 2.5 & 0.60 & 26 & 23.4 & 5.2 & 0.63 & 31 & I \\ 
J033313.8+312005 & O$_{\mathrm{8}}$ E$_{\mathrm{H14}}$ N$_{\mathrm{C7}}$ & 2.0 & 0.39 & 0.46 & 42 & 5.5 & 0.9 & 0.68 & 39 & C \\ 
J033314.4+310710 & P$_{\mathrm{71}}$ E$_{\mathrm{HC48}}$ & 7.0 & 1.2 & 0.30 & 40 & 33.0 & 3.0 & 0.45 & 40 & C \\ 
J033320.3+310721 & O$_{\mathrm{148}}$ E$_{\mathrm{HC}}$ N$_{\mathrm{7}}$ & 5.4 & 1.2 & 0.57 & 45 & 32.1 & 4.2 & 0.62 & 40 & C \\ 
J033327.3+310710 & P$_{\mathrm{H7148}}$ E$_{\mathrm{C}}$ & 1.2 & 0.37 & 0.29 & 30 & 7.9 & 2.1 & 0.65 & 30 & I \\ 
J034350.9+320324$^{\mathrm{B}}$ & P$_{\mathrm{7148}}$ O$_{\mathrm{H}}$ E$_{\mathrm{C}}$ & 2.9 & 0.7 & 0.37 & 35 & 20.9 & 5.3 & 0.73 & 35 & I \\ 
J034356.9+320304 & P$_{\mathrm{H7148}}$ E$_{\mathrm{C}}$ & 2.7 & 1.7 & 0.65 & 30 & 28.4 & 6.9 & 0.69 & 33 & I \\ 
J034357.3+320047 & P$_{\mathrm{H}}$ O$_{\mathrm{7148}}$ E$_{\mathrm{C}}$ & 5.6 & 2.2 & 0.62 & 36 & 20.9 & 8.5 & 0.65 & 24 & I \\ 
J034359.4+320035 & O$_{\mathrm{H7148}}$ E$_{\mathrm{C}}$ & $-$ & $-$ & $-$ & $-$ & $-$ & $-$ & $-$ & $-$ & C \\ 
J034402.4+320204 & P$_{\mathrm{7}}$ E$_{\mathrm{HC481}}$ & 2.1 & 0.61 & 0.27 & 30 & 27.7 & 5.2 & 0.52 & 30 & C \\ 
J034443.3+320131 & O$_{\mathrm{HC7148}}$ & 1.8 & 1.0 & 0.63 & 31 & 27.3 & 12.2 & 0.62 & 22 & I \\ 
J034741.6+325143 & P$_{\mathrm{C71}}$ O$_{\mathrm{H}}$ & $-$ & $-$ & $-$ & $-$ & 5.1 & 3.2 & 0.65 & 19 & I \\ 
J041354.7+281132 & P$_{\mathrm{7148}}$ N$_{\mathrm{HC}}$ & 0.28 & 0.19 & 0.56 & 25 & $-$ & $-$ & $-$ & $-$ & II \\ 
J041412.3+280837 & P$_{\mathrm{71}}$ N$_{\mathrm{HC}}$ & $-$ & $-$ & $-$ & $-$ & $-$ & $-$ & $-$ & $-$ & II \\ 
J041534.5+291347 & P$_{\mathrm{71}}$ N$_{\mathrm{HC}}$ & $-$ & $-$ & $-$ & $-$ & $-$ & $-$ & $-$ & $-$ & II \\ 
J041851.4+282026 & P$_{\mathrm{71}}$ E$_{\mathrm{C}}$ N$_{\mathrm{H}}$ & $-$ & $-$ & $-$ & $-$ & $-$ & $-$ & $-$ & $-$ & II \\ 
J041942.5+271336 & P$_{\mathrm{H7148}}$ E$_{\mathrm{C}}$ & 1.7 & 0.68 & 0.62 & 36 & 9.4 & 4.0 & 0.64 & 23 & I \\ 
J041959.2+270958 & O$_{\mathrm{H7148}}$ E$_{\mathrm{C}}$ & 2.2 & 0.67 & 0.62 & 41 & 39.3 & 4.1 & 0.61 & 45 & I \\ 
J042110.0+270142$^{\mathrm{B}}$ & P$_{\mathrm{7}}$ E$_{\mathrm{HC1}}$ & $-$ & $-$ & $-$ & $-$ & 13.3 & 1.5 & 0.42 & 35 & II \\ 
J042111.4+270108$^{\mathrm{B}}$ & P$_{\mathrm{71}}$ O$_{\mathrm{H}}$ E$_{\mathrm{C}}$ & $-$ & $-$ & $-$ & $-$ & 9.8 & 3.2 & 0.63 & 26 & C \\ 
J042656.3+244335 & P$_{\mathrm{C718}}$ O$_{\mathrm{H}}$ N$_{\mathrm{4}}$ & 0.85 & 0.36 & 0.29 & 25 & 1.5 & 0.9 & 0.43 & 15 & II \\ 
J042757.3+261918 & P$_{\mathrm{H71}}$ E$_{\mathrm{C48}}$ & 0.55 & 0.21 & 0.58 & 34 & 16.6 & 2.3 & 0.66 & 41 & I \\ 
J042838.9+265135 & P$_{\mathrm{H7148}}$ E$_{\mathrm{C}}$ & 6.1 & 0.64 & 0.49 & 60 & 5.6 & 3.8 & 0.69 & 20 & I \\ 
J042905.0+264904 & P$_{\mathrm{7}}$ O$_{\mathrm{1}}$ N$_{\mathrm{HC}}$ & $-$ & $-$ & $-$ & $-$ & $-$ & $-$ & $-$ & $-$ & II \\ 
J042907.6+244350 & P$_{\mathrm{71}}$ N$_{\mathrm{HC}}$ & $-$ & $-$ & $-$ & $-$ & $-$ & $-$ & $-$ & $-$ & II \\ 
J042923.6+243302 & P$_{\mathrm{H71}}$ E$_{\mathrm{C}}$ & $-$ & $-$ & $-$ & $-$ & 27.2 & 5.0 & 0.66 & 36 & I \\ 
J043150.6+242418 & P$_{\mathrm{71}}$ N$_{\mathrm{HC}}$ & $-$ & $-$ & $-$ & $-$ & $-$ & $-$ & $-$ & $-$ & II \\ 
J043215.4+242903 & P$_{\mathrm{718}}$ N$_{\mathrm{HC4}}$ & $-$ & $-$ & $-$ & $-$ & $-$ & $-$ & $-$ & $-$ & II \\ 
J043232.0+225726 & P$_{\mathrm{7148}}$ N$_{\mathrm{HC}}$ & 0.39 & 0.34 & 0.61 & 23 & $-$ & $-$ & $-$ & $-$ & II \\ 
J043316.5+225320 & P$_{\mathrm{H7148}}$ N$_{\mathrm{C}}$ & 0.67 & 0.56 & 0.55 & 23 & 2.2 & 2.2 & 0.52 & 13 & II \\ 
J043535.0+240822 & P$_{\mathrm{718}}$ E$_{\mathrm{HC}}$ N$_{\mathrm{4}}$ & 0.22 & 0.20 & 0.66 & 25 & 7.1 & 1.7 & 0.45 & 25 & II \\ 
J043556.7+225436 & N$_{\mathrm{HC71}}$ & $-$ & $-$ & $-$ & $-$ & $-$ & $-$ & $-$ & $-$ & II \\ 
J043953.9+260309 & P$_{\mathrm{HC7148}}$  & 7.9 & 1.8 & 0.65 & 49 & 87.5 & 14.7 & 0.67 & 38 & I \\ 
J044138.8+255626 & P$_{\mathrm{7148}}$ N$_{\mathrm{HC}}$ & 0.13 & 0.10 & 0.65 & 29 & $-$ & $-$ & $-$ & $-$ & II \\ 
\hline                                   %inserts single line
\end{tabular}
}
\tablefoot{ \\
Sources marked with B superscript indicate a visible binary or multiple system based on PACS 70 $\mu$m data. \\
\tablefoottext{a}{Emission classification: P = peaking on-source, O = offset peaking, E = extended emission, N = non-detection. 
Subscript identifiers: H = HCO$^{+}$ 4--3, C = C$^{18}$O 3--2, 7 = 70 $\mu$m, 1 = 160 $\mu$m, 4 = 450 $\mu$m, 8 = 850 $\mu$m} \\
\tablefoottext{b}{Fluxes for 850 $\mu$m were derived using a 23$\arcsec$ beam.} \\
\tablefoottext{c}{$R_{\mathrm{obs}}$ was derived from the FWHM of a circular 2D Gaussian fit to the source emission.} \\
\tablefoottext{d}{Spatially and spectrally integrated intensities for HCO$^{+}$ 4--3 were derived using a 15$\arcsec$ beam.}
}
\end{table*}

\subsection{Concentration factors}
\label{sec:concfac}

To quantitatively distinguish embedded sources from non-embedded disks or background features, we can set constraints on the
concentration of HCO$^{+}$ 4--3 and 850 $\mu$m emission. In this section we aim to bring a quantitative definition to identify truly
embedded, early-stage sources based on the degree of concentration in their emission. Previous studies of SCUBA 850 $\mu$m
emission \citep{Johnstone2001, Walawender2005} define the following concentration factor: 

\begin{equation}
 C_{850} = 1 - \frac{1.13 B^{2} S_{850}}{\pi R^{2}_{\mathrm{obs}} F_{0}}
 \label{eq:conc850}
\end{equation}

\noindent where $B$ is the diameter of the instrument beam, $R_{\mathrm{obs}}$ is the radius of the source envelope in the same units as the beam size,
$S_{850}$ is the total 850 $\mu$m flux within $R_{\mathrm{obs}}$ in units of Jansky, and $F_{0}$ is the peak flux of the envelope
in Jansky beam$^{-1}$. The 850 $\mu$m concentration factor gives an initial insight into the nature of the core. 
\citet{vanKempen2009} suggested that concentrated sources ($C_{850} > 0.75$) likely contain protostars while low concentration
sources ($C_{850} < 0.40$) are unlikely to be embedded YSOs. In our calculations we determine $R_{\mathrm{obs}}$ by fitting a circular 2D Gaussian 
to the emission profile. $R_{\mathrm{obs}}$ is set equal to the FWHM of the Gaussian fit. In some cases masking was necessary to
exclude contaminating emission or emission from nearby sources, in which case a mask radius was chosen with respect to the source position, 
and all points outside the mask were set to a representative background value. If the FWHM of the 2D Gaussian fit exceeded the masking radius, 
$R_{\mathrm{obs}}$ was set equal to the mask radius.
To determine the flux, maps were converted to units of Jansky using a beam correction factor of $\pi (B/2) ^{2}$ where $B$ = 23$\arcsec$.
This is larger than the main-beam size of the JCMT at this wavelength ($\sim$15$\arcsec$) due to the data reduction methods and error beam corrections applied by \citet{diFrancesco2008}.
The SCUBA 850 $\mu$m concentration factors, total flux, peak flux, and envelope radii are listed in Table~\ref{tab:classification}.

To identify whether these sources are truly embedded, we can apply the same formula to the HCO$^{+}$ 4--3 emission.
Following a method similar to that described by \citet{vanKempen2009}, Eq.~\ref{eq:conc850} can be translated to a spectral map concentration parameter
that is analogous to the calculation of the 850 $\mu$m concentration factor:

\begin{equation}
 C_{\mathrm{HCO^{+}}} = 1 - \frac{1.13 B^{2} S_{\mathrm{HCO^{+}}}}{\pi R^{2}_{\mathrm{obs}} S_{\mathrm{HCO^{+}}}^{\mathrm{peak}}}
 \label{eq:concHCO}
\end{equation}

\noindent where $B$ and $R_{\mathrm{obs}}$ retain their previous definitions with $B$ = 15$\arcsec$, the HARP beam size. 
$S_{\mathrm{HCO^{+}}}^{\mathrm{peak}}$ is the peak integrated
intensity within the beam, and $S_{\mathrm{HCO^{+}}}$ is the spatially and spectrally integrated intensity
$\int T_{\mathrm{{mb}}} \mathrm{d}V$ within $R_{\mathrm{obs}}$. To calculate $S_{\mathrm{HCO^{+}}}$, each
integrated intensity map first has a beam correction applied in the same manner as the SCUBA maps, with a correction factor of $\pi (B/2)^{2}$ where $B$ = 15$\arcsec$.
The resulting concentration factors are given in Table~\ref{tab:classification}. 

Concentration factors at 850$\mu$m have been previously 
calculated using \textsc{clumpfind} \citep{Williams1994} in Perseus \citep[and references therein]{Kirk2006}. About half of the sources have concentration factors consistent to
within 20\%, but some cases differ from the values reported here by up to 50\%. These discrepancies can be attributed to the different methods used to 
determine $R_{\mathrm{obs}}$. \textsc{clumpfind} extends the radius out to a pre-determined sensitivity limit, often resulting in larger $R_{\mathrm{obs}}$ than
that derived from the 2D Gaussian fits used here. Applying masks to exclude emission from nearby sources also has an effect on the $R_{\mathrm{obs}}$ 
determined from a 2D Gaussian fit. The 850 $\mu$m flux values reported in \citet{diFrancesco2008} agree with the $S_{850}$ flux values presented in this 
paper to within 10\%, which is within the flux uncertainties.

\subsection{Evolutionary stages}
\label{sec:evstage}

\begin{figure*}[!htbp]
  \centering
  \resizebox{\hsize}{!}{
  \includegraphics[width=0.9\textwidth]{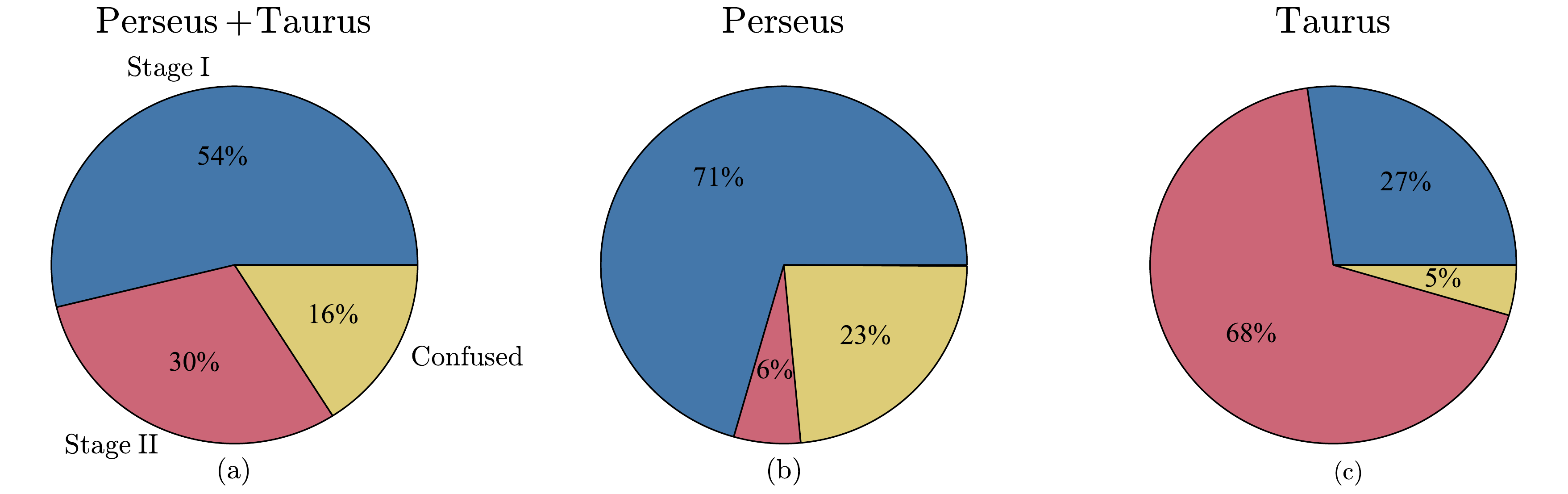}
  }
  \caption{Statistics for stage classification. Stage I embedded objects are shown in light blue, Stage II objects are shown in light red, Confused objects are shown in yellow.
  \textit{(a)} The entire sample observed in HCO$^{+}$ 4--3 and C$^{18}$O 3--2 from the Perseus and Taurus clouds: 56 sources.
  \textit{(b)} Statistics for only the Perseus cloud: 34 sources. \textit{(c)} Statistics for only the Taurus cloud: 22 sources.}
  \label{fig:emb_stats}
\end{figure*}

Given our concentration factor definitions, we can apply a similar scheme for stage identification as \citet{vanKempen2009}.
The classification of an object depends on molecular line observations of HCO$^{+}$ 4--3 and C$^{18}$O 3--2 as well as
emission in the continuum, in particular the 850 $\mu$m SCUBA observations. Multi-wavelength observations are necessary
for a proper classification of the Stage. Detection and concentration analysis of only HCO$^{+}$ 4--3 is not sufficient.
For Stage I, both HCO$^{+}$ 4--3 and the 850 $\mu$m continuum must be detected within 5--10$\arcsec$ of the source position and show compact emission. 
The C$^{18}$O 3--2 line may be compact or extended, but it must also be detected within a 5--10$\arcsec$ offset.
Specifically, the following criteria are adopted for stage classification:

\begin{itemize}
 \item Stage I:
  \begin{itemize}
      \item HCO$^{+}$ 4--3 peaking on-source and/or contained entirely within the 15$\arcsec$ beam, with $C_{\mathrm{HCO^{+}}} > 0.6$ and $\int T_{\mathrm{{mb}}} \mathrm{d}V > 0.4$ K km s$^{-1}$ \textit{and};
      \item SCUBA emission at the source position that is not point-like \textit{and}; 
      \item detection of C$^{18}$O 3--2 at the source position.
  \end{itemize}

 \item Stage II:
  \begin{itemize}
      \item non-detection of HCO$^{+}$ 4--3 and SCUBA down to rms levels, or $C_{\mathrm{HCO^{+}}}$ and $C_{\mathrm{850}} < 0.4$ \textit{or};
      \item non-detection of C$^{18}$O 3--2 down to rms levels, or no variation of C$^{18}$O 3--2 on scales of 30$\arcsec$ \textit{or};
      \item point-like emission in all continuum bands.
  \end{itemize}

 \item Confused:
  \begin{itemize}
      \item HCO$^{+}$ 4--3, SCUBA, and C$^{18}$O 3--2 peaking at an offset of 20$\arcsec$ or more from source position.
  \end{itemize}
\end{itemize}

In order to be classified as Stage I, a source must meet all of the Stage I criteria;
partially meeting the requirements is not sufficient.
The concentration factor criteria adopted here are less strict than
those defined in \citet{vanKempen2009}. 

To motivate our criteria for HCO+, it is useful to compare with models
of the line emission as a function of evolutionary stage. This has been done by \citet[][see Figure 6 in their paper]{Hogerheijde1997} using the same type of \citet{Shu1977} collapse model 
that was used to model the changes in continuum SEDs with evolution. Both the spatial extent and integrated envelope intensities of multiple 
HCO$^{+}$ lines were modeled using the full non-LTE excitation and radiative
transfer code \textsc{RATRAN} \citep{Hogerheijde2000}. For a constant HCO$^{+}$ abundance of $1.2 \times 10^{-8}$
they find that HCO$^{+}$ 4--3 emission during the Stage I lifetime ($\lesssim$ few$\times$10$^{5}$ yr)
can be expected on spatial scales of $\sim 30-60\arcsec$ for sources in Taurus. This number will
be up to a factor of two lower for sources in the more distant Perseus cloud, but is consistent with the scales on which 
the majority of Stage I sources in this paper are fitted by concentration factor analysis.
As the Shu collapse model evolves in time toward Stage II, the HCO$^{+}$ 4--3 FWHM of 
emission from the source drops drastically, near to the 14$\arcsec$ beam size with which the predictions
were convolved, which is consistent with the beam size of the JCMT.

The same models
show that the expected integrated envelope intensities for HCO$^{+}$ 4--3  within Stage I lifetimes 
are on the order of 10-20 K km s$^{-1}$ and drops to $\sim$5 K km s$^{-1}$. 
These values are in the same range or
somewhat larger than our observed intensities, but the model values
depend on the adopted HCO+ abundance.

Another comparison can be made with models of HCO$^{+}$ 4--3 and C$^{18}$O 3--2
emission from the deeply embedded source IRAS 2A located in NGC 1333 
using \textsc{RATRAN} as presented in San-Jose Garcia, et al. (subm.). They also use a 
constant abundance profile for HCO$^{+}$ 4--3 whereas that for C$^{18}$O 3--2 takes
freezeout onto grains into account at lower temperatures. The same analysis as described in 
Section~\ref{sec:concfac} is applied to both lines. The 2D Gaussian
fits result in $R_{\mathrm{obs}}$ values of 21$\arcsec$ and 26$\arcsec$ for 
HCO$^{+}$ 4--3 and C$^{18}$O 3--2, respectively. Concentration factors are $C = 0.58$
for both lines. These are taken as indicative values for an embedded YSO
in the analysis of our observed sample.

Based on the expected scale and intensity of the
HCO$^{+}$ 4--3 and 850 $\mu$m emission, a concentration factor
$C > 0.6$ is sufficient to classify a source as a Stage I
embedded object. The integrated intensity limit on HCO$^+$ 4--3 of
0.4 K km s$^{-1}$ is derived from observations of Stage II
sources with disks where, due to beam dilution, the disk emission
in a 15$\arcsec$ beam is generally well below this limit, 
of order 0.1 K km s$^{-1}$ \citep[e.g.,][]{Thi2004}.

Applying these definitions to the observed sample, 30 out of 56
sources (54\%) are Stage I deeply embedded protostars, 17 (30\%) are Stage II classical T Tauri stars with a surrounding disk,
and 9 sources (16\%) remain confused based on a morphology
in gas and dust emission that does not cleanly fit the criteria for Stage I or Stage II.  
These statistics include all sources in both Perseus and
Taurus. Sources in Serpens could not be classified as they were only
observed in C$^{18}$O 3--2. 

Separating the classifications by cloud reveals that Perseus
contains a higher percentage of embedded sources than
Taurus.  Out of the 34 sources in Perseus, 24 (71\%) are bonafide
Stage I, 2 (6\%) are Stage II, and 8 (23\%) are confused sources.  Of
22 sources in Taurus, 6 (27\%) are classified as Stage I, 15 (68\%)
are Stage II, and 1 (5\%) is confused. These statistics are represented
in Figure~\ref{fig:emb_stats}.

All sources with emission at all wavelengths that fall under the
on-source centrally concentrated observational category outlined in
Section~\ref{sec:res_morph} (i.e., P$_{\mathrm{HC7148}}$) are
identified as bonafide Stage I sources using the concentration factor
method. For the entire sample there is only a single object,
J043316.5+225320, that shows HCO$^{+}$ 4--3 peaking on-source
(P$_{\mathrm{H}}$) but falls under the Stage II category due to low
concentration factors for 850 $\mu$m and HCO$^{+}$ 4--3, and
non-detection in C$^{18}$O 3--2. Thus, centrally concentrated
on-source HCO$^{+}$ 4--3 emission is a good qualitative indicator that
the object is a truly embedded YSO.  Five of the eight confused sources
have widespread emission in both molecular lines and SCUBA dust
continuum emission (E$_{\mathrm{HC48}}$) and often contain a peak
within the extended emission that is offset from the source position.
The three remaining confused sources are classified as such for
different reasons $-$ further explanation can be found in
Appendix~\ref{app:A}.

Stage II sources either exhibit a morphology that is extended in
HCO$^{+}$ 4--3 and C$^{18}$O 3--2 with no clear peak in emission, thus
resulting in a low concentration factor, or there is no detection in
either molecular line (N$_{\mathrm{HC}}$). Non-detection of both
molecular lines suggests an evolved PMS star with a disk. One example
is the case of IRS63 in Oph, where previous interferometric data of
HCO$^{+}$ with the Sub-Millimeter Array (SMA) have revealed a disk
\citep{Lommen2008,Brinch2013}. The integrated intensity in HCO$^{+}$
3--2 when convolved with the JCMT beam is 0.28 K km s$^{-1}$ and
originates mostly from the disk \citep{vanKempen2009}. 
Due to the weak HCO$^{+}$ 4--3 contribution and small size of the
emitting region in disks, the signal is beam-diluted below the
sensitivity limit of the single-dish JCMT observations.

It should be noted that there are two caveats when relying on the
presence of a central peak in the HCO$^{+}$ spatial map for Stage I
identification: HCO$^{+}$ may be optically thick which may prevent a peak;
and CO, the precursor of HCO+, may be frozen out in the lowest
luminosity and/or coldest sources. Models of NGC 1333 IRAS 2A, one
of the strongest HCO$^{+}$ sources, demonstrate that both issues
do not largely affect the results.

\subsection{HCO$^{+}$ infall signatures}
The presence of a blue-dominated asymmetry in the line profile 
of a HCO$^{+}$ single pointing observation is often used to provide strong evidence of infall 
of dense material from the outer envelope \citep{Gregersen1997,Mardones1997,Myers2000}. 
Three sources in Taurus are identified as having such an asymmetry in their central spectra
contained within the 15$\arcsec$ JCMT beam (see Section~\ref{sec:res_spec}).
All three are found to be Stage I sources according to the concentration
factor method presented here.

Other sources show interesting line profile shapes
with a blue-dominated excess and self-absorption at the source velocity of the spectrum.
These include J032837.1+311328, J032904.0+311446 (HH 7-11), J032951.8+313905, 
J033121.0+304530, J033218.0+304946, and J034443.3+320131. 
Five of the six sources are classified as Stage I in this paper, with 
J032904.0+311446 (HH 7-11) classified as confused due to the widespread 
HCO$^{+}$ 4--3 emission in the field, which may explain its irregular line profile.
Overall, spectrally resolved line profile asymmetries are not frequently observed 
and therefore infall signatures cannot be used for widespread confirmation
of the Stage I nature YSOs. Multiwavelength spatial maps provide the strongest technique 
thus far to characterize YSOs.

\subsection{HCO$^{+}$ single pointing vs. mapping}
\label{sec:singlept}
Other attempts are being made to determine the evolutionary stage of
YSOs with HCO$^{+}$. A study by \citet{Heiderman2015} has 
classified YSOs based solely on the detection of HCO$^{+}$ 3--2 
at the source position without the aid of spatial mapping.
It is useful to test the robustness of this method of classification
 by comparing our own central spectra detections in
HCO$^{+}$ 4--3 to the stages determined via the concentration
factors. Based on the HCO$^{+}$ 4--3 spectra extracted from the
source position, shown in Appendix~\ref{app:B}, and the spectral
characteristics listed in Table~\ref{tab:data_props}, our sample
consists of 43 detections out of the 56 objects in Perseus and
Taurus. 

Based on the concentration factor criteria of
HCO$^{+}$ 4--3, 32 out of 56 sources in our sample are Stage I, leaving many
sources that would be misidentified using the detection-only
method. Some of these are Stage II sources in our assignment, others have
low concentration factors or have emission that is
offset by 20$\arcsec$ or more away from the source position. 

Based on the comparison of central spectrum detection
to concentration factor classification in our sample, there are likely
to be misassignments on the order of 20--30\% in the identification of
Stage I embedded sources if only the detection of HCO$^{+}$ is applied
as an indicator of protostellar stage. A single 
pointing detection cannot properly characterize the YSO, thus
a detection-only sample is likely to introduce a bias that 
will overestimate the number of Stage I sources and thus also their
lifetime. This point is discussed further in Section~\ref{sec:lifetimes}.

\begin{figure}[!htbp]
  \centering
  \includegraphics[width=0.45\textwidth]{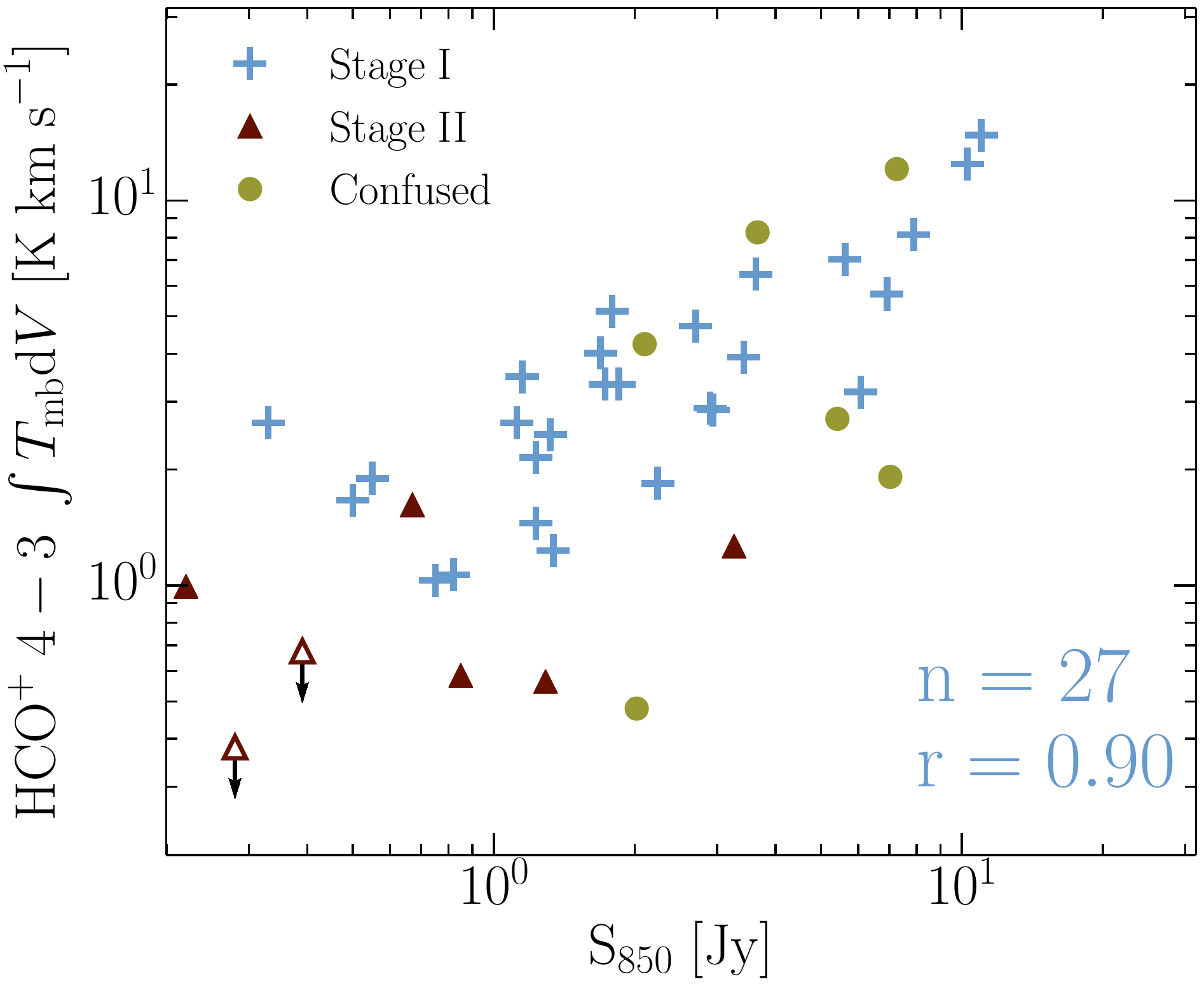} \\
  \includegraphics[width=0.45\textwidth]{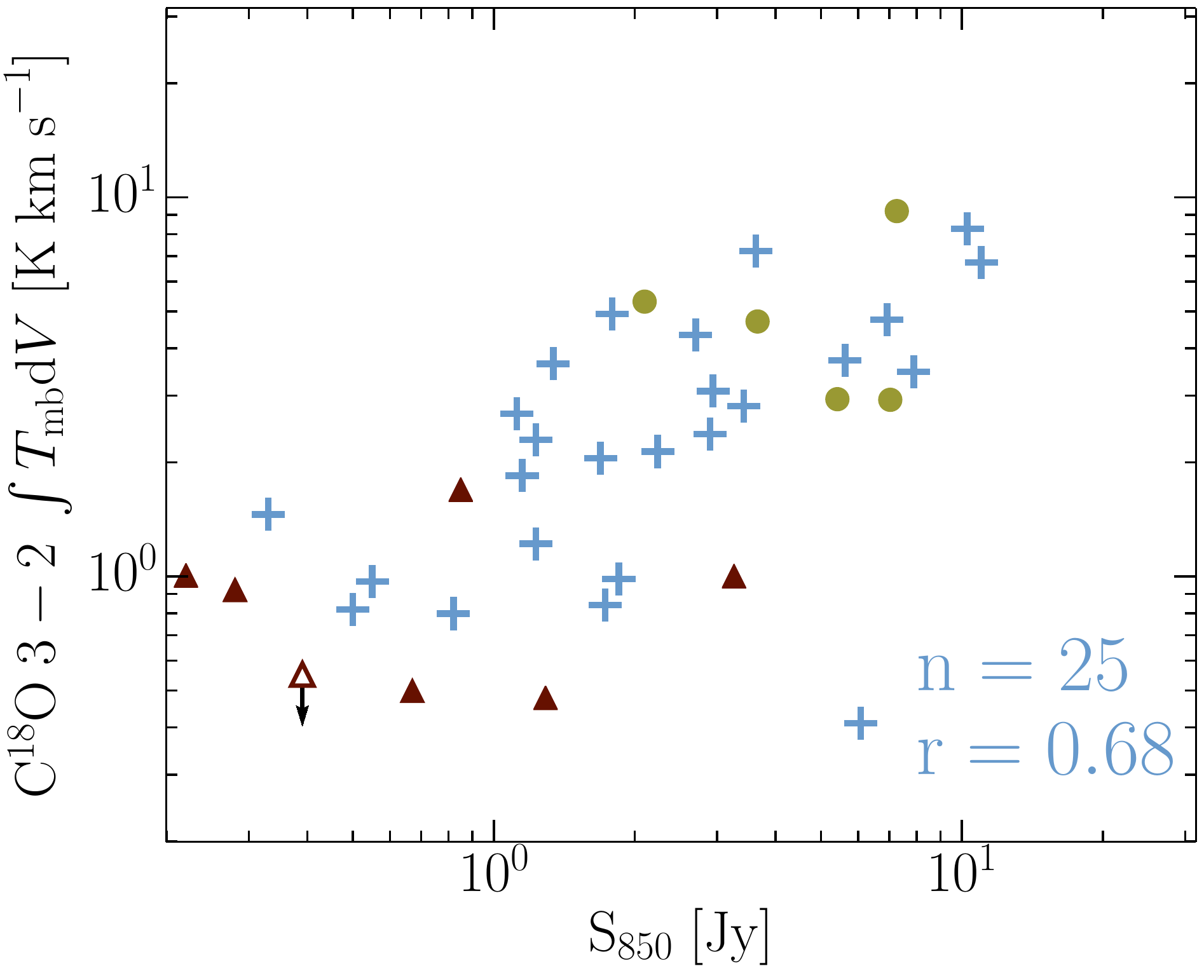} \\
  \includegraphics[width=0.45\textwidth]{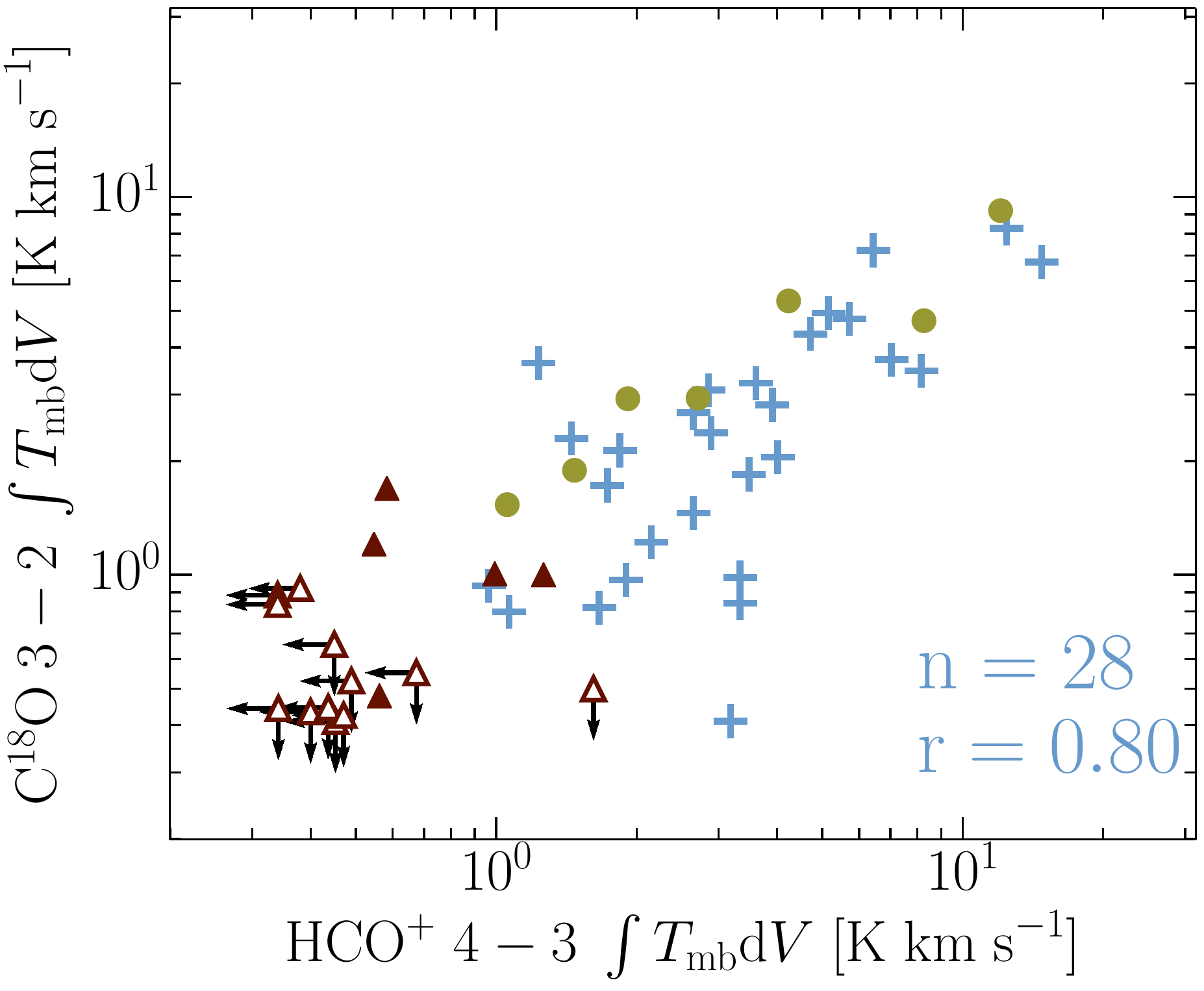} \\
  \caption{Molecular line intensity and dust continuum flux correlations. Integrated intensities for spectral lines are taken from the central spectrum of each source.
  Flux values at 850 $\mu$m are taken from concentration factor analysis. Stage I (blue cross), Stage II (red triangle), and Confused (yellow circle) sources are shown. Upper limits are shown in hollow markers with black arrows indicating the bounds.
  Sample size and Pearson correlation coefficient for Stage I sources are given in the lower-right corner of each plot.
  Serpens sources are excluded.
  \textit{Top:} HCO$^{+}$ 4--3 versus 850 $\mu$m flux
  \textit{Middle:} C$^{18}$O 3--2 versus 850 $\mu$m flux
  \textit{Bottom:} C$^{18}$O 3--2 versus HCO$^{+}$ 4--3.}
  \label{fig:corr_plots}
\end{figure}

\subsection{Correlations}
\label{sec:fluxcorr}
Trends between the two observed molecular lines and 850 $\mu$m dust
emission may give insight into the relative amounts of high (column)
density gas and cool dust that are present in Stage I and Stage II
sources. Figure~\ref{fig:corr_plots} shows three different correlation
plots between these tracers. Pearson correlation coefficients are
calculated for each comparison using only Stage I detections.
The significance of each correlation is calculated in terms of 
$\sigma_{\rm{P}} = |r| \sqrt{n-1}$, for a correlation coefficient $r$ 
and sample size $n$ \citep{Marseille2010}. Molecular line integrated 
intensities are taken from the
central spectra of the source and are therefore contained within a
15$\arcsec$ beam whereas the 850 $\mu$m flux values are taken from
concentration factor analysis as described in
Section~\ref{sec:concfac} and apply to a somewhat larger 23$\arcsec$
beam.

The top panel shows the HCO$^{+}$ 4--3 integrated intensities versus
the 850 $\mu$m flux for all sources where both tracers are
detected. These values are found to be strongly correlated, with a significance
of 4.6$\sigma_{\rm{P}}$. Such a correlation can be
expected, particularly for Stage I sources. More envelope material,
as traced by the cooler dust emission at 850 $\mu$m, suggests that
there will be a larger number of molecules present in the gas phase
for embedded protostars, assuming that the gas and dust are well
coupled and that the abundance of the molecule is constant throughout
this region. All Stage II sources with 3$\sigma_{I}$ detections in
HCO$^{+}$ 4--3 have low intensity ($\leq1.3$ K km s$^{-1}$). Though
the numbers are small, it is interesting to note that 
for positive detections the HCO$^{+}$
4--3 intensity remains relatively flat while 850 $\mu$m flux varies
by an order of magnitude for Stage II sources. This cannot be used as
a threshold for separation of Stage I and Stage II, however, as there
are Stage I sources that share these low HCO$^{+}$ 4--3
intensities. Spatial maps are therefore a requirement for proper
identification. 

Note that for Stage I sources, some fraction of the continuum emission
($\sim$20--80\%, \citealt{Hogerheijde1997,Jorgensen2009}) originates
from the disk rather than the envelope whereas the majority of the
line emission originates in the envelope. The variable disk continuum
contribution will tend to lower the correlation.

C$^{18}$O 3--2 integrated intensity and 850 $\mu$m flux also show
a positive correlation with a 3.3$\sigma_{\rm{P}}$ significance.
Flux for Stage II sources here generally remain low in both C$^{18}$O 3--2 
and 850 $\mu$m. Several Stage I sources also share these low
intensity and flux values, but only Stage I or confused sources 
are found with C$^{18}$O 3--2 intensity $\gtrsim 
2.0$ K km s$^{-1}$.  The C$^{18}$O 3--2 versus HCO$^{+}$ 4--3
trend in the bottom panel of Figure~\ref{fig:corr_plots} is
well correlated (4.2$\sigma_{\rm{P}}$). In this plot the Stage II
sources are most tightly grouped around low HCO$^{+}$ 4--3 and
C$^{18}$O 3--2 intensities, with only a few Stage I or
confused sources of comparable low intensity for both lines. These
plots indicate that HCO$^{+}$ 4--3 is indeed the best tracer for
Stage I identification and that an HCO$^{+}$ 4--3 intensity
threshold of ${\gtrsim} 0.6$ K km s$^{-1}$ will result in a majority of
Stage I sources being correctly identified. However, spatial mapping is still
needed to properly classify those sources with HCO$^{+}$ 4--3
intensity above this threshold but with low spatial concentration.

\subsection{Comparison to other studies}
\label{sec:otherstudies}
\subsubsection{Perseus sources}
\label{sec:ospers}
\begin{table}[!htbp]
\caption{Properties of Perseus sources -- HCO$^{+}$ 3--2, and H$_{2}$O obseravations in other studies}
\resizebox{9cm}{!}{
\begin{tabular}{lccc}
\hline \hline
Object & Stage\tablefootmark{a} & Stage I?\tablefootmark{b} & PACS H$_{2}$O?\tablefootmark{c} \\
\hline
J032522.3+304513 & I & - & Y \\ 
J032536.4+304523 & I & - & Yc\tablefootmark{d} \\ 
J032637.4+301528 & I & - & N \\ 
J032837.1+311328 & I & - & Y \\ 
J032900.6+311200 & I & Y & - \\ 
J032901.6+312028 & C & - & Ye\tablefootmark{e} \\ 
J032904.0+311446 & C & Y & - \\ 
J032910.7+311820 & I & Y & Yc\tablefootmark{d} \\ 
J032917.2+312746 & I & Y & - \\ 
J032918.2+312319 & I & Y & - \\ 
J032923.5+313329 & I & Y & - \\ 
J032951.8+313905 & I & Y & N \\ 
J033121.0+304530 & I & Y & Y \\ 
J033218.0+304946 & I & Y & N \\ 
J033313.8+312005 & C & N & - \\ 
J033314.4+310710 & C & Y & N \\ 
J033320.3+310721 & C & Y & - \\ 
J033327.3+310710 & I & Y & Y \\ 
J034356.9+320304 & I & - & Ye\tablefootmark{e} \\ 
J034443.3+320131 & I & - & Ye\tablefootmark{e} \\ 
\hline
\end{tabular}
}
\tablefoot{
\tablefoottext{a}{Classification in this paper.}
\tablefoottext{b}{Classification from \citet{Heiderman2015} based on single pointing detection of HCO$^{+}$ 3--2.}
\tablefoottext{c}{Detection of the H$_{2}$O (2$_{12}$ -- 1$_{01}$) 179.527$\mu$m line from {\it Herschel} PACS observations \citep{Karska2014}.}
\tablefoottext{d}{Source map is contaminated by emission from other nearby sources.}
\tablefoottext{e}{Emission is extended and associated with the target source.}
}
\label{tab:per_comp}
\end{table}

\begin{table*} 
\caption{Properties of Taurus sources -- color criteria, HCO$^{+}$ 3--2, and H$_{2}$O obseravations in other studies}
\centering
\begin{tabular}{lcccc}
\hline
\hline
Object & Stage\tablefootmark{a} & Embedded/Visible\tablefootmark{b} & HCO$^{+}$ 3--2?\tablefootmark{c} & PACS H$_{2}$O?\tablefootmark{d} \\ 
\hline
J041354.7+281132 & II & Embedded & N & $-$ \\ 
J041959.2+270958 & I & Embedded & Y & Y \\ 
J042110.0+270142 & II & Embedded & Y & $-$ \\ 
J042111.4+270108 & C & $-$ & $-$ & Y \\
J042656.3+244335 & II & Embedded & Y & $-$ \\ 
J042757.3+261918 & I & Embedded & Y & Y \\ 
J042907.6+244350 & II & $-$ & $-$ & Y \\
J042923.6+243302 & I & Embedded+Visible & Y & $-$ \\ 
J043150.6+242418 & II & Visible & N & $-$ \\ 
J043215.4+242903 & II & Visible & N & $-$ \\ 
J043232.0+225726 & II & Embedded & N & $-$ \\ 
J043316.5+225320 & II & Embedded & N & $-$ \\ 
J043535.0+240822 & II & Embedded & Y & Y \\ 
J043953.9+260309 & I & Embedded & Y & $-$ \\ 
\hline
\end{tabular}
\tablefoot{
\tablefoottext{a}{Classification in this paper.}
\tablefoottext{b}{Classification taken from \citet{Hogerheijde1997} based on a flux- and color-limited sample.}
\tablefoottext{c}{Detection of HCO$^{+}$ 3--2 within a 19$\arcsec$ beam \citep{Hogerheijde1997}.}
\tablefoottext{d}{Detection of the H$_{2}$O (2$_{12}$ -- 1$_{01}$) 179.527$\mu$m line from {\it Herschel} PACS observations (Mottram et al., in prep.).}
}
\label{tab:tau_comp}
\end{table*}

The source sample observed in HCO$^{+}$ 3--2 and classified by
\citet{Heiderman2015} has 13 sources in common with our
sample. Though it is a small subset from Perseus, comparing their
classification to our assignments provides a first look at how
well the single pointing detection method matches with the spatial map
concentration factor method. Note that for single pointings the only
requirement for a Stage I source classification is based on the
detection criterion $\int T_{\mathrm{{mb}}} \mathrm{d}V \geq 0.68 $ K
km s$^{-1}$ for the HCO$^{+}$ 3--2 line in a 28$\arcsec$ beam, which is
more conservative than our integrated intensity criterion for  HCO$^{+}$ 4--3. There is no
effort to identify the source specifically at another stage.
Table~\ref{tab:per_comp} shows the common sources and designations
from each study.

For every source in Table~\ref{tab:per_comp} that is assigned a Stage
I classification in this paper, \citet{Heiderman2015} also report a confirmed Stage I
source from their HCO$^{+}$ 3--2 data. Note that they define an earlier Stage 0, 
but make no distinction between Stage 0 and Stage I sources observationally. 
No such distinction is made in this paper. Stage I encompasses
the entire embedded phase of the YSO. Four
differences arise from this comparison. J032904.0+311446 (HH 7-11
MMS6), J033314.4+310710 (B1-SMM3), and J033320.3+310721 (B1-b) are identified as confused
sources in this work due to their emission morphology (see
Section~\ref{sec:evstage}) while they are identified as Stage I
sources in \citet{Heiderman2015}. J033313.8+312005 is
also identified as confused in this work, though it may be closer to 
Stage II (see Appendix~\ref{app:A}).
\citet{Heiderman2015} report no detection of HCO$^{+}$ 3--2 in their observations 
of this source and conclude that it is not Stage I.
The two methods are both effective and
mostly consistent for this given subset, but spatial mapping is needed
for information about the environment surrounding the source in order
to confirm the Stage I assignment.

\citet{Karska2014} present {\it Herschel} PACS spectral maps of 22
sources in the Perseus molecular cloud as part of the ``William
\textit{Herschel} Line Legacy'' (WILL) survey (Mottram et al., in
prep.), including 13 in common with the sample studied in this
paper. Table~\ref{tab:per_comp} indicates that the H$_{2}$O (2$_{12}$
-- 1$_{01}$) 179.527$\mu$m transition was detected within the PACS
central ($\sim$9.5$\arcsec \times $9.5$\arcsec$) spaxel for four Stage
1 sources while three show no detection. The confused source
J033314.4+310710 (B1-SMM3) also lacks detection. A further two Stage I
sources, J034356.9+320304 (IC 348-MMS) and J034443.3+320131, and one
confused source, J032901.6+312028, all show H$_{2}$O (2$_{12}$ --
1$_{01}$) emission extending outside of the PACS central spaxel, but
still associated with the source.  Water emission from
J032536.4+304523 (L1448-N) and J032910.7+311820 are both confused
by nearby sources. In this paper J032910.7+311820 is marked as a
binary source, with dominant emission offset to the northeast of the
target source, which is consistent with the location of contaminating
H$_{2}$O (2$_{12}$ -- 1$_{01}$) emission.  There is no such similarity
for J032536.4+304523 (L1448-N). Overall there does not seem to be a
clear relationship between the Stage I classification of sources
presented here and the nature of H$_{2}$O (2$_{12}$ -- 1$_{01}$)
emission from the same sources.

\subsubsection{Taurus sources}
\label{sec:ostau}
Several Taurus sources in this paper have previously been classified
based on a simpler identification scheme.  The sample presented in
Table~\ref{tab:tau_comp} was taken from \citet{Tamura1991}, where they
used a flux- and color-limited sample of IRAS point sources with the
following selection criteria: log ($F_{\nu}$(25
$\mu$m)/$F_{\nu}$(60 $\mu$m)) $< -0.25$, and $F_{\nu} >
5$ Jy at either 60 or 100$\mu$m.  These color criteria were used to
determine if the object was optically visible or
embedded. \citet{Hogerheijde1997} performed single pointing
observations of HCO$^{+}$ 3--2 towards these sources and analyzed only those with the
strongest line intensities. Table~\ref{tab:tau_comp} provides the
stage classification given in this paper, the color criteria
classification, and whether or not HCO$^{+}$ 3--2 was detected
within a 19$\arcsec$ beam.  All detections marked `Y' have integrated
intensities greater than the \citet{Heiderman2015} threshold, and all but
J042656.3+244335 have $\int T_{\mathrm{{mb}}} \mathrm{d}V > 1.0 $ K km
s$^{-1}$. Therefore, these sources would be identified as embedded
Stage I sources. All sources marked `N' indicate no detection in 
HCO$^{+}$ 3--2 and would be classified as Stage II sources
using the \citet{Heiderman2015} threshold.

It is immediately clear that the color criteria definitions are highly
inconsistent with the method of stage classification presented here.
Out of 13 sources, nearly half would be incorrectly identified as
embedded when they are in fact Stage II evolved PMS stars with disks.
The detection of HCO$^{+}$ 3--2 is a better match, but discrepancies
remain. J042110.0+270142, J042656.3+244335, and J043535.0+240822 are
indeed detected above the 3$\sigma_{I}$ limit in the central spectra of
the HCO$^{+}$ 4--3 data presented here, but their spatial maps
reveal that the emission is not well concentrated. These three sources
are assigned a Stage II classification despite detection in HCO$^{+}$
4--3.  As with the \citet{Heiderman2015} comparison, the comparison
for a small sample in Taurus reinforces the need for spatial maps to
characterize the environment and prevent improper identification of
truly embedded sources. Similar to the central spectra versus
spatial map test in Section~\ref{sec:evstage}, the difference in
Taurus assignments shows that there is approximately 20--30\% error in
classification if only single pointing observations are used.

Six sources in Taurus overlap with the WILL sample observed with {\it
  Herschel}. Five of these have PACS detections of the 179.527$\mu$m
water line on-source or within one spaxel offset (see
Table~\ref{tab:tau_comp}, also Karska et al., in prep.). J042110.0+270142 and J042757.3+261918 are
Stage I embedded YSOs, J042907.6+244350 and J043535.0+240822 are Stage
II objects, and J042111.4+270108 remains confused. J043232.0+225726 is
the sixth source in common, a Stage II object, but was not observed
with PACS.  No water lines were detected in the {\it Herschel} HIFI
observations of this source (Mottram et al., in prep.). As with the sample in Perseus, there is
no clear trend between the Stage of the YSO and the presence of water
emission.

There are two possible reasons for this lack of correlation. First,
water traces the currently shocked gas where the jet or wind impacts
on the envelope, rather than the quiescent envelope or
entrained outflow material \citep{Kristensen2012,Mottram2014}. Second,
some water emission can arise from the disk rather than the shocks
\citep{RiviereMarichalar2012,Fedele2013}.

\subsubsection{Near-IR veiling as a diagnostic}
\label{osnirspec}
An independent diagnostic of evolutionary stage can be provided by
optical and near infrared spectroscopy of the young stars themselves
rather than their surroundings. This technique is limited to the more
evolved stages when the young (PMS/proto) star is actually visible. Large veiling
indices $r_\lambda=F_{\rm excess}/F_{\rm star}$ are thought to be
indicative of high mass accretion rates.  At K-band, $r_K$ indices $>
2$ identify sources as likely Stage I objects whereas $r_K<1$ is
characteristic of Stage II sources \citep{Greene1996}.

As a proof of concept, we compare the sample of \citet{vanKempen2009} 
in Ophiuchus with that of \citet{Greene1996} and \citet{Doppmann2005}.
The 11 overlapping sources are: GSS 30, WL12, WL 17, Elias
29, WL 19, WL 6, IRS 43, IRS 44, IRS 51, IRS 54, IRS 63. Except for
IRS 51 and 63, all sources have infrared veiling indices $r_K > 2.0$
whereas IRS 51 and 63 have 1.8 and 1.7 respectively. All sources in
the overlapping sample are classified as Stage I by \citet{vanKempen2009}
except IRS 51, which is listed as an obscured disk, and WL 19 which is
classified as confused. \citet{Doppmann2005} derive a veiling of 3.7 for
WL 19 establishing its nature as a likely Stage I object. This good
agreement provides further support for our classification scheme.

For Taurus, high resolution optical spectra have been obtained by
\citet{White2004} and \citet{Connelley2010}. The
sources which overlap with our sample are summarized in Table~\ref{tab:tau_nir}. The
fact that \citet{White2004} could detect and obtain high resolution
optical spectra for only half of the Class I Taurus sources already
suggests a bifurcation in source types, with the optically visible
sources closer to Stage II. The low veiling excesses at $R$-band found
for most sources by \citet{White2004}, coupled with their low
accretion rates (see below), indeed suggests that all these
overlapping sources in Table ~\ref{tab:tau_nir} are Stage II objects.  This is largely
consistent with our classification, except that their sources with the
highest $r_R$ values --IRAS04248+2612 and GV Tau-- are found to be
Class I by our classification. The comparison with \citet{Connelley2010}
shows a mixed picture, partly because of the large uncertainties
in many of their low $r_K$ values which could make them consistent
with both Stage I and II based on a division line around $r_K=1-2$.

Another diagnostic of evolutionary stage could be mass accretion rates
derived from spectroscopic features such as H$\alpha$ equivalent
widths. Indeed, \citet{Greene1996}
estimate the accretion rates for their Class I sources in Oph to be
$\sim 10^{-6}$ M$_\odot$ yr$^{-1}$, higher than those for Stage II
objects at $\sim (0-3) \times 10^{-7}$ M$_\odot$ yr$^{-1}$. However,
\citet{White2004} find no difference in mass accretion rates
between their Class I and II sources. Table~\ref{tab:tau_nir} shows that the few
Stage I sources in our classification do not have systematically
higher mass accretion rates than the other sources. Although 
source selection can play a role, a similar lack of
difference between embedded and optically visible sources was found by
\citet{Salyk2013} who compared mass accretion rates derived from
Pfund $\beta$ lines. The known variability and episodic nature of the
accretion process \citep[e.g.,][]{Kenyon1987, Evans2009,
Vorobyov2015} may be largely responsible for this lack of a
systematic trend.

Overall, our finding that a high fraction of Class I sources in Taurus
display properties more similar to Stage II objects is consistent with
earlier conclusions based on optical spectroscopy.

\begin{table}
\caption{Veiling indices and mass accretion rates derived from optical spectroscopy in Taurus}
\resizebox{9cm}{!}{
\begin{tabular}{lcccc}
\hline
\hline
Object & Stage\tablefootmark{a} & \multicolumn{2}{c}{Veiling index} & log $\dot{M}_{\rm acc}$\tablefootmark{d} \\
 &  & $r_K$\tablefootmark{b,c} & $r_{8400}$\tablefootmark{d} &  \\
 &  &  &  &   [M$_{\odot}/\rm yr$] \\
\hline
J034741.6+325143 & I &        high &            - & - \\
J041354.7+281132 & II &     0.12$^{+1.08}_{-0.0}$ &       - & - \\
J041851.4+282026 & II &          - &              <0.19 &  -7.36 \\
J041959.2+270958 & I &     0.60$^{+1.20}_{-0.24}$ &      - & -  \\
J042110.0+270142 & II &     0.24$^{+0.00}_{-0.24}$ &      - & - \\
J042111.4+270108 & C &     0.00$^{+1.44}_{-0.00}$ &      - & - \\
J042656.3+244335 & II &       high &               - & - \\
J042757.3+261918 & I &     0.00$^{+2.76}_{-0.00}$ &  0.60$\pm$0.21 &-8.97 \\
J042905.0+264904 & II &         - &            0.57$\pm$0.26 &-6.79 \\
J042923.6+243302 & I &         - &              1.1$\pm$0.4 &-6.71 \\
J043215.4+242903 & II &     0.12$^{+2.52}_{-0.00}$ &  0.26$\pm$0.08 &-7.54 \\
J043150.6+242418 & II &         -  &           0.24$\pm$0.06 &-7.65 \\
J043232.0+225726 & II &     0.54$^{+1.08}_{-0.30}$ &     - & - \\
J043556.7+225436 & II &         -  &              <0.07 &  -8.50 \\
J043535.0+240822 & II &       high &               - & - \\
J044138.8+255626 &  II &         -  &              <0.04 &  -8.11 \\
\hline
\end{tabular}
}
\tablefoot{
\tablefoottext{b}{Classification in this paper.}
\tablefoottext{b}{Sources with $r_K$>2 are considered Stage I, those with $r_K<1$ as Stage II.}
\tablefoottext{c}{\citet{Connelley2010}.}
\tablefoottext{d}{\citet{White2004}.}
}
\label{tab:tau_nir}
\end{table}

\subsection{Stage I lifetime}
\label{sec:lifetimes}

\citet{Heiderman2015} find that 84\% of sources in their
observed Class 0+I and Flat SED sample are true Stage I embedded objects based on their 
HCO$^{+}$ 3--2 single pointing criteria. They propose an updated 
timescale for Stage I sources (Stage 0+I in their notation) of 0.54 Myr, assuming a Class II lifetime 
of 2 Myr (see Section 4.4 and Table 3 in their paper). 
In Section~\ref{sec:singlept} of this paper it was shown that the HCO$^{+}$ single pointing method
may still result in $\sim$30\% overestimate of the number of Stage I objects,
implying that the updated value of Stage I lifetime may be overestimated by as much as 30\%.

Reevaluation of the lifetime for the embedded phase
based on Stage I identification with HCO$^{+}$ spatial mapping depends on how one treats the confused sources.
If all confused sources are assumed to be Stage I, then the Perseus and Taurus sample together have 
30\% of Class 0+I sources misidentified. On the contrary, if all confused
sources are assumed to be Stage II, then the sample would have 46\% of sources
misidentified. Using the updated \citet{Heiderman2015} value as a reference, 
new estimates for the Stage I lifetime would then be 0.38 Myr and 0.29 Myr, respectively.
It is more realistic to assume that confused sources are distributed between the two Stages,
but correct classification using spatial mapping will still result in 
overall shorter timescales for Stage I objects.

%__________________________________________________________________

\section{LOMASS Database}
\label{sec:lomass}

Over the past decade, the molecular astrophysics group at the Leiden
Observatory has observed a comprehensive dataset of various molecular
lines toward a large number of Class 0+I YSOs as part of
various research projects.  Such single-dish data provides a wealth of
information which can be more easily exploited if collected together
in a coherent way in one place. Therefore the {\it Single-dish
  Submillimeter Spectral Database of Low-mass YSOs hosted at Leiden
  Observatory}
(LOMASS)\footnote{\url{http://lomass.strw.leidenuniv.nl}} database has
been established as a public web archive to fulfill this aim and
provide a service to the community.

%Over the past decade, the molecular astrophysics group at the Leiden
%Observatory has obtained a comprehensive dataset of various molecular
%lines toward a large number of Class 0 and Class I young stellar
%objects (YSOs) through various research projects. 
%The demand for collecting those data at one coherent place become a necessity in order 
%to fully benefit the wealth of previously obtained `old' data.
%Therefore,  {\it Single-dish Submillimeter Spectral Database of Low-mass YSOs 
%hosted at Leiden Observatory} (LOMASS)\footnote{\url{http://lomass.strw.leidenuniv.nl}} 
%database is established as a public web archive to fulfill this aim and providing a service to the community. 

This database not only contains data from the observing projects
conducted at the Leiden Observatory, but also molecular data on the
same or other new sources which have been obtained from public data
archives during the course of various projects.  To date, observations
from the JCMT and Atacama Pathfinder EXperiment (APEX, especially the
CHAMP+ instrument) are included. Other relevant
molecular spectral data of low mass YSOs from other telescopes, such
as \textit{Herschel}, the IRAM~30-m, etc. could be also included in the future.

%The database not only contains data from the observing projects conducted at the Leiden Observatory, molecular data
%from other projects on the same sources or other new sources have been downloaded from the
%raw public data archives of telescopes. 
%So far, data from the James Clerk Maxwell
%Telescope (JCMT) and Atacama Pathfinder EXperiment (APEX, especially
%the CHAMP+ instrument) are included. Our target is to include more molecular spectral 
%data of YSOs from other telescopes, such as $Herschel$, IRAM~30-m, etc.
%We acknowledge using the following programs and languages for building this database: 
%for data reduction, \verb1GILDAS-CLASS1 package; for plotting, \verb1python-matplotlib1 package;
%for database, \verb1MySQL1; for web interface, \verb1php1 and \verb1html1.

Depending on availability, the database presents spectral maps and
single pointing observations. Users can search via the web interface
using one of two options - by source name or by molecular line
name. In cases where the source names are not familiar, it is possible to generate a source list
which provides source coordinates.

%Depending on the availability, maps and single pointing observations are
%presented. Two types of searching options are included; either by `source name' or by `molecular line name'.
%The commonly used names are listed as source names, however, in case those names are not familiar, 
%it is possible to generate a source list, where it is possible to check with coordinate.

Once a source is selected, a page provides metadata about the source
and available observations.  Examples include the source
coordinates, $V_{\mathrm{lsr}}$, telescope and instrument names, the
beam efficiency that was used during data reduction and the rest
frequency of the observations.  Other source properties such as
$L_\mathrm{bol}$, $T_\mathrm{bol}$, $M_\mathrm{env}$ and
$n_{\mathrm{1000AU}}$ etc. are included, if available, with references
to the papers these are drawn from. The "Downloads Corner" provides
raw data in \textsc{class} format, as well as reduced data in
\textsc{class} and/or \textsc{fits} format.

%In each source's page, the metadata about source, telescope, and molecular line are listed. These are 
% such as RA, Dec (J2000), \Vlsr, telescope name, instrument name, beam efficiency that was used in the 
% reduction, and rest frequency of the observations. Some other metadata, such as \Menv, \Lbol, \Tbol, and \n1000AU etc. are included, if available.
%The "Downloads Corner" provides raw data in CLASS format, as well as reduced data in CLASS and/or fits format. 

In the case of single pointing observations, some of the observed
quantities from the spectra are automatically extracted or calculated
and then listed in each page. Examples include the observation date
and rms noise of spectrum, together with the FWHM, integrated
intensity and peak intensity of the specified molecular
transition. The velocity range used in the integrated intensity
calculation is also indicated.

%In the single pointing observations, some of the observed quantities from the spectra are automatically 
% extracted or calculated and then listed in each page. These are mainly, the observation date, and $rms$, 
% together with the $FWHM$, integrated intensity, and peak intensity of the peak of the specified molecular 
% transition. The velocity range is also indicated where the integrated intensity is calculated.

For mapping observations, both integrated intensity and spectral maps
are provided. In addition, the spectrum from the central region is
included to aid quick comparison with other sources. The observed quantities for single pointing
observations mentioned above are also given for map central spectra.

%In the case of mapping observations, an integrated intensity and spectral maps are provided. In addition, 
% the spectrum from the central region is included in order to compare intensity with the other sources, 
% together with providing the above mentioned observed quantities for single pointing observations. 
%The intensities in the integrated intensity map are calculated using the velocity range limits also 
% provided in the table of central spectrum details.

This database of single-dish spectral data for low-mass YSOs will
provide a valuable and easily accessible reference for higher
resolution interferometry data, especially the Atacama Large Millimeter/submillimeter Array (ALMA).

%This database will include more data in time and these single-dish spectral data yield information on the chemical and
%physical properties of low-mass YSOs and provide a reference for
%higher resolution interferometry data, especially ALMA.

%__________________________________________________________________

\section{Conclusions}
\label{sec:concl}

We have presented molecular line emission maps of HCO$^{+}$ 4--3 and
C$^{18}$O 3--2 observed with HARP on the JCMT, as well as SCUBA
450/850 $\mu$m and {\it Herschel} PACS 70/160 $\mu$m dust continuum
maps to characterize the emission morphology of 56 candidate YSOs in
the Perseus and Taurus clouds. An additional nine sources in Serpens
observed in C$^{18}$O 3--2 with HARP are also presented in this paper.
The footprints of the 2$\arcmin \times$2$\arcmin$ spatial maps enable
qualitative categorization of the morphology and quantitative analysis
of the concentration of emission from each source.  

Four types of emission morphology are identified: on-source
concentrated, offset concentrated, extended and non-detection.
Calculation of the concentration factor for each source in HCO$^{+}$
4--3 and 850 $\mu$m determines the stage classification where Stage I
corresponds to embedded YSOs and Stage II corresponds to an evolved PMS
star with a disk.

Conclusions of this work are:

\begin{itemize}
 \item Previous methods of classifying the evolutionary stages of YSOs 
 cannot always correctly identify Stage I sources in the embedded phase. 
 Using the concentration factor of HCO$^{+}$ 4--3
 as a way to identify truly embedded objects, we find that approximately 30\% of Stage I sources in Perseus and Taurus are incorrectly classified. The 56 source
 sample between these two star forming regions consists of 32 truly embedded Stage I sources and 17 more evolved Stage II PMS stars with disks.
 \item The Perseus cloud hosts a higher percentage of Stage I sources compared to Taurus in our sample. Perseus has a population that is 71\% (24 of 34 sources)
 embedded Stage I objects while Taurus consists of 68\% (15 of 22 sources) more evolved Stage II objects.
 \item Nine sources in the sample cannot be identified as either Stage I or Stage II. Their emission in HCO$^{+}$ 4--3 may be strong or weak
 but their morphologies show a central peak that is offset from the source position by 20$\arcsec$ in most cases. Interferometric data is needed to further
 identify their nature.
 \item Using only the detection of HCO$^{+}$ at the source position as a criterion for identifying truly embedded sources is not as robust as classification
 with the spatial concentration. The single pointing method can result in up to 30\% of sources being misidentified as embedded.
 \item Stage I lifetimes may be overpredicted on the order of 30\%. The current timescale estimate for Stage I is 0.54 Myr. After correcting
 misidentified sources using HCO$^{+}$ spatial mapping the Stage I lifetime is 0.38 Myr, but may be as low as 0.29 Myr depending on how one treats the confused sources in this study.
\end{itemize}

The correct classification of YSOs in star forming regions such as
Perseus and Taurus will lead to more complete sample of YSO
evolutionary stages used to determine stage lifetimes.  A
comprehensive understanding of the stage of these YSOs will help
future surveys selectively target embedded YSOs or evolved PMS stars
with disks. Additional data with interferometers such as the Plateau
de Bure Interferometer (PdBI), Submillimeter Array (SMA), and Atacama
Large Millimeter/submillimeter Array (ALMA) will allow detailed
investigation of the ongoing chemistry and kinematics of such objects.

\begin{acknowledgements}

The authors are grateful to Amanda Heiderman and Neal Evans for their useful 
collaboration and discussion on embedded protostellar environments. 
Many thanks to the referee, Charles Lada, for his very useful comments and suggestions.
Astrochemistry in Leiden is supported by the Netherlands Research School for 
Astronomy (NOVA), by a Royal Netherlands Academy of Arts and Sciences (KNAW) 
professor prize, by a Spinoza grant and grant 614.001.008 from the Netherlands 
Organisation for Scientific Research (NWO).

The authors are indebted to the various observers who have collected data for their papers and kindly 
provided them to be included in the LOMASS database, as well as staff at the JAC. 
Acknowledgment is given to the following programs and languages for building this database:
\textsc{gildas-class} and \textsc{python} for data reduction and format manipulation; 
\textsc{python-matplotlib} for plotting; \textsc{MySQL} for construction of the database;
\textsc{php} and \textsc{html} for the web interface.

\end{acknowledgements}

\bibliographystyle{aa}
\bibliography{carney}

\clearpage
\clearpage

%\end{document}  % uncomment to latex faster

\begin{appendix}

    \section{Notes on Individual Sources}
    \label{app:A}
    
    \noindent {\bf J032536.4+304523 (L1448-N):} Note that the C$^{18}$O 3--2 map is incomplete, with bad pixels flagged
    in a rectangular region from (0$\arcsec$, 0$\arcsec$) to (40$\arcsec$, -50$\arcsec$). The flagged data
    is not immediately apparent in the spatial maps due to interpolation, but it is easily seen in the spectral map.
    (See entry in the LOMASS database.) \\
    
    \noindent {\bf J032738.2+301358 (L1455-FIR 2):} Another sub-mm object, L1455-IRS 1, lies at offset (20$\arcsec$, -60$\arcsec$) 
    and dominates emission in the map. Concentrated HCO$^{+}$ 4--3 emission from the source is still apparent in the spatial map. \\
    
    \noindent {\bf J032743.2+301228 (L1455-IRS 4):}  Another sub-mm object, L1455-IRS 1, lies at offset (-60$\arcsec$, 40$\arcsec$) 
    and dominates emission in the map. There is widespread emission of HCO$^{+}$ 4--3 seen in the spectral map from the northwest
    corner to the southeast corner. However, concentrated HCO$^{+}$ 4--3 emission from the source is still apparent 
    above this strong background. \\
    
    \noindent {\bf J032834.5+310705:} The source has a high 850 $\mu$m concentration, but cannot be properly 
    analyzed due to contaminating absorption features in the molecular line maps.\\
        
    \noindent {\bf J032859.5+312146:} HCO$^{+}$ 4--3 spectral map indicates a slight broadening of the line off-source, 
    there appear to be no other nearby YSOs or dense cores that might contaminate the field.\\
    
    \noindent {\bf J032900.6+311200:} HCO$^{+}$ 4--3 spectral map indicates a slight broadening of the line off-source, 
    there appear to be no other nearby YSOs or dense cores that might contaminate the field.\\
    
    \noindent {\bf J032901.6+312028:} HCO$^{+}$ 4--3 spectral map indicates a slight broadening of the line off-source. 
    The field of view contains three YSO or dense core objects, JCMTSF J032901.3+312031, 
    SSTc2d J032901.6+312021, and [RAC97] VLA 43 that lie at an offset of about 8$\arcsec$ to the southwest.\\
    
    \noindent {\bf J032903.3+311555:} HCO$^{+}$ 4--3 spectral map indicates a slight broadening of the line off-source. The field contains 
    Herbig Haro objects to the southeast and southwest, and a YSO (2MASS J03290289+3116010)
    at an offset of (-10$\arcsec$, -5$\arcsec$) that may contribute to the extended 
    HCO$^{+}$ 4--3 emission.\\
        
    \noindent {\bf J032904.0+311446 (HH 7-11 MMS6):} shows strong, extended HCO$^{+}$
    4--3 throughout the field of view indicating the presence of large
    scale high density material that could host an embedded protostar, but
    the lack of a clear peak at the source position prevents definite
    Stage I classification. The J032904.0+311446 (HH 7-11 MMS6) spectral map
    does show evidence of line wings, suggesting that the extent of 
    the emission may be due to outflows from the source or nearby objects.\\
    
    \noindent {\bf J034350.9+320324:} The map shows off-source emission in HCO$^{+}$ 4--3 
    due to a binary companion at offset (0,+15$\arcsec$). \\
    
    \noindent {\bf J032910.7+311820:} The map shows off-source emission in HCO$^{+}$ 4--3 
    due to a binary companion at offset (+10$\arcsec$,+10$\arcsec$). \\
        
    \noindent {\bf J032951.8+313905:} HCO$^{+}$ 4--3 spectral map indicates a slight broadening of the line off-source, 
    there appear to be no other nearby YSOs or dense cores that might contaminate the field.\\
    
    \noindent {\bf J033313.8+312005:} The source has a high concentration in HCO$^{+}$ 4--3, but weak integrated intensity
    just at the threshold for consideration as a Stage I object. It has a low concentration in SCUBA 850 $\mu$m emission, and the
    C$^{18}$O 3--2 map contained emission and absorption in the line profiles, preventing proper characterization. 
    The source is marked as confused due to the C$^{18}$O 3--2 contamination, but other tracers are more consistent with Stage II properties.\\
    
    \noindent {\bf J033314.4+310710 (B1-SMM3):} A weak outflow wing can be seen in the HCO$^{+}$ 4--3 central spectrum. 
    Two objects -- SSTc2d J033316.7+310755 and B1-IRS -- lie at an offset of about (+50$\arcsec$, +40$\arcsec$), thus
    it is likely that the extended emission found in this map and line wings in the spectral profile are due to
    emission and outflows from these other objects in the field of view. \\
        
    \noindent {\bf J033320.3+310721 (B1-b):} A weak outflow wing can be seen in the HCO$^{+}$ 4--3 central spectrum. 
    The map shows extended emission with a weak peak offset by about 15$\arcsec$. A 
    much stronger peak at an offset of about (-50$\arcsec$, +30$\arcsec$) is due to emission from 
    SSTc2d J033316.7+310755 and B1-IRS. No peaks in emission are located at the source
    position. In addition, the 850 $\mu$m emission has a low concentration factor, thus the object remains confused.\\
    
    \noindent {\bf J034359.4+320035:} The HCO$^{+}$ 4--3 map has emission peaking significantly off-source as
    a result of another object at an offset of about (-20$\arcsec$, +10$\arcsec$) 
    as seen in all bands of molecular line and dust continuum emission. The source 
    is categorized in nearly all maps as offset (O$_{\mathrm{H7148}}$ E$_{\mathrm{C}}$).
    The nearby object SSTc2d J034357.6+320045 \citep[Pers-45 from][]{Jorgensen2007} 
    is found at these coordinates.  HCO$^{+}$ 4--3 is detected at the target source position, as
    seen in the central spectrum, but its spectral map
    shows that this emission cannot be disentangled from the secondary object.
    Contamination prevents calculation of the concentration factor
    in 850 $\mu$m and HCO$^{+}$ 4--3. \\
       
    \noindent {\bf J034402.4+320204:} The field of view contains HCO$^{+}$ 4--3 
    that is extended on scales of 40$\arcsec$ offsets. Though 
    there is a peak within this extended emission, it has a low concentration 
    factor that prevents a definite Stage I classification.\\   
    
    \noindent {\bf J042110.0+270142:} The map shows off-source emission in HCO$^{+}$ 4--3 
    due to a binary companion at offset (-15$\arcsec$,+25$\arcsec$). Emission from 
    J042111.4+270108 is seen at an offset of about (40$\arcsec$,-30$\arcsec$). \\
    
    \noindent {\bf J042111.4+270108:} The map shows extended HCO$^{+}$ 4--3 throughout the field
    due to a third object -- J042110.0+270142 -- at offset (-25$\arcsec$,+40$\arcsec$).
    There is an off-source peak at the location of the binary companion at
    offset (-25$\arcsec$,+25$\arcsec$). With no complementary SCUBA dust emission it cannot be
    properly classified. \\
        
    \clearpage
    
    \section{Central Spectra}
    \label{app:B}
 
      \begin{minipage}{\textwidth}
	\includegraphics[width=\textwidth]{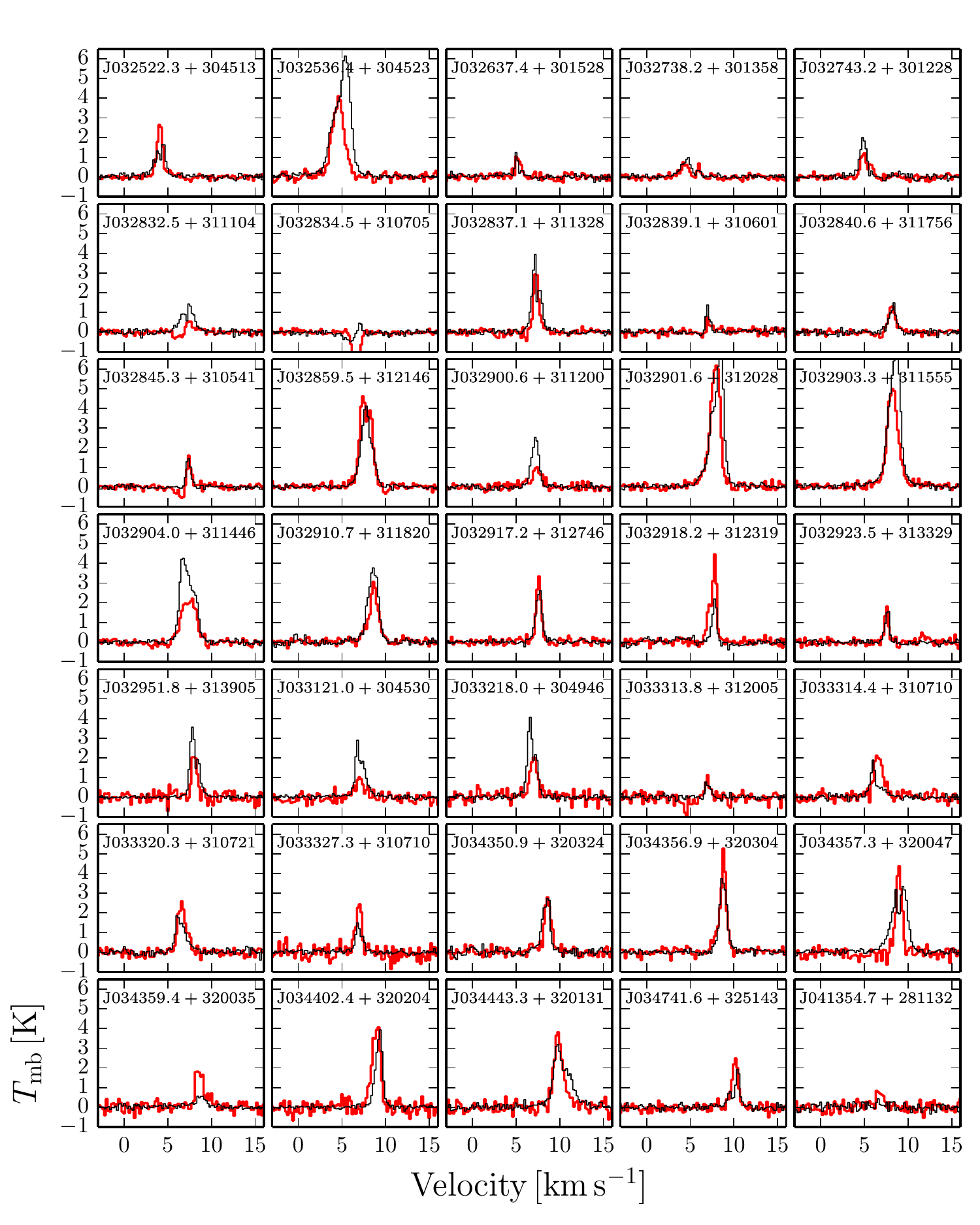}
	\captionof{figure}{Central spectra for HCO$^{+}$ 4--3 (black) and C$^{18}$O 3--2 (red) in a 15$\arcsec$ beam extracted from the source locations listed in Table~\ref{tab:data_props}.}
	\label{fig:ApB1}
      \end{minipage}
      
      \clearpage
      
      \begin{figure*}[!htbp]
	\includegraphics[width=\textwidth]{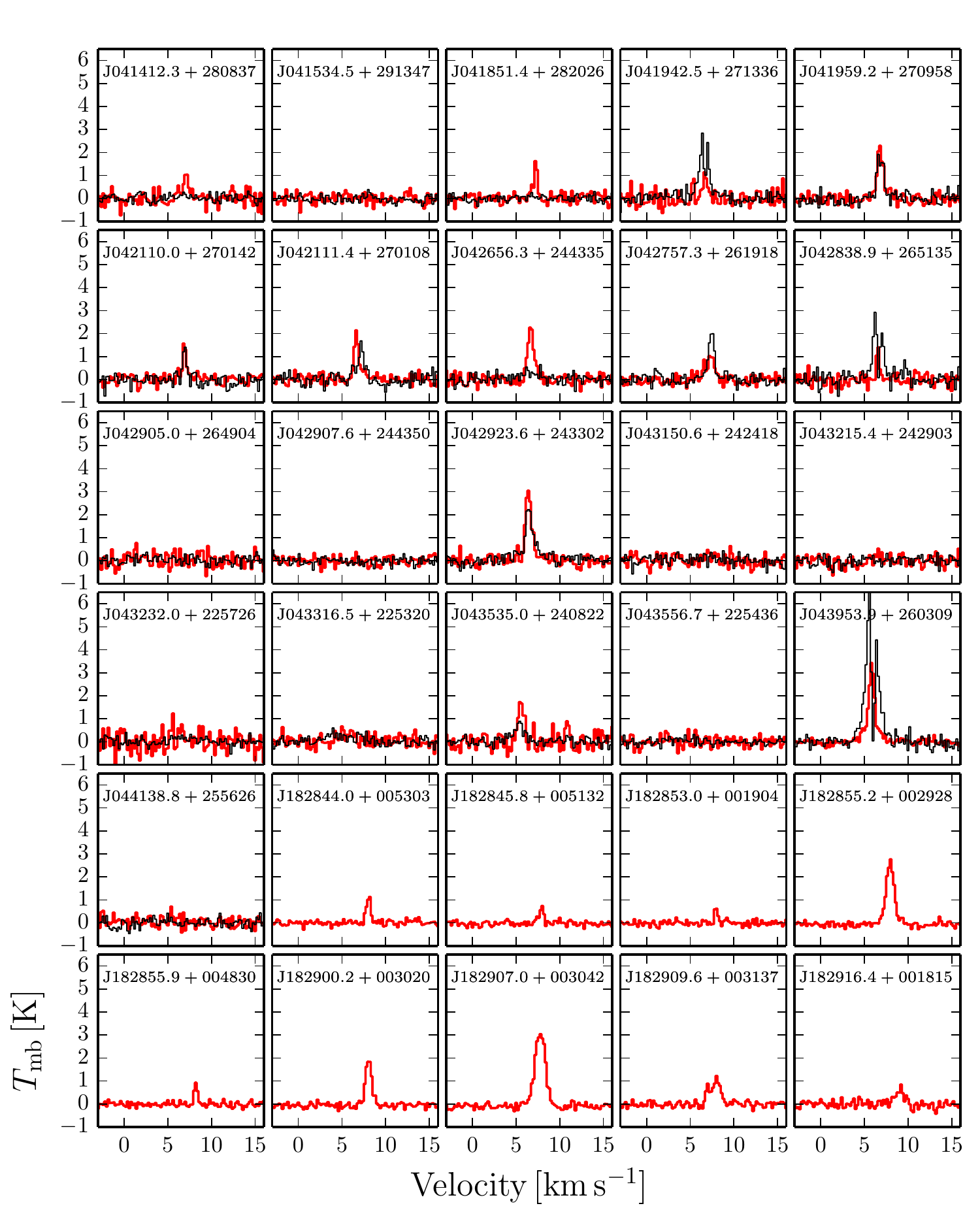}
	\caption{Central spectra for HCO$^{+}$ 4--3 (black) and C$^{18}$O 3--2 (red) in a 15$\arcsec$ beam extracted from the source locations listed in Table~\ref{tab:data_props}.}
	\label{fig:ApB2}
      \end{figure*}

       \clearpage
       
  \section{Emission Maps}
  \label{app:C}  
    
    %%%%%%%%%%%%%%%%%%%%%%%%%%%%%%%%%%%%%%%%%%%%%%%%%%%%%%%%%%%%%%%%%%%%%%%%%%%%%%%
    %%% PERSEUS MAPS
    
    \begin{sidewaysfigure}
      \vspace{-9.5cm}
      \resizebox*{0.98\hsize}{!}{\includegraphics{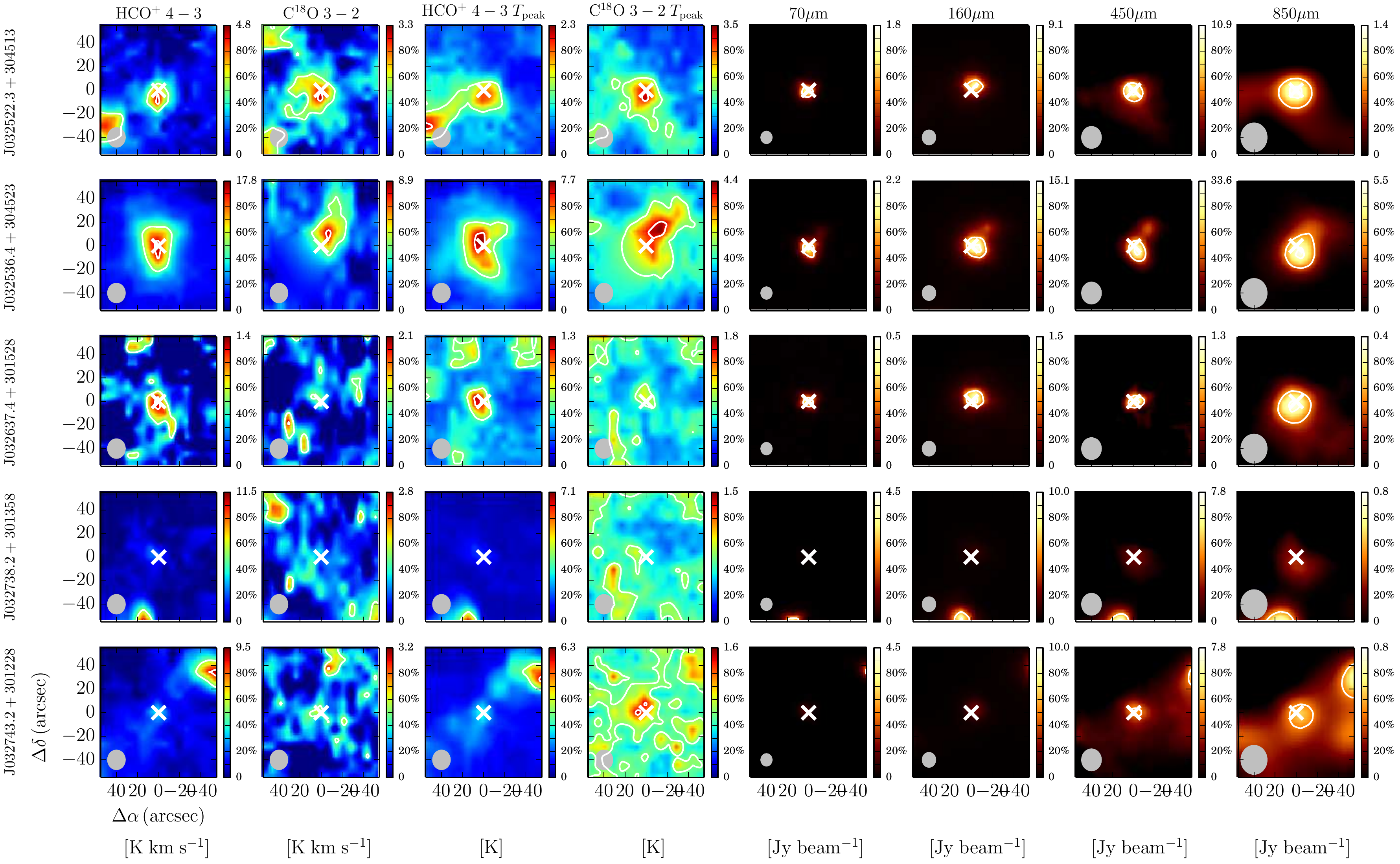}}%[viewport=4 4 1277 1277,clip]
      \caption{Perseus sources. $2\arcmin\times2\arcmin$ maps of JCMT HARP HCO$^{+}$ 4--3 and C$^{18}$O 3--2, \textit{Herschel} PACS 70$\mu \rm{m}$ and 160$\mu \rm{m}$ continuum, and
      JCMT SCUBA 450$\mu \rm{m}$ and 850$\mu \rm{m}$ continuum. Each map has solid white contours at 50\% and 90\% of the peak emission. Color scaling is
      normalized to peak emission and shown as percentage values. Source location based on Table~\ref{tab:data_props} is indicated by a white cross at the center of each image. There are two molecular line maps. The first shows the spectrally integrated intensity over the region. The second shows the peak temperature
      over the region. Filled gray circles in the lower left give the relevant beam sizes.}
      \label{fig:ApC1}
    \end{sidewaysfigure}

    \clearpage
    
    \begin{sidewaysfigure*}
      \centering
      \resizebox*{\hsize}{!}{\includegraphics{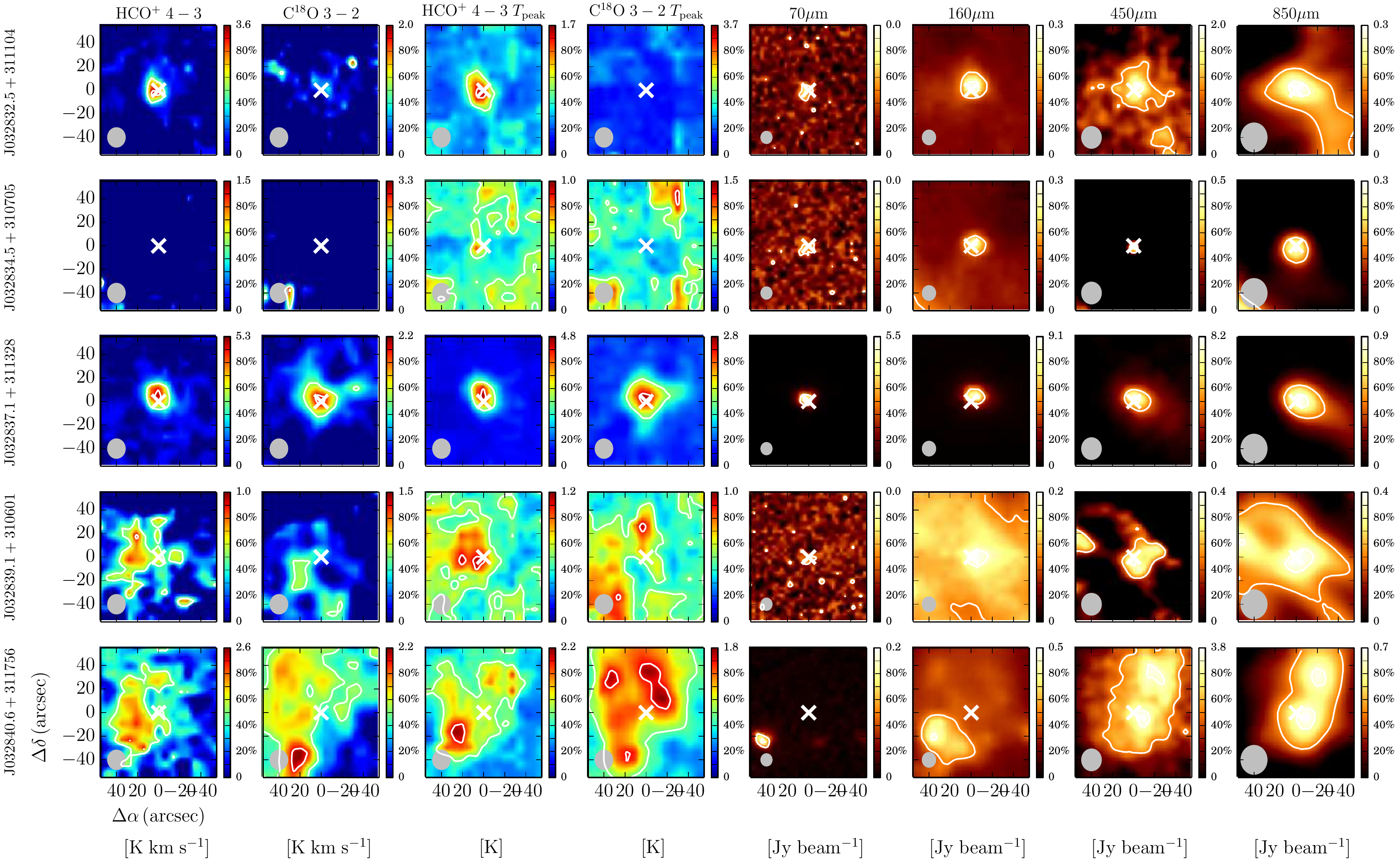}}
      \caption{Perseus sources. See Figure~\ref{fig:ApC1} for caption.}
      \label{fig:ApC2}
    \end{sidewaysfigure*}
    
    \clearpage

    \begin{sidewaysfigure*}
      \centering
      \resizebox*{\hsize}{!}{\includegraphics{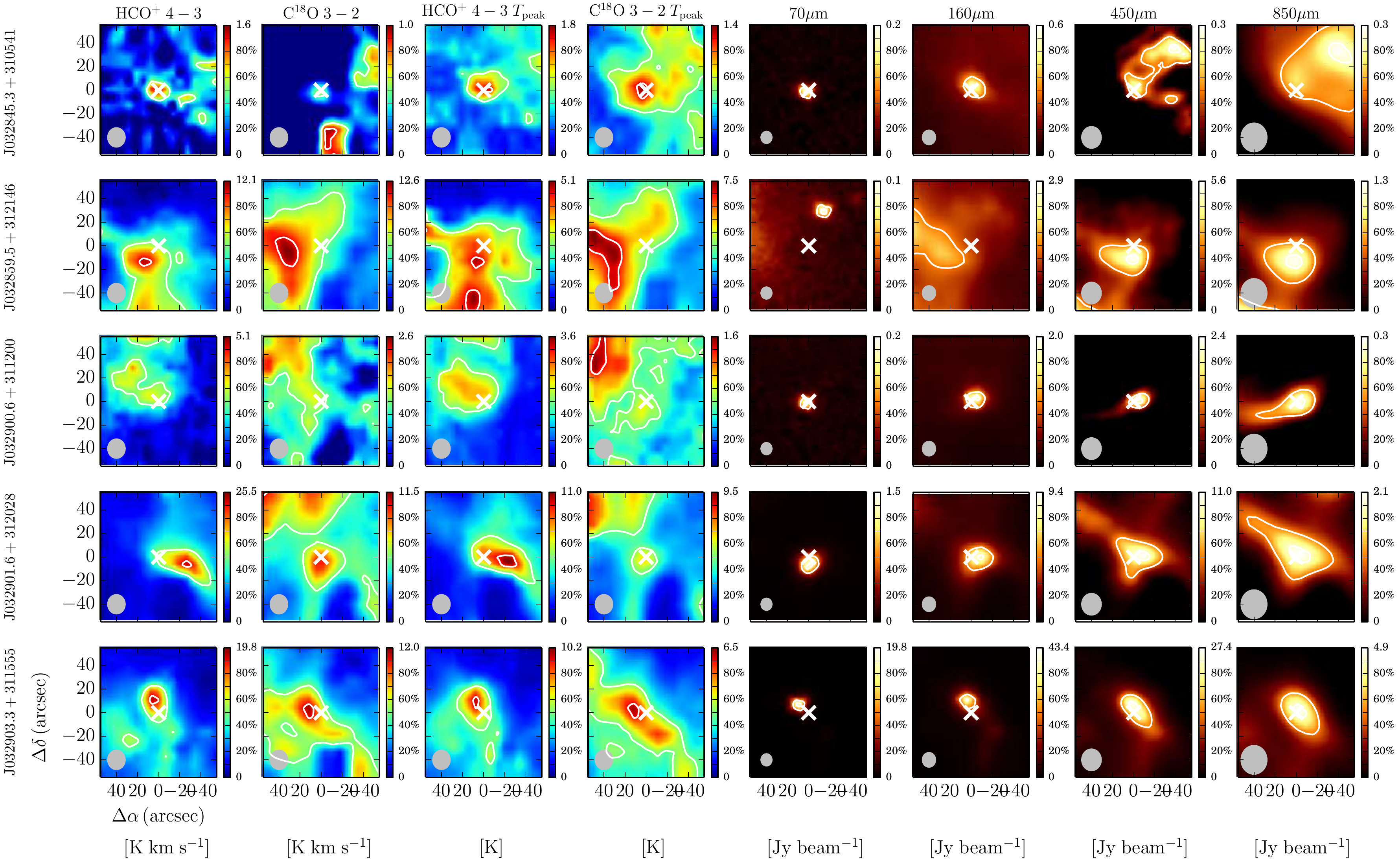}}
      \caption{Perseus sources. See Figure~\ref{fig:ApC1} for caption.}
      \label{fig:ApC3}
    \end{sidewaysfigure*}
    
    \clearpage
 
    \begin{sidewaysfigure*}
      \centering
      \resizebox*{\hsize}{!}{\includegraphics{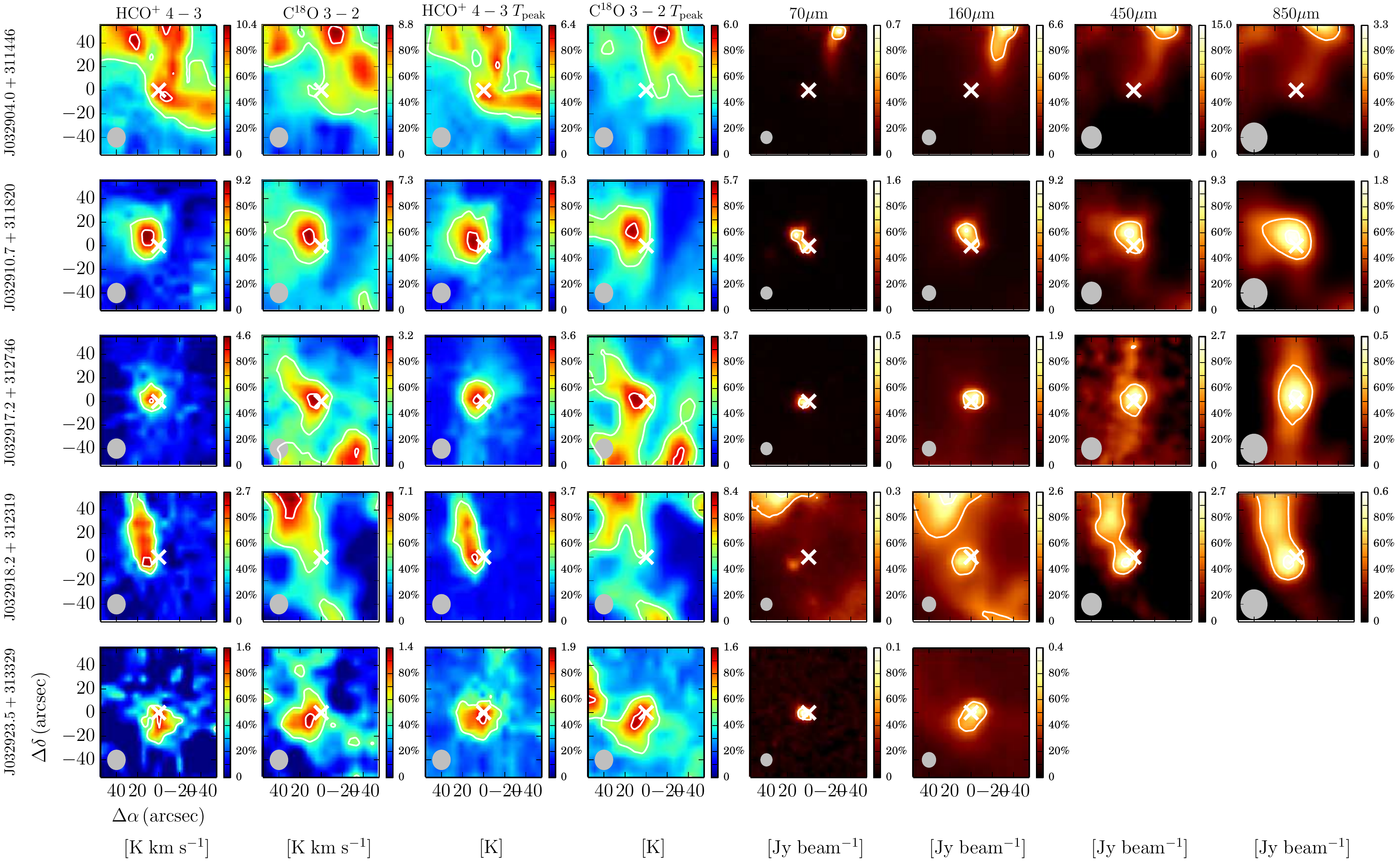}}
      \caption{Perseus sources. See Figure~\ref{fig:ApC1} for caption.}
      \label{fig:ApC4}
    \end{sidewaysfigure*}
    
    \clearpage

    \begin{sidewaysfigure*}
      \centering
      \resizebox*{\hsize}{!}{\includegraphics{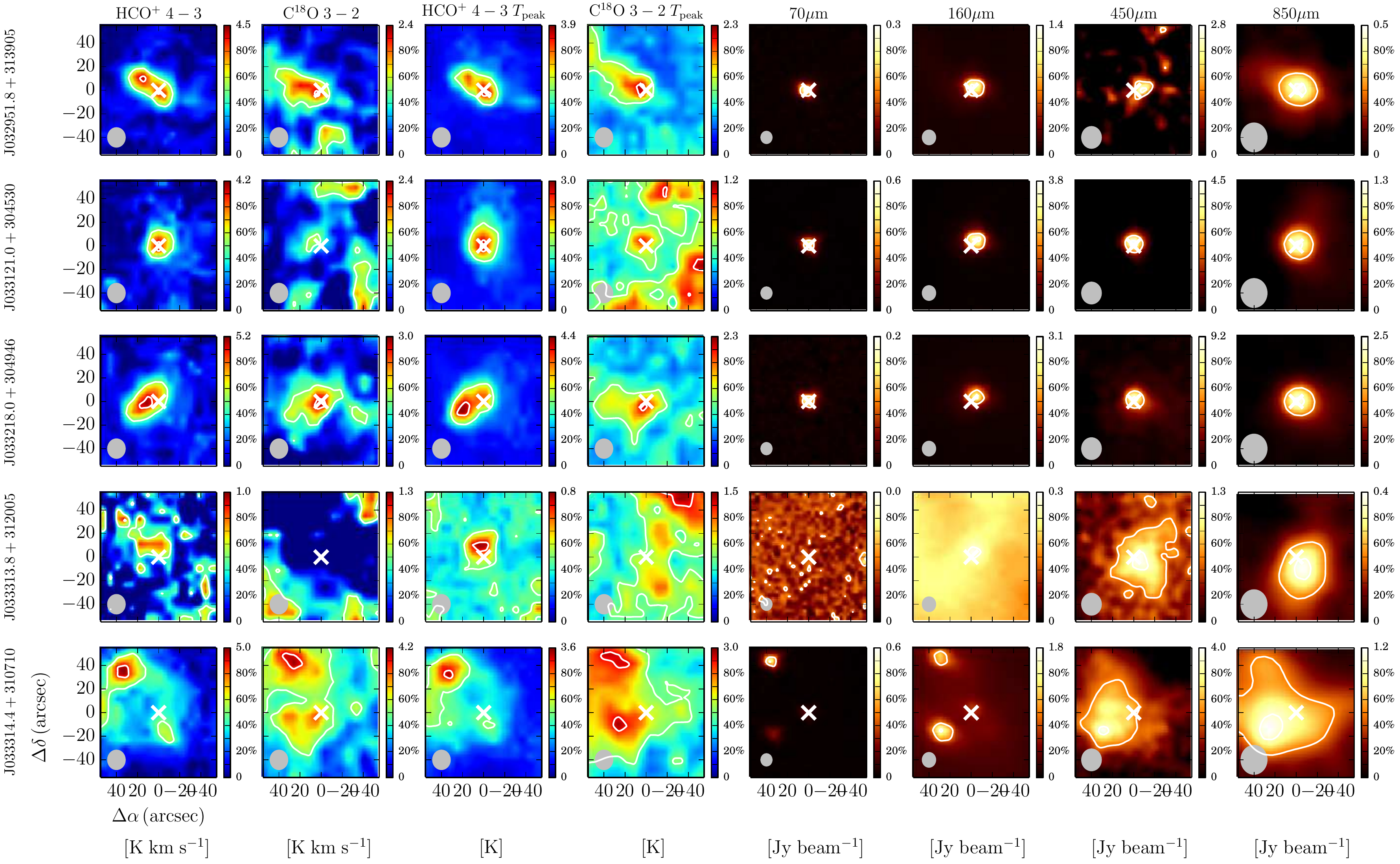}}
      \caption{Perseus sources. See Figure~\ref{fig:ApC1} for caption.}
      \label{fig:ApC5}
    \end{sidewaysfigure*}
    
    \clearpage

    \begin{sidewaysfigure*}
      \centering
      \resizebox*{\hsize}{!}{\includegraphics{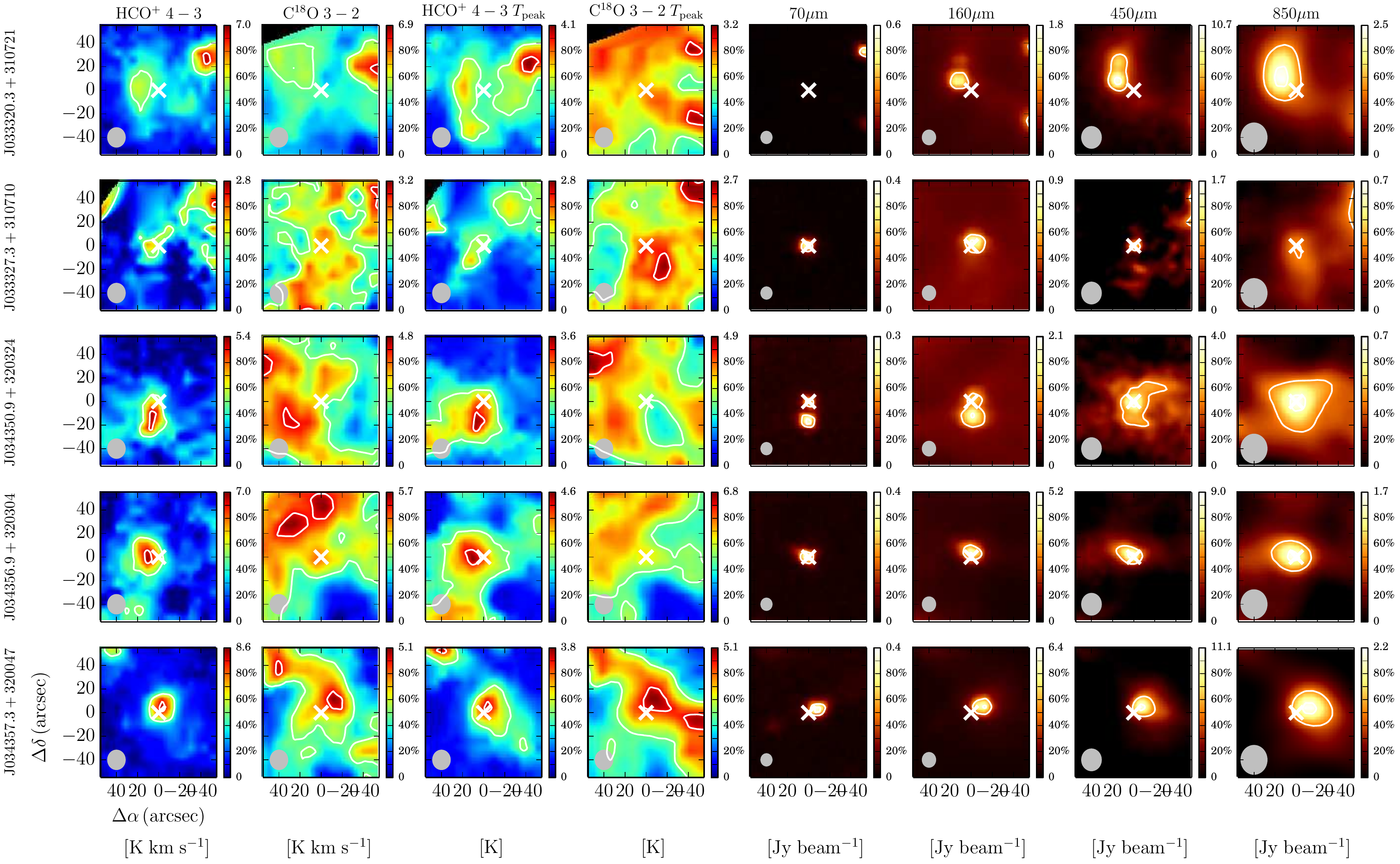}}
      \caption{Perseus sources. See Figure~\ref{fig:ApC1} for caption.}
      \label{fig:ApC6}
    \end{sidewaysfigure*}
    
    \clearpage

    \begin{sidewaysfigure*}
      \centering
      \resizebox*{\hsize}{!}{\includegraphics{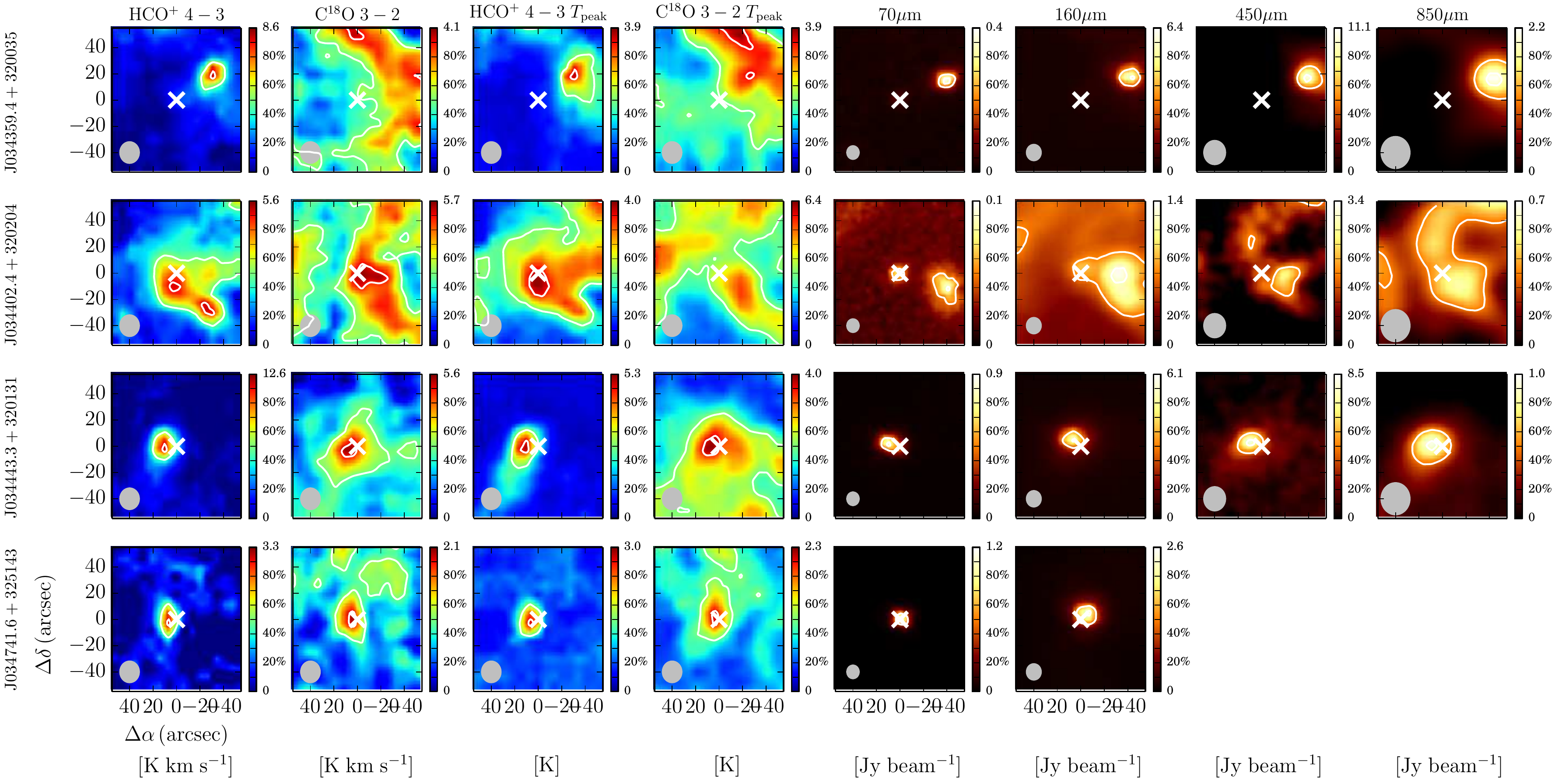}}
      \caption{Perseus sources. See Figure~\ref{fig:ApC1} for caption.}
      \label{fig:ApC7}
    \end{sidewaysfigure*}
    
    \clearpage

    %%%%%%%%%%%%%%%%%%%%%%%%%%%%%%%%%%%%%%%%%%%%%%%%%%%%%%%%%%%%%%%%%%%%%%%%%%%%%%%
    %%% TAURUS MAPS
    
    \begin{sidewaysfigure*}
      \centering
      \resizebox*{\hsize}{!}{\includegraphics{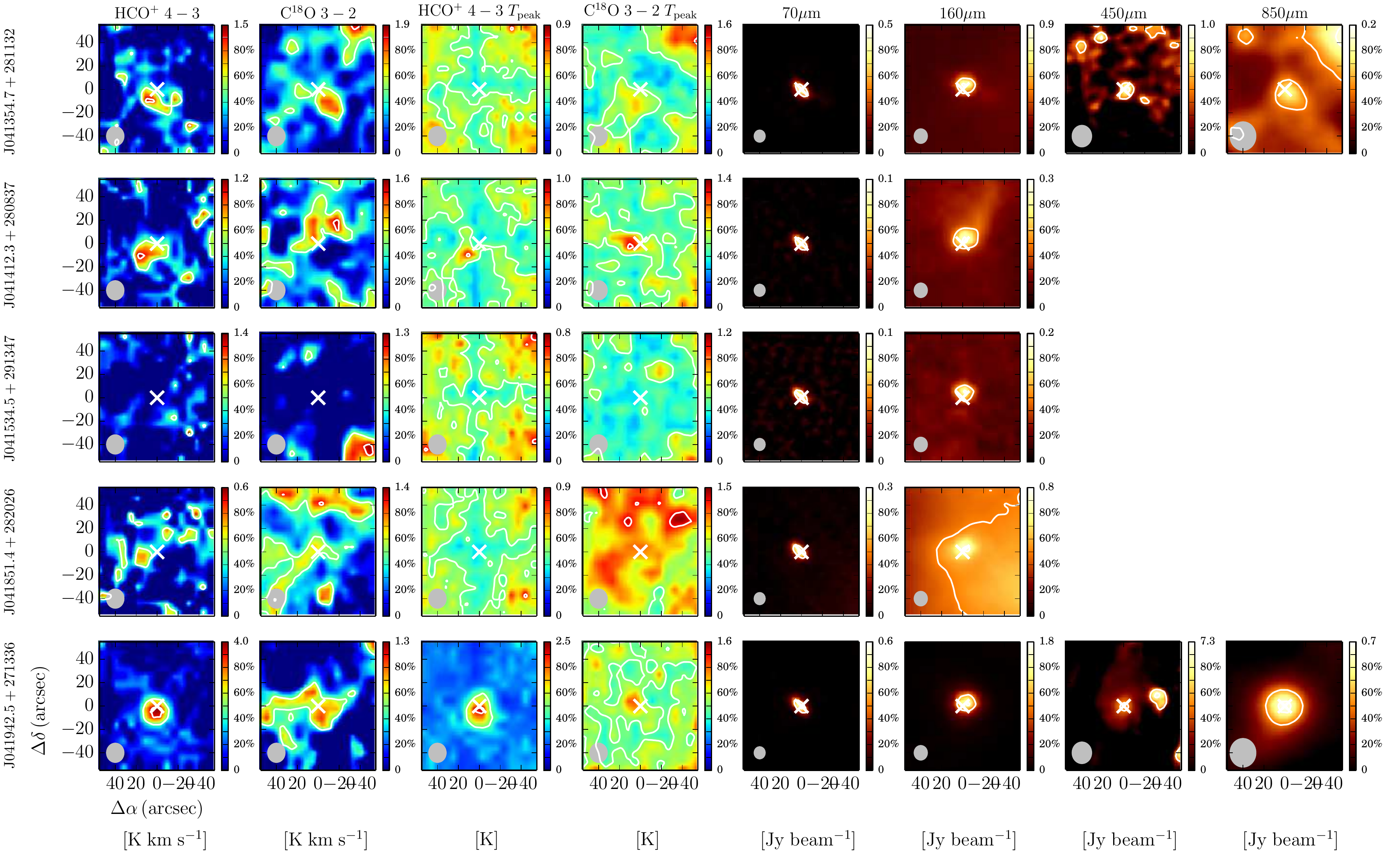}}
      \caption{Taurus sources. See Figure~\ref{fig:ApC1} for caption.}
      \label{fig:ApC8}
    \end{sidewaysfigure*}
    
    \clearpage
    
    \begin{sidewaysfigure*}
      \centering
      \resizebox*{\hsize}{!}{\includegraphics{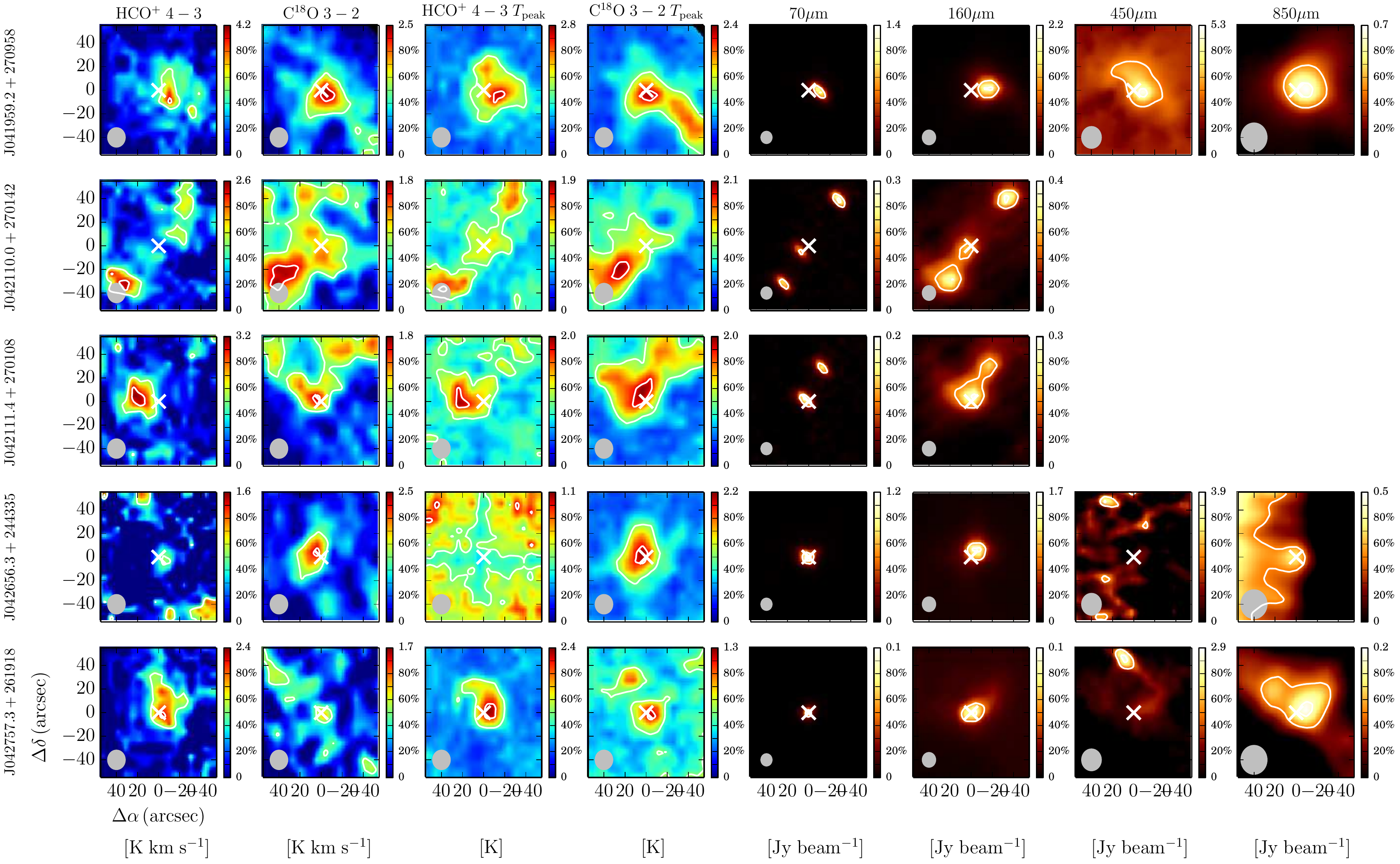}}
      \caption{Taurus sources. See Figure~\ref{fig:ApC1} for caption.}
      \label{fig:ApC9}
    \end{sidewaysfigure*}
    
    \clearpage
    
    \begin{sidewaysfigure*}
      \centering
      \resizebox*{\hsize}{!}{\includegraphics{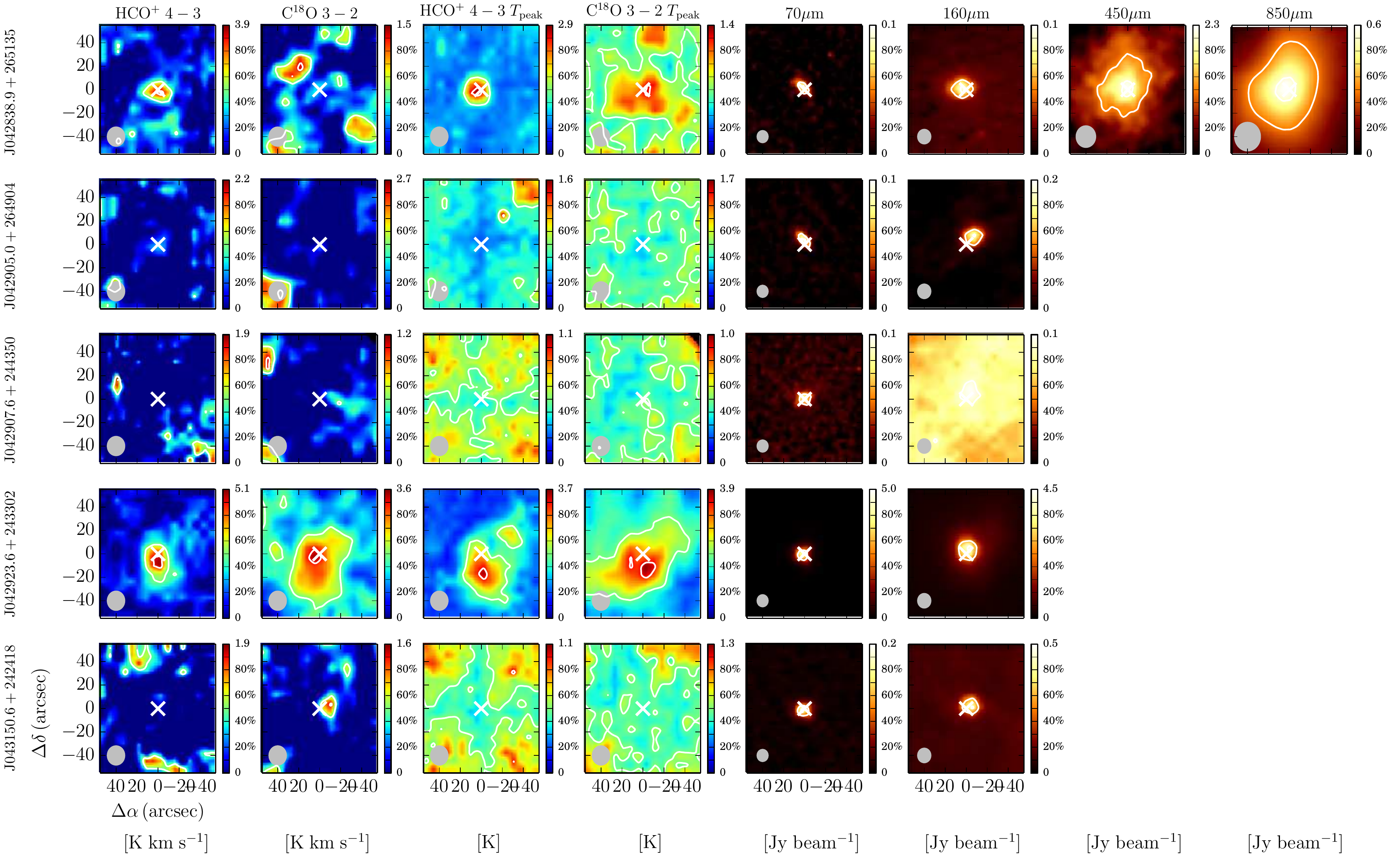}}
      \caption{Taurus sources. See Figure~\ref{fig:ApC1} for caption.}
      \label{fig:ApC10}
    \end{sidewaysfigure*}
    
    \clearpage
    
    \begin{sidewaysfigure*}
      \centering
      \resizebox*{\hsize}{!}{\includegraphics{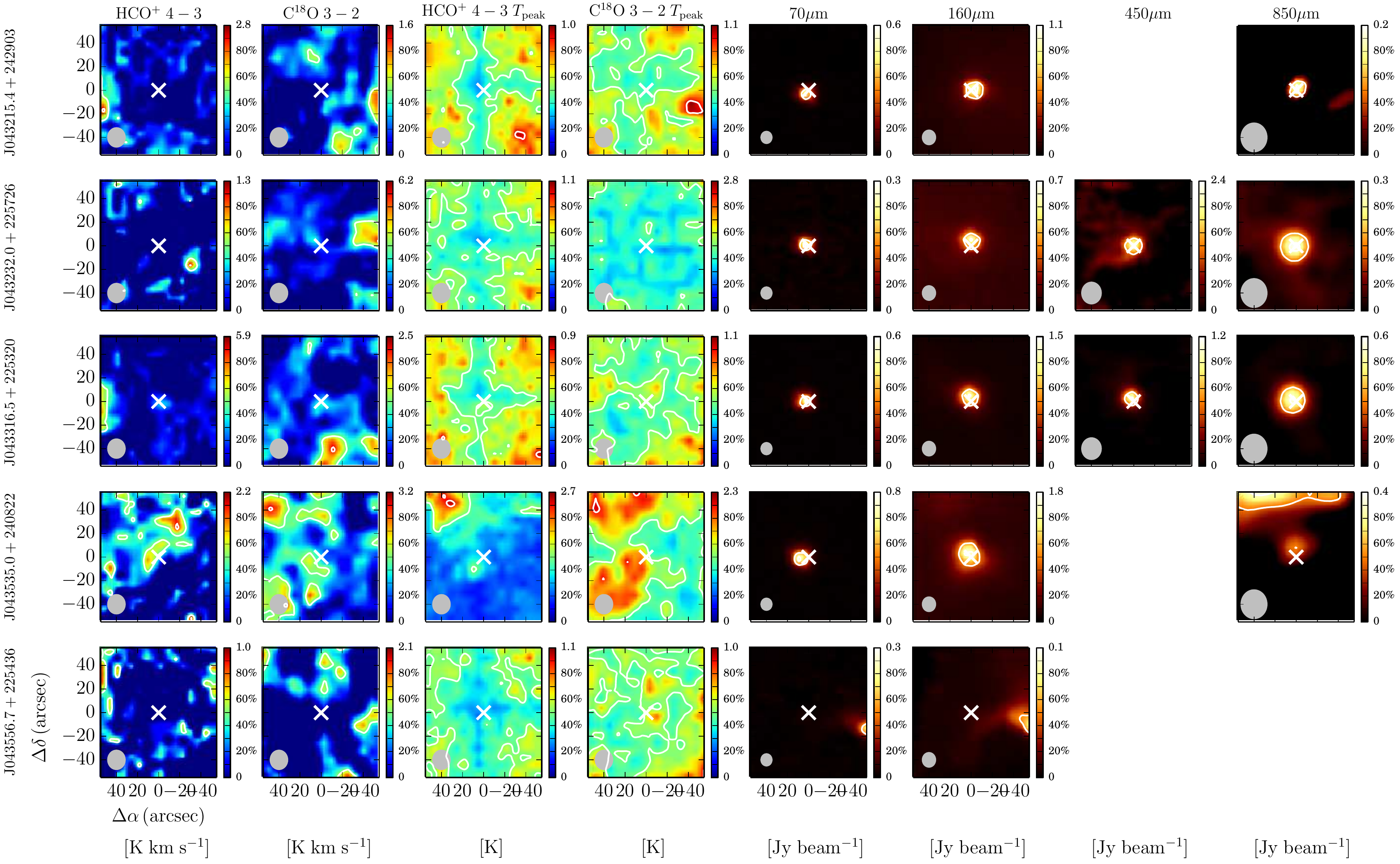}}
      \caption{Taurus sources. See Figure~\ref{fig:ApC1} for caption.}
      \label{fig:ApC11}
    \end{sidewaysfigure*}
    
    \clearpage
    
    \begin{sidewaysfigure*}
      \centering
      \resizebox*{\hsize}{!}{\includegraphics{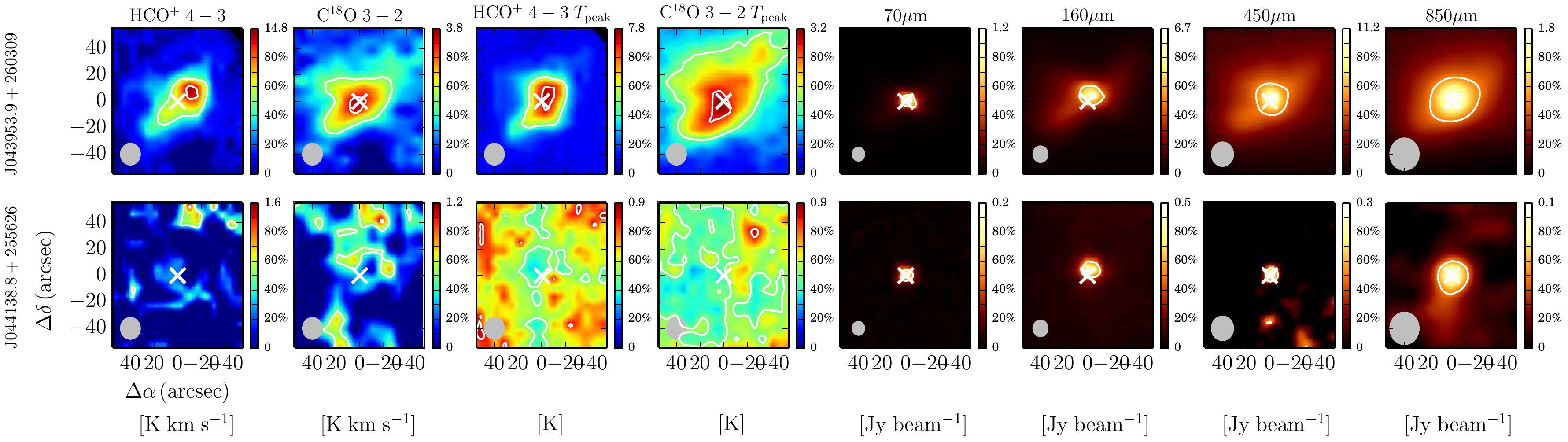}}
      \caption{Taurus sources. See Figure~\ref{fig:ApC1} for caption.}
      \label{fig:ApC12}
    \end{sidewaysfigure*}
    
    \clearpage

    %%%%%%%%%%%%%%%%%%%%%%%%%%%%%%%%%%%%%%%%%%%%%%%%%%%%%%%%%%%%%%%%%%%%%%%%%%%%%%%
    %%% SERPENS MAPS
    
    \begin{sidewaysfigure*}
      \centering
      \resizebox*{14cm}{!}{\includegraphics{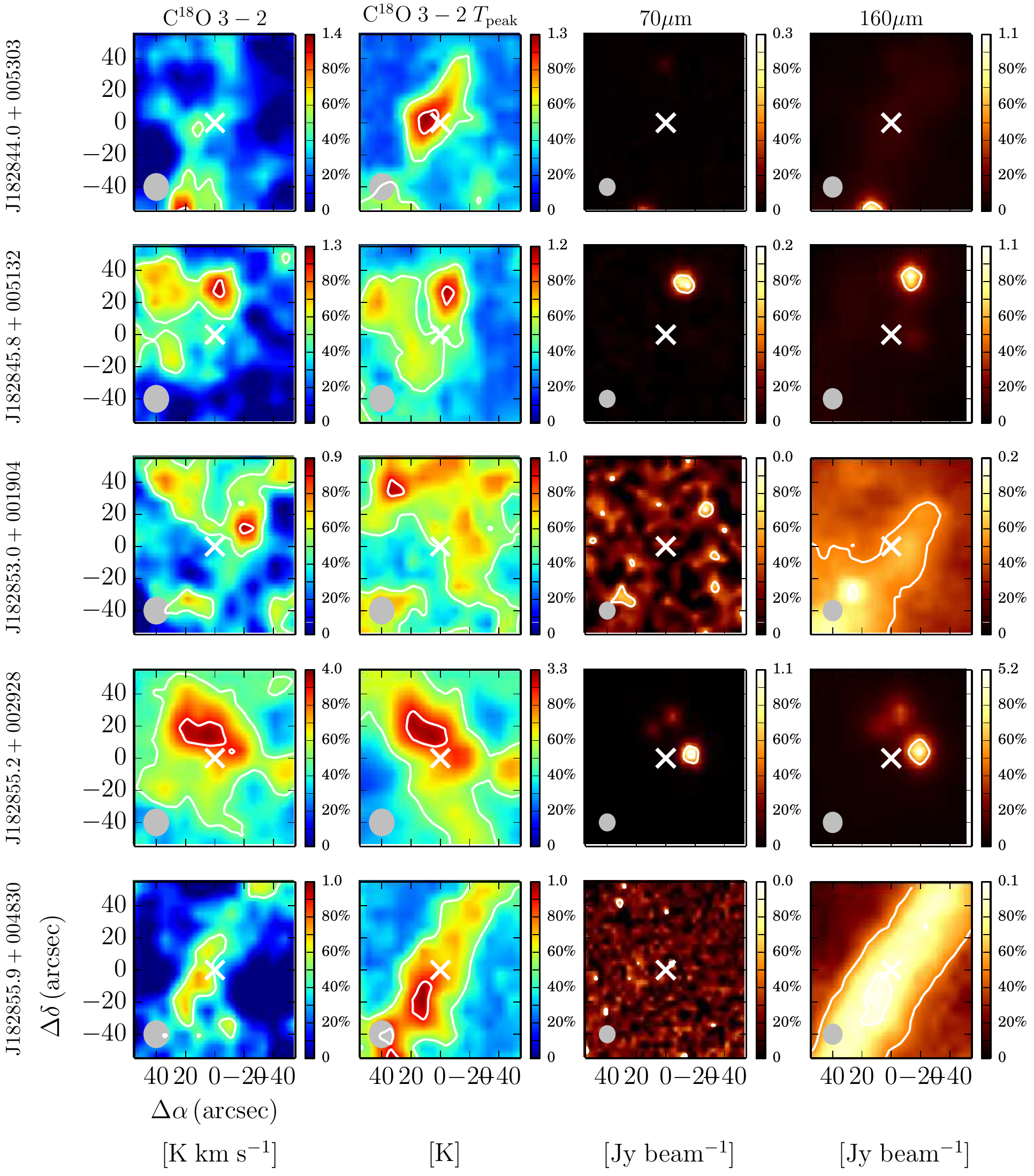}}
      \caption{Serpens sources. See Figure~\ref{fig:ApC1} for caption.}
      \label{fig:ApC13}
    \end{sidewaysfigure*}
    
    \clearpage
    
    \begin{sidewaysfigure*}
      \centering
      \resizebox*{14cm}{!}{\includegraphics{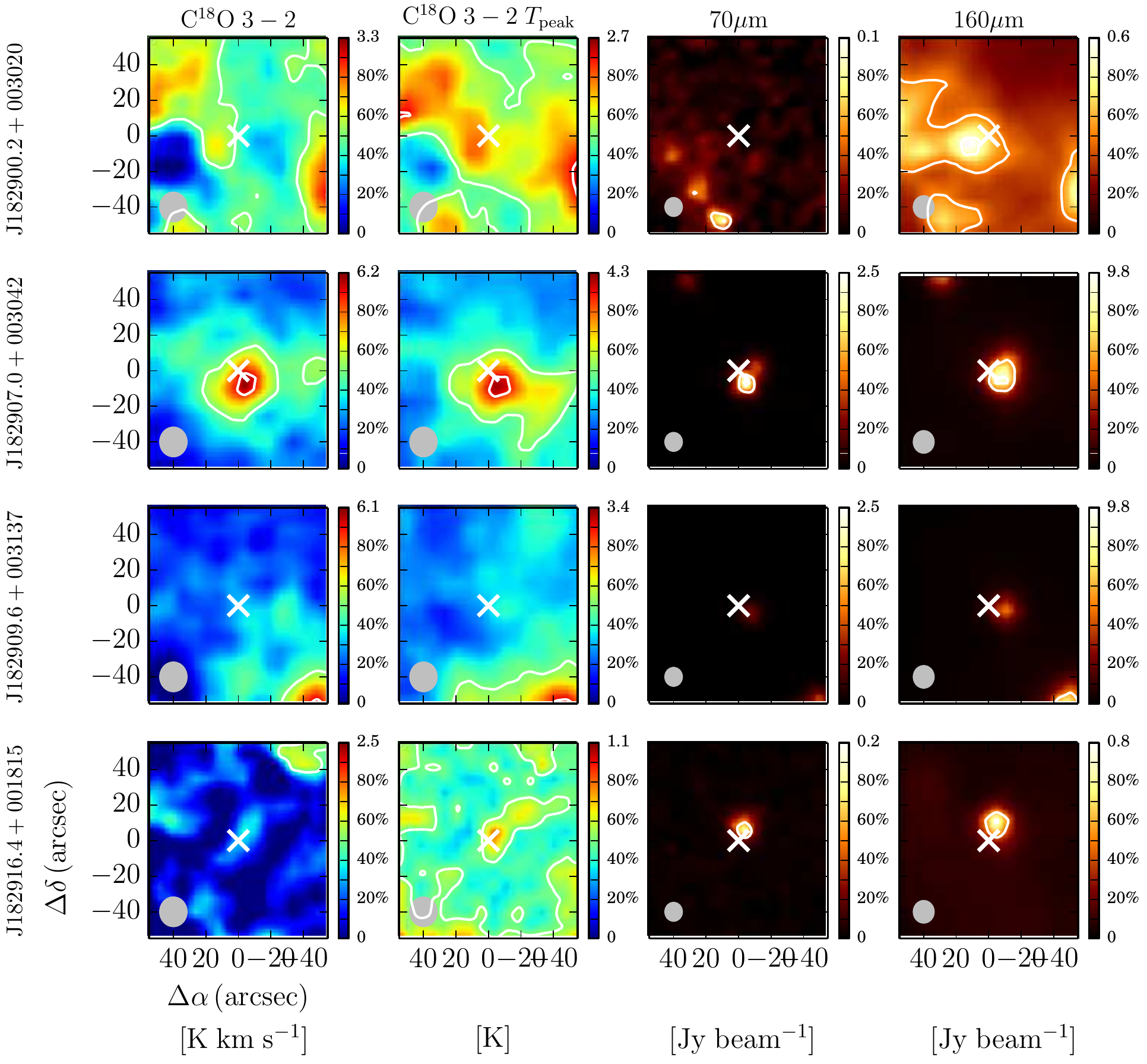}}
      \caption{Serpens sources. See Figure~\ref{fig:ApC1} for caption.}
      \label{fig:ApC14}
    \end{sidewaysfigure*}

\end{appendix}
%-------------------------------------------------------------------

\clearpage

\end{document}